\documentclass{iopart}

\usepackage{cite}
\usepackage{latexsym}
\usepackage{iopams}
\usepackage{amscd}
\usepackage{mathrsfs}
\usepackage{graphicx}
\usepackage{url}
\usepackage{hyperref}

\newtheorem{theorem}{Theorem}[section]
\newtheorem{lemma}[theorem]{Lemma}
\newtheorem{proposition}[theorem]{Proposition}
\newtheorem{corollary}[theorem]{Corollary}
\newtheorem{remark}[theorem]{\bf Remark}
\newtheorem{definition}[theorem]{Definition}

\newcommand{\rank}{{\rm rank}}

\newcommand{\sgn}{{\rm sgn}}
\newcommand{\id}{{\rm id}}

\newcommand{\CC}{\mathbb{C}}
\newcommand{\RR}{\mathbb{R}}
\newcommand{\PP}{\mathbb{P}}

\newcommand{\mb}[1][]{\mathbf}
\newcommand{\m}[1]{\mb{m_{#1}}}
\newcommand{\x}{\mb{x}}

\newcommand{\X}{\mb{X}}

\renewcommand{\bs}[1][]{\boldsymbol}
\newcommand{\bt}{\boldsymbol{\tau}}
\newcommand{\eqn}{\begin{eqnarray}}
\newcommand{\feqn}{\end{eqnarray}}

\renewcommand{\PM}[1]{\langle #1 \rangle}

\begin{document}
\title[Geometry of TDOA maps]{A comprehensive analysis of
the geometry of \\
TDOA maps in localisation problems
\footnote{This is an author-created, un-copyedited version of an article published in Inverse Problems. IOP Publishing Ltd is not responsible for any errors or omissions in this version of the manuscript or any version derived from it. The Version of Record is available online at doi:10.1088/0266-5611/30/3/035004.}}
\author[M. Compagnoni, R. Notari, F. Antonacci, A. Sarti]
{Marco Compagnoni$^\chi$, Roberto Notari$^\chi$
\footnote{M.Compagnoni and R.Notari should be equally
considered as first coauthors of the present work.}\\
Fabio Antonacci$^\varsigma$, Augusto Sarti$^\varsigma$}

\address{$^{\chi}$ Dipartimento di Matematica,\\
$^{\varsigma}$ Dipartimento di Elettronica, Informazione e
Bioingegneria,\\
Politecnico di Milano, Piazza L. Da Vinci 32, I-20133 Milano, Italia}
\eads{\mailto{marco.compagnoni@polimi.it}, \mailto{roberto.notari@polimi.it},
\mailto{fabio.antonacci@polimi.it}, \mailto{augusto.sarti@polimi.it}}

\begin{abstract}
In this manuscript we consider the well-established problem of
TDOA-based source localization and propose a comprehensive
analysis of its solutions for arbitrary sensor measurements and
placements.
More specifically, we define the TDOA map from the physical
space of source locations to the space of range measurements
(TDOAs), in the
specific case of three receivers in 2D space. We then study the identifiability of the model, giving a complete analytical characterization of the image of this map and its invertibility.
This analysis has been conducted in a completely mathematical fashion, using many different tools which make it valid for every sensor configuration. These results are the first step towards the solution of more general problems involving,
for example, a larger number of sensors, uncertainty in their placement, or lack of synchronization.
\end{abstract}

\section{Introduction}
The localization of radiant sources based on a spatial distribution of
sensors has been an important research topic for the past
two decades, particularly in the area of space-time audio
processing. Among the many solutions that are available
in the literature, those based on Time Differences Of Arrival
(TDOAs) between distinct sensors of a signal emitted by the source are the most widespread and popular. Such solutions,
in fact, are characterized by a certain flexibility, a reasonably
modest computational cost with respect to other solutions
and a certain robustness against noise. Popular TDOA-based
solutions are
\cite{Beck2008, Abel1987,Do2007,Foy1976,Gillette2008a,
Huang2004,Huang2000,Huang2001,Militello2007,
Pourmohammad2010,Reddi1993,Schau1987,Schmidt1972,
Teng2010,Yu2004,Zannini2010, Antonacci2006}.

Let us consider the problem of planar source localization
in a homogeneous medium with negligible reverberation.
From elementary geometry, the locus of putative source locations that are compatible with a TDOA measurement between two sensors in
positions $\mathbf{m}_i$ and $\mathbf{m}_j$ is one
branch of a hyperbola of foci $\mathbf{m}_i$ and
$\mathbf{m}_j$, whose aperture depends on the range
difference (TDOA $\times$ speed of sound).
A single TDOA measurement is, therefore, not sufficient
for localizing a source, but it narrows down the set of
locations that are compatible with that measurement by
reducing its dimensionality.

Multiple measurements do
enable localization but measurement errors cause the
corresponding hyperbola branches not to meet at a single
point, thus ruling out simple geometric intersection as
a solution to the localization problem \cite{Canclini2013}.
This is why research has focused on techniques that are
aimed at overcoming this problem while achieving
robustness. Examples are Maximum Likelihood (ML)
\cite{Chen2002,Foy1976,Torrieri1984}; Least Squares (LS)
\cite{Abel1987}; and Constrained Least Squares (CLS)
\cite{Huang2004a,Schau1987}, which offer accurate results
for the most common configurations of sensors.

There are many situations, however, in which it is necessary
to minimize the number of sensors in use, due to specific
sensor placement constraints, or cost limitations. In these
cases it becomes important to assess how the solutions to
the localization problem ``behave" (and how many there are)
as the measurements or the sensor geometry vary.
This problem has been partially addressed in the case of the localization of a radio-beacon receiver in LORAN navigation systems \cite{Schmidt1972} and
in the context of the Global Positioning System (GPS), where measurements
are of Time Of Arrivals (TOAs) instead of TDOAs (see \cite{Bancroft1985,Kraus1987,Abel1991,Chauffe1994,Leva1995,
Hoshen1996,Grafarend2002,Awange2002,Coll2009,Coll2012}). In particular, these studies provide the solution for the case of planar (2D) source localization with three receivers (i.e. with two TDOAs) and they recognize the possibility of dual solutions in some instances, as two different source positions could correspond to the same pair of TDOA measurements.

Recently, in \cite{Spencer2007} the author focused on the assessment of the ill-posedness of the localization problem in the case of 2D minimal sensor configurations, i.e. on quantify how changes in the measurements propagate onto changes in the estimated source location. In particular, in the same quoted manuscript it has been introduced the space of TDOA measurements and it has been shown that in this space there exist small regions associated with dual solutions corresponding to large regions in physical space. This assessment, however, is performed in a simulative fashion and for one specific sensor geometry, and it would be important to extend its generality further.

What we propose in this manuscript is a generalization of the
discussion contained in \cite{Spencer2007} based on a fully
analytical and mathematically rigorous approach. We encode the TDOA localization problem into a map, called the TDOA map, from the space of source locations to the space of TDOA measurements and we offer a complete characterization
of such a map. Not only it is our goal to analytically derive results shown in
\cite{Spencer2007} (irrespective of the geometry of the acquisition
system), but also to complete the characterization of the TDOA
map by analyzing the properties of its image and pre-image,
finding closed-form expressions for the boundaries of the regions
of interest. We observe that this approach to the problem fits into the research field of structural identifiability of complex systems (see for example \cite{Bellman1970,Miao2011}), where one is interested in studying if the parameters of a model (in our case, the coordinates of the source) can be fully retrieved from the experimental data. A similar analysis of the source localization problem has been proposed and investigated very recently also in \cite{Alameda2013,Cheng2013}, the latter in the context of the TOA--based target tracking.

We believe that characterizing the TDOA map to its fullest extent,
even in the simplest case of three calibrated and synchronous
sensors, is a necessary step for developing new mathematical
tools for a wide range of more general problems. One immediate
consequence of this gained knowledge is the possibility to
study how to optimize sensor placement in terms of robustness
against noise or measuring errors. More importantly, this study
paves the way to new venues of research. For example, it enables
the statistical analysis of error propagation in TDOA-based
localization problems; and it allows us to approach more
complex scenarios where the uncertainty lies with sensor
synchronization or spatial sensor placement. This prospective
investigation, in fact, is in line with the recently revamped
interest of the research community in self-calibrating and
self-synchronizing spatial distributions of sensors
\cite{Chen2002,Raykar2005,Redondi2009}.

Our analysis starts from \cite{Compagnoni2012}, where a
different perspective on the localization problem is offered
through the adoption of the Space--Range Differences
(SRD) reference frame, where the wavefront propagation
is described by a (propagation) cone whose vertex lies on
the source location.
As range difference measurements (TDOA $\times$
propagation speed) are bound to lie on the surface of the
propagation cone, localizing a source in the SRD space
corresponds to finding the vertex of the cone that best fits
the measured data. The SRD reference frame is also used
in \cite{Bestagini2013} to offer geometric interpretations
to the underlying principles behind the most common
TDOA-based localization solutions.
Although not explicitly claimed, the localization problem is
described in \cite{Compagnoni2012,Bestagini2013} in terms
of null surfaces and planes in the 3D Minkowski space. This
suggests us that exterior algebra can give us powerful tools
for approaching our problem as well. We therefore begin our
analysis by showing how the SRD reference frame can be
better represented within the framework of exterior algebra,
and we show how the newly gained tools allow us to derive
a global analytical characterization of the TDOA map.
Working with exterior algebra in the Minkowski space is not
unheard of in the literature of space-time signal processing.
In \cite{Coll2009,Coll2012}, for example, this representation
is used for approaching source localization in the GPS context.

\begin{figure}[htb]
\begin{center}
\resizebox{7cm}{!}{
  \includegraphics
  {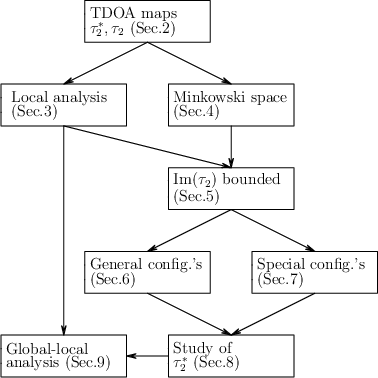}}
  \caption{\label{work_organization}Organization of the manuscript.}
\end{center}
\end{figure}

The manuscript is organized as shown in
Fig. \ref{work_organization}.
Section 2 introduces the concept of TDOA map. Two are
the TDOA maps defined: $\bs{\tau_2}$, where the TDOAs
are referred to a common reference microphone; and
$\bs{\tau_2^*}$, which considers the TDOAs between
all the pairs of microphones. The two maps are, in fact,
equivalent in absence of measurement errors.
This is why most of the techniques in the literature work
with $\bs{\tau_2}$. However, in the presence of measurement
noise, adopting $\bs{\tau_2^*}$ helps gain robustness.
For this reason we decided to consider both $\bs{\tau_2}$
and $\bs{\tau_2^*}.$ In order to introduce our mathematical
formalisms with some progression, in the first part of the
manuscript our analysis will concern $\bs{\tau_2}$.
Section 3 focuses on the local analysis of the TDOA map
$\bs{\tau_2}$. In practice, we show what can be accomplished
using ``conventional" analysis tools (analysis of the Jacobian
matrix). This analysis represents the first step towards the
study of the invertibility of $\bs{\tau_2}$.
In Section 4 we move forward with our representation by
defining the TDOA mapping in the Space - Range Difference
(SRD) reference frame. This is where we show that the
Minkowski space is the most natural representation for a
mapping that ``lives" in the SRD reference frame.
Section 5 describes the early properties of $\bs{\tau_2}$,
with particular emphasis on the fact that its image is
contained in a compact polygonal region.
Section 6 offers a complete description of the mapping
$\bs{\tau_2}$ for the case of non-aligned microphones.
In particular, Subsection 6.1 shows that the preimage
(inverse image) of $\bs{\tau_2}$ can be described in
terms of the non-negative roots of a degree-2 equation,
while 6.3 describes $\mathrm{Im}(\bs{\tau})$ and the
cardinality of the pre-image. Finally, Subsection 6.4
shows the pre-image regions in $\bs{\tau_2}$ and the
bifurcation curve $\tilde{E}$ that divides the region
of cardinality 1 from the regions of cardinality 0 or 2.
Similar results are derived for the case of aligned
microphones in Section 7. In Section 8 we use the previous results on $\bs{\tau_2}$ to describes the image and the preimages of the map $\bs{\tau^*_2}$. Section 9 discusses the impact of this work and offers an example aimed at showing
that the global analysis on $\bs{\tau_2}$ (or $\bs{\tau_2^*}$)
gives new insight on the localization problem, which
could not be derived with a local approach. Finally, Section 10 draws some
conclusions and describes possible future research directions that can take advantage of the analysis presented in this manuscript.

In order to keep the manuscript as self-contained as
possible, in Appendix A we give an overview on
exterior algebra on a vector space. For similar reasons, we also included
an introduction to plane algebraic geometry in Appendix B.
These two Sections, of course, can be skipped by the
readers who are already familiar with these topics.
Finally, in Appendix C we included the code for computing
the cartesian equation of the bifurcation curve $\tilde{E}$.


\section{From the physical model to its mathematical description}\label{sec:p2m}

As mentioned above, we focus on the case of coplanar
source and receivers, with synchronized receivers in
known locations and with anechoic and homogenous
propagation.  The physical world can therefore be identified
with the Euclidean plane, here referred to as the $x$--plane.
This choice \cite{Compagnoni2012,Bestagini2013} allows us
to approach the problem with more progression and visualization
effectiveness.

After choosing an orthogonal Cartesian co-ordinate system,
the Euclidean $x$--plane can be identified with $\RR^2$.
On this plane, $\m{i}=(x_{i},y_{i}),\ i=0,1,2$ are the positions
of the microphones and $\mb{x} = (x,y)$ is the position of the
source $S$.  The corresponding displacement vectors are
\begin{equation}
\mb{d_i}(\mb{x})=\mb{x}-\m{i},\qquad
\mb{d_{ji}}=\m{j}-\m{i},\qquad i,j=0,1,2,
\end{equation}
whose moduli are $d_i(\mb{x})$ and $ d_{ji}$, respectively.
Generally speaking, given a vector $\mb{v}$, we denote its
norm $||\mb{v}||$ with $v$ and with $\tilde{\mb{v}}=
\frac{\mb{v}}{v}$ the corresponding unit vector.

Without loss of generality, we assume the speed of propagation
in the medium to be equal to $1$. For each pair of different
microphones, the measured TDOA $\hat{\tau}_{ji}(\x)$ turns out
to be equal to the pseudorange (i.e. the range difference)
\begin{equation}\label{TDOA}
{\tau}_{ji}(\mb{x})=d_{j}(\mb{x})-d_{i}(\mb{x}),\quad i,j=0,1,2,
\end{equation}
plus a measurement error $\epsilon_{ji}:$
\begin{equation}\label{TDOAm}
\hat{\tau}_{ji}(\mb{x})=\tau_{ji}(\x)+\epsilon_{ji},\quad i,j=0,1,2.
\end{equation}
A wavefront originating from a source in $\x$ will produce a
set of measurements
$ (\hat{\tau}_{10}(\mb{x}),\hat{\tau}_{20}(\mb{x}),\hat{\tau}_{21}(\mb{x}))$.
As the measurement noise is a random variable, we are concerning with a stochastic model.
\begin{definition}
The complete TDOA model is
\begin{equation} \label{complete-TDOA-model}
\bs{\hat{\tau}_2^*}(\mb{x})=(\hat{\tau}_{10}(\mb{x}),\hat{\tau}_{20}(\mb{x}),
\hat{\tau}_{21}(\mb{x})).
\end{equation}
The deterministic part of this model is obtained by setting $ \epsilon_{ji} = 0$ in
$\bs{\hat{\tau}_2^*}(\mb{x})$, which gives us the complete TDOA
map:
\begin{equation} \label{complete-TDOA-map}
\begin{array}{cccc}
\bs{\tau_2^*} : & \RR^2 & \to & \RR^3 \\ &
\mb{x} & \to & (\tau_{10}(\mb{x}), \tau_{20}(\mb{x}),\tau_{21}(\mb{x})).
\end{array}
\end{equation}
The target set is referred to as the $\tau^*$--space.
\end{definition}
In this manuscript we approach the deterministic problem,
therefore we only consider the complete TDOA map.
Using the above definition, localization problems can be readily
formulated in terms of $ \bs{\tau_2^*} .$ For example, given a set of measurements, we are interested to know if there exists a source that has produced them, if such a source is unique, and where it is. In a mathematical setting, these questions are equivalent to:
\begin{itemize}
\item given $ \bs{\tau_2^*}\in\RR^3,$ does there exist a source in the
$x$--plane such that $\bs{\tau_2^*}(\mb{x}) = \bs{\tau^*}$, i.e.
$ \bs{\tau^*} \in \mbox{Im}(\bs{\tau_2^*})$?
\item If $\mb{x}$ exists, is it unique, i.e. $ \vert \bs{\tau_2^*}^{-1}(\bt) \vert = 1$?
\item If so, is it possible to find the coordinates of $ \mb{x}$? i.e. given
$\bs{\tau^*}$, can we find the only $\mb{x}$ that solves the equation
$\bs{\tau_2^*}(\mb{x}) = \bs{\tau^*}$?
\end{itemize}
With these problems in mind, we focus on the study of the image of the
TDOA map $ \bs{\tau_2^*}$ and of its global properties. In particular, we
are interested in finding the locus of points where the map becomes
$1$--to--$1$. Moreover, as solving the localization problem consists of finding the
inverse image of $\bs{\tau^*} \in \mbox{Im}(\bs{\tau_2^*})$, we aim at
giving an explicit description of the preimages, also called the fibers, of
 $\bs{\tau_2^*} $.

The complete model $\bs{\hat{\tau}_2^*}(\x)$ takes into account each
one of the three TDOA that can be defined between the sensors. This,
in fact, becomes necessary when working in a realistic (noisy) situation
\cite{Hing2007}. We should keep in mind, however, that there is a linear
relationship between the pseudoranges (\ref{TDOAm}), which allows us
to simplify the deterministic problem.
\begin{definition}
Let $(\tau_{10},\tau_{20},\tau_{21})$ be the coordinates of the
$\tau^*$--space. Then, $\mathcal{H}$ is the plane of equation
$\tau_{10}-\tau_{20}+\tau_{21}=0$.
\end{definition}
\begin{lemma} \label{im-complete-tau}
The image $ \mbox{Im}(\bt_2^*) $ is contained in $\mathcal{H}$.
\end{lemma}
\noindent\emph{Proof.}
For each $\x\in\RR^2$ we have
\begin{equation}\label{tau-dependance}
\tau_{10}(\mb{x}) - \tau_{20}(\mb{x}) + \tau_{21}(\mb{x}) = 0
\end{equation}
from the definition (\ref{TDOA}) of pseudoranges.
\hfill$\square$\vspace{1mm}

In the literature, Lemma \ref{im-complete-tau} is usually presented
by saying that there are only two linearly independent pseudoranges
and $(\tau_{10}(\mb{x}),\tau_{20}(\mb{x}))$, for example, are sufficient
for completely encoding the deterministic TDOA model. This suggests
us to define a reduced version of the above definition:
\begin{definition}
The map from the position of the source in the $x$-plane to the
linearly independent pseudoranges
\begin{equation}
\begin{array}{cccc}
\bs{\tau_2}: & \RR^2          & \longrightarrow & \RR^2\\
 & \mb{x}   & \longrightarrow & \quad
(\tau_{10}(\mb{x}),\tau_{20}(\mb{x}))
\end{array}
\end{equation}
is called the TDOA map. The target set is referred to as the
$\tau$--plane.
\end{definition}
In $\bs{\tau_2}$ we consider only the pseudoranges involving receiver
$\m{0}$, which we call reference microphone.
If $ p_i: \mathcal{H} \to \RR^2 $ is the projection that takes care of
forgetting the $ i$--th coordinate, we have that $\bs{\tau_2^*}$ is
related to $\bs{\tau_2}$ by $ \bs{\tau_2} = p_3 \circ \bs{\tau_2^*}$.
As $p_i$ is clearly $1$--to--$1$, it follows that all the previous questions
about the deterministic localization problem can be equivalently
formulated in terms of $\bs{\tau_2}$ and its image Im($\bs{\tau_2}$) (see Figure \ref{fig:taucomimage} in Section \ref{sec:symmetry} for an example of Im($\bs{\tau_2^*}$) and its projection Im($\bs{\tau_2}$) via $p_3$).
Analogous considerations can be done if we consider $ p_1 \circ \bs{\tau_2^*}$ or $ p_2 \circ \bs{\tau_2^*}$, that is equivalent to choose $ \mb{m_2} $ or $ \mb{m_1} $ as reference point, respectively.

In Sections from \ref{sec:Jacobian} to \ref{sec:special-conf}, we will
focus on the study of $\bs{\tau_2}$ and we will complete the analysis
of $\bs{\tau_2^*}$ in Section \ref{sec:symmetry}. For reasons of notational
simplicity, when we study the map $\bs{\tau_2}$ we will drop the second
subscript and simply write $\tau_h(\mb{x}) = \tau_{h0}(\mb{x})$, $h=1,2$.
Moreover, as we focus on the deterministic model, in the rest of the manuscript
we will interchangeably use the terms pseudorange and TDOA.


\section{Local analysis of $ \bs{\tau_2} $}
\label{sec:Jacobian}

In this Section, we present a local analysis of the TDOA map
$ \bs{\tau_2} $. From a mathematical standpoint, this is the first
natural step towards studying of the invertibility of $\bs{\tau_2}$.
In fact, as stated by the Inverse Function Theorem, if the Jacobian
matrix $J(\mb{x})$ of $\bs{\tau_2}$ is invertible in $\mb{x}$, then
$\bs{\tau_2}$ is invertible in a neighborhood of $\mb{x}$.
Studying the invertibility of a map through linearization (i.e. studying
its Jacobian matrix) is a classical choice when investigating the
properties of a complex (non--linear) model. In the case
of acoustic source localization, for example,
\cite{Raykar2005,Compagnoni2012} adopt this method to study
the accuracy of various statistical estimators for the TDOA model.
As a byproduct of our study, at the end of the section we will discuss
how the accuracy in a noisy scenario is strictly related to the
existence of the so-called \emph{degeneracy locus}, which is the
locus where the rank of $ J(\mb{x}) $ drops.

The component functions $ \tau_i(\mb{x}) $ of $ \bs{\tau_2} $ are
differentiable in $\RR^2 \setminus \{\m{0}, \m{1}, \m{2} \}$,
therefore so is $ \bs{\tau_2}.$  The $i$--th
row of $J(\mb{x})$ is the gradient $\nabla\tau_i(\mb{x})$, i.e.
\begin{equation} \label{grad}\fl\qquad\qquad
\nabla\tau_i(\mb{x}) = \left( \frac{x-x_{i}}{d_i(\mb{x})}
-\frac{x-x_{0}}{d_0(\mb{x})}\;, \;\frac{y-y_{i}}{d_i(\mb{x})}
-\frac{y-y_{0}}{d_0(\mb{x})} \right) =
\mb{\tilde{d}_i}(\mb{x})-\mb{\tilde{d}_0}(\mb{x}).
\end{equation}
\begin{definition}\label{rette}
Let us assume that $ \m{0}, \m{1}, \m{2} $ are not collinear.
Let $r_0,r_1,r_2$ be the lines that pass through two
of such three points, in compliance with the notation
$\m{i}\notin r_i$, $i=0,1,2$.
Let us split each line in three parts as $r_0 = r_0^- \cup r_0^0
\cup r_0^+ $, where $r_0^0$ is the segment with endpoints
$\m{1}$ and $\m{2}$, $r_0^-$ is the half--line originating from $\m{2}$
and not containing $\m{1}$, and $r_0^+$ is the half--line originating
from $ \m{1} $ and not containing $\m{2}$. Similar splittings are done
for $r_1, r_2$, with $r_1^+, r_2^+$ having $\m{0}$ as
endpoint.

Let us now assume that $\m{0},\m{1},\m{2}$ belong to the line $r$. Then, $r^0$ is the smallest segment containing all three points and $r^c$ is its complement in $r$.
\end{definition}
\begin{figure}[htb]
\resizebox{5.7cm}{!}{
  \includegraphics
  {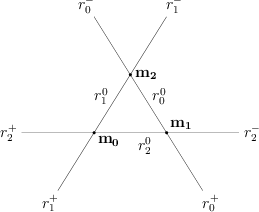}}
\hspace{10mm}
\resizebox{6cm}{!}{
  \includegraphics
  {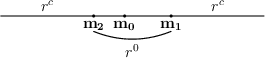}}
  \caption{\label{rette-acustica}A general and a collinear configuration of the microphones $\mathbf{m}_i,\;i=0,1,2$. Left-hand side: line $r_i^+,\;r_i^-$ and $r_i^0$ refer to the portions of the line joining the microphones $j$ and $k,\;j,k\neq i$. }
\end{figure}

\begin{theorem} \label{n=2} Let $ J(\mb{x}) $ be the Jacobian matrix
of $ \bs{\tau_2} $ at $ \mb{x} \not= \m{0}, \m{1}, \m{2}.$ Then,
\begin{enumerate}
\item if $ \m{0}, \m{1}, \m{2} $ are not collinear, then
$$ \rank(J(\mb{x})) = \left\{ \begin{array}{cl} 1 & \mbox{if }
\mb{x} \in \left( \cup_{i=0}^2 (r_i^- \cup r_i^+) \right),
\\ 2 & \mbox{otherwise};
\end{array} \right. $$
\item if $ \m{0}, \m{1}, \m{2} $ are collinear, then
$$ \rank(J(\mb{x})) = \left\{ \begin{array}{cl} 0 & \mbox{if }
\mb{x} \in r^c, \\ 1 & \mbox{if } \mb{x} \in r^0, \\ 2 &
\mbox{otherwise.} \end{array} \right. $$
\end{enumerate}
\end{theorem}

\noindent\emph{Proof.} Assume $ \mb{x} \not= \m{i},$ for
$ i=0,1,2.$ As explained in Section \ref{sec:p2m}, the
$x$--plane is equipped with the Euclidean inner product,
therefore we can use the machinery of Appendix A. As
claimed in Proposition \ref{hodge-2-dim}, $\ast(\det(J(\mb{x})))
= \nabla\tau_1(\mb{x}) \wedge \nabla\tau_2(\mb{x})$.
Hence, we work in the exterior algebra of the $ 2$--forms.
From eq. (\ref{grad}) and the general properties of $2$--forms,
we obtain
\begin{equation}\label{2forme}
\eqalign{
\nabla\tau_1(\mb{x}) \wedge \nabla\tau_2(\mb{x})
= (\mb{\tilde{d}_1}(\mb{x}) -\mb{\tilde{d}_0}(\mb{x})) \wedge
(\mb{\tilde{d}_2}(\mb{x}) -\mb{\tilde{d}_0}(\mb{x})) = \cr
= \mb{\tilde{d}_1}(\mb{x}) \wedge \mb{\tilde{d}_2}(\mb{x})-
\mb{\tilde{d}_0}(\mb{x}) \wedge \mb{\tilde{d}_2}(\mb{x}) -
\mb{\tilde{d}_1}(\mb{x}) \wedge\mb{\tilde{d}_0}(\mb{x}).}
\end{equation}

Let us first assume that $ \mb{\tilde{d}_1}(\mb{x}),
\mb{\tilde{d}_2}(\mb{x}) $ are linearly independent or,
equivalently, that $ \mb{x} \notin r_{0}.$ In this case there
exist $ a_1, a_2 \in \RR $ such that
$$ \mb{\tilde{d}_0}(\mb{x}) = a_1 \mb{\tilde{d}_1}(\mb{x}) + a_2
\mb{\tilde{d}_2}(\mb{x}).$$ After simplifying equation (\ref{2forme}), we get
\begin{equation}\label{eq:Jacobianot2}
\det(J(\mb{x})) = \nabla\tau_1(\mb{x}) \wedge
\nabla\tau_2(\mb{x}) = (-a_1 - a_2 + 1)\, \mb{\tilde{d}_1}(\mb{x})
\wedge \mb{\tilde{d}_2}(\mb{x}),
\end{equation}
therefore $\det(J(\mb{x})) = 0$ if, and only if, $a_1 + a_2 = 1$,
because the linear independence of $ \mb{\tilde{d}_1}(\mb{x}),
\mb{\tilde{d}_2}(\mb{x}) $ implies $ \mb{\tilde{d}_1}(\mb{x})
\wedge \mb{\tilde{d}_2}(\mb{x}) \not= 0$.
Furthermore, from $ \mb{\tilde{d}_0}(\mb{x}) = a_1
\mb{\tilde{d}_1}(\mb{x}) + a_2 \mb{\tilde{d}_2}(\mb{x}),$ we obtain
$$ 1 = \| \mb{\tilde{d}_0}(\mb{x}) \|^2 = a^2_1 + a^2_2 + 2 a_1 a_2
\mb{\tilde{d}_1}(\mb{x}) \cdot \mb{\tilde{d}_2}(\mb{x}).$$
After simple calculations, the previous equality becomes $$ 2 a_1 a_2 (
\mb{\tilde{d}_1}(\mb{x}) \cdot \mb{\tilde{d}_2}(\mb{x}) - 1 ) = 0 \; ,
$$ therefore either $ a_1 = 0 $ or $ a_2 = 0,$ because the third factor
is different from zero.
If $ a_1 = 0$, then $a_2 = 1 $ and
$ \mb{\tilde{d}_0}(\mb{x}) = \mb{\tilde{d}_2}(\mb{x})$, i.e. $ \mb{x} \in r_{1}^+ \cup r_{1}^-.$ Otherwise, if $ a_2 = 0$, then $a_2 = 1 $ and $
\mb{\tilde{d}_0}(\mb{x}) = \mb{\tilde{d}_1}(\mb{x}),$ i.e. $
\mb{x} \in r_{2}^+ \cup r_{2}^-.$

On the other hand, if $ \mb{x} \in r_{0},$ then $ \mb{\tilde{d}_1}(\mb{x}) =
\mb{\tilde{d}_2}(\mb{x}) $ if $ \mb{x} \in r_{0}^+ \cup r_{0}^-,$
and $ \mb{\tilde{d}_1}(\mb{x}) = -\mb{\tilde{d}_2}(\mb{x}) $ if $
\mb{x} \in r_{0}^0.$ Therefore, the equality (\ref{2forme}) becomes
$$ \nabla\tau_1(\mb{x}) \wedge\nabla\tau_2(\mb{x}) =
\left\{ \begin{array}{cl}
0 & \mbox{ if } \mb{x} \in r_{0}^+ \cup r_{0}^-, \\
-2 \mb{\tilde{d}_0}(\mb{x}) \wedge \mb{\tilde{d}_2}(\mb{x}) &
\mbox{ if } \mb{x} \in r_{0}^0.
\end{array} \right.$$
In conclusion, if $ \m{0}, \m{1},\m{2} $ are not collinear, then
$\det(J(\mb{x})) = 0 $ for each $\mb{x} \in \cup_{i=0}^2 (r_{i}^+
\cup r_{i}^-),$ proving the first claim. If, on the other hand,
$ \m{0}, \m{1}, \m{2} $ lie on the line $r$, then $\det(J(\mb{x}))=0$
for all $\mb{x} \in r$. Furthermore, $\mb{\tilde{d}_0}(\mb{x}) =
\mb{\tilde{d}_1}(\mb{x}) = \mb{\tilde{d}_2}(\mb{x}) $ if and only if
$\mb{x} \in r^c $, therefore $\nabla\tau_1(\mb{x}) =
\nabla\tau_2(\mb{x}) = (0,0),$ i.e. $ J(\mb{x})$ is the null matrix.
\hfill$\square$\vspace{1mm}

Theorem \ref{n=2} has an interesting geometric interpretation.

\begin{definition}\label{HA} Let $ \tau \in \RR$. The set
\begin{equation}
A_{i}(\tau)= \{ \mb{x} \in \RR^2 | \, \tau_{i}(\mb{x}) = \tau \}
\end{equation}
is the level set of $\tau_{i}(\mb{x})$ in the $x$--plane.
\end{definition}

\begin{lemma} \label{LemmaTriang} If $ \vert \tau \vert > d_{i0},$
then $ A_{i}(\tau) = \emptyset.$ Moreover, if $ 0 < \vert \tau
\vert < d_{i0},$ then $ A_{i}(\tau) $ is the branch of hyperbola
with foci $ \m{0},\m{i} $ and parameter $ \tau,$ while
$$
A_{i}(\tau) =
\cases{
r_{j}^+ & if $\tau = d_{i0}$,\\
r_{j}^- & if $\tau = -d_{i0}$,\\
a_{j} & if $\tau = 0$,\\}
$$
where $ j \not= i, \{ i, j \} = \{ 1, 2 \},$ and $ a_{j} $ is the
line that bisects the line segment $ r_{j}^0.$
\end{lemma}

\noindent\emph{Proof.} By definition, we have $ \tau_{i}(\mb{x}) =
d_i(\mb{x}) - d_0(\mb{x})$, therefore the first claim follows from
the classical inequalities between the sides of the triangle of
vertices $ \mb{x}, \m{i}, \m{0}$. The second claim follows from
a classical result: given any hyperbola with foci $\m{i},\m{0}$ and
parameter $c \in \RR^+,$ the two branches are defined by either
one of the two equations
$$d_i(\mb{x}) - d_0(\mb{x}) = c \qquad \mbox{ and } \qquad
d_i(\mb{x}) - d_0(\mb{x}) = -c \quad. $$
The last claim is a straightforward computation.
\hfill$\square$

\noindent Fig. \ref{curve-livello}(a) shows the hyperbola branches
with foci $ \mb{m_0}, \mb{m_i}$. By definition of level set, each
point in the domain of $ \tau_{i} $ lies on exactly one branch
$A_{i}(\tau) $ for some $ \tau \in [-d_{i0}, d_{i0}] $  (by abuse of
notation, we consider $ A_{i}(0), A_{i}(\pm d_{i0})$ as branches of
hyperbolas as well).
This means that, given $\bt=(\tau_1,\tau_2)$, the source is
identified as the intersection points $A_1(\tau_1)\cap A_2(\tau_2)$.
As a direct consequence, the quality of the localization depends on
the type of intersection: in a noisy scenario, an error on the
measurements $\bt$ changes the shape of the related hyperbolas,
therefore the localization accuracy is strictly related to the
incidence angle between the hyperbolas branches (see
\cite{Bestagini2013} for a similar analysis of the localization
problem).

\noindent{\bf Notation:} We denote the tangent line to a curve
$C$ at a smooth point $\x\in C$ as $T_{\x,C}$.
\begin{remark}\rm\label{rem-trivial}
$(1)$ $ \nabla\tau_i(\mb{x}) = \mb{0} $ if, and only if,
$ \mb{x} \in r_{j}^+ \cup r_{j}^-,$ with $ j \not= 0, i.$
In fact, $ \nabla\tau_i(\mb{x}) =
\mb{0} $ is equivalent to $ \mb{\tilde{d_i}(x)} =
\mb{\tilde{d_0}(x)},$ i.e. $ \mb{x} \in r_{j}^+ \cup
r_{j}^-$. Hence $A_{i}(\pm d_{i0}) $ is nowhere smooth.

\noindent $(2)$ Assume that $\mb{x}\notin r_{j}^+\cup r_{j}^-$.
Then, it is well-known that $ \nabla\tau_i(\mb{x}) $ is orthogonal to
the line $ T_{\mb{x},A_i(\tau_i)}$ and that it bisects the angle
$\widehat{\m{0}\x\m{i}}$, where $ \m{0},\m{i} $ are the foci of the
hyperbola. Consequently, the tangent line is parallel to the vector
$\mb{\tilde d_i(x) + \tilde d_0(x)}$ and, quite clearly,
$\nabla\tau_i(\mb{x}) = \mb{\tilde d_i(x) - \tilde d_0(x)} $ is
orthogonal to the previous vector (as we can see in
Fig. \ref{curve-livello}(b), if we draw the unit vectors
$\mb{\tilde{d_i}}(\mb{x})$ and $\mb{\tilde{d_0}}(\mb{x})$,
their sum lies on the tangent line $T_{\mb{x},A_i(\tau_i)}$
while their difference is the gradient $\nabla\tau_i(\mb{x})$).

\begin{figure}[htb]
\begin{center}
\resizebox{7.5cm}{!}{
  \includegraphics
  {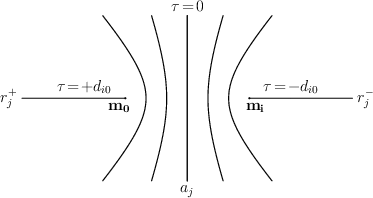}}
  \hspace{15mm}
\resizebox{2.5cm}{!}{
  \includegraphics
  {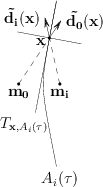}}\\ \phantom{a} \hskip2cm $(a)$ \hskip5.5cm $(b)$
  \caption{\label{curve-livello}$(a)$ Level sets $ A_i(\tau)$; $(b)$ Gradient and tangent line $T_{\mb{x}, A_i(\tau)}$.}
\end{center}
\end{figure}
\end{remark}

\begin{proposition} \label{n=2-geom}
Let $ \mb{x} \in A_1(\tau_1) \cap A_2(\tau_2).$ Then,
\begin{enumerate}
\item if $ \m{0},\m{1},\m{2} $ are not collinear, then $ T_{\mb{x}, A_1(\tau_1)} \not= T_{\mb{x}, A_2(\tau_2)},$ or equivalently, $ A_1(\tau_1) $
and $ A_2(\tau_2) $ meet transversally at $ \mb{x} $ if, and only
if, $ \mb{x} \in \RR^2 \setminus \{ \cup_{i=0}^2 (r_{i}^+ \cup
r_{i}^-) \};$
\item if $ \m{0},\m{1},\m{2} $ lie on $r$,
then $A_1(\tau_1)\cap A_2(\tau_2)$ is finite if, and only
if, $ \mb{x} \in \RR^2 \setminus r^c$. Furthermore $A_1(\tau_1)$
and $A_2(\tau_2) $ meet transversally at $\mb{x}$ if, and only if,
$\mb{x} \in \RR^2 \setminus r$.
\end{enumerate}
\end{proposition}

\noindent\emph{Proof.} The loci $ A_1(\tau_1) $ and $ A_2(\tau_2) $
meet transversally at $ \mb{x},$ i.e. $ T_{\mb{x}, A_1(\tau_1)}
\not= T_{\mb{x}, A_2(\tau_2)}$ if, and only if, $ \nabla\tau_1(\mb{x}) $
and $ \nabla\tau_2(\mb{x}) $ are linearly independent.
That last condition is equivalent to $ \det(J(\mb{x}))\not= 0$.
The claim concerning transversal intersection is therefore
equivalent to Theorem \ref{n=2}. Finally, if $ \mb{x} \in r^c$,
then either $ A_1(\tau_1) \subset A_2(\tau_2) $ or $ A_2(\tau_2)
\subset A_1(\tau_1).$
\hfill$\square$

\begin{figure}[htb]
\begin{center}
\resizebox{6cm}{!}{
  \includegraphics
  {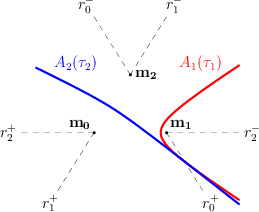}}
\end{center}
\caption{\label{rette-acusticav2}The hyperbola branches intersect
tangentially on the degeneracy locus. This configuration can lead to poor
localization accuracy when TDOAs are affected by measurement errors.}
\end{figure}

In Fig. \ref{rette-acusticav2} we showed a case of tangential
intersection of $A_1(\tau_1)$ and $ A_2(\tau_2)$. From
Proposition \ref{n=2-geom}, we gather new insight on source
localization in realistic scenarios. The above discussion, in fact,
allows us to predict the existence of unavoidable poor localization
regions centered on each half--line forming the degeneracy locus.
We will return on this topic in Section \ref{sec:Impact}.


\section{The $ 3$--dimensional Minkowski space}\label{sec:Minkowski}

As discussed in Section \ref{sec:Jacobian}, TDOA--based localization
is mathematically equivalent to computing the intersection points of
some hyperbola branches. This can be treated as an algebraic problem
in the $x$--plane by simply considering the full hyperbolas.
In this case, however, it is not easy to manipulate the system of two
quadratic equations and remain in full control of all the intersection
points. In particular, there could appear extra (both real and complex)
intersection points with no meaning for the problem, and there is no
systematic way to select the ones that are actually related to the
localization.

\begin{figure}[htb]
\begin{center}
\resizebox{11.2cm}{!}{
  \includegraphics
  {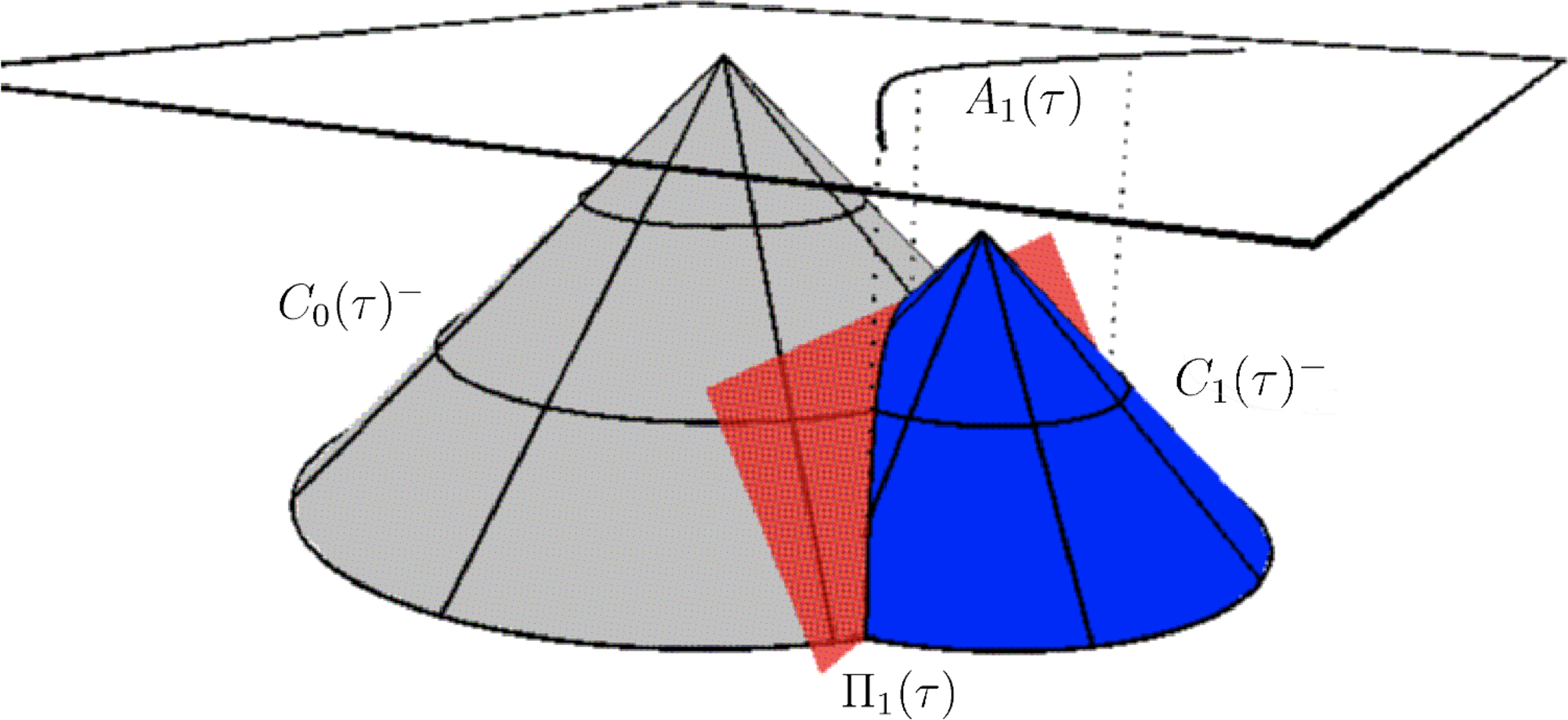}}\\
  \caption{\label{fig:ConiPianoIperbole}The intersection of the two negative half--cones $ C_0(\bs{\tau})^- $ and $ C_1(\bs{\tau})^- $ is a curve contained in the plane $ \Pi_1(\bs{\tau}).$ The curve projects onto the hyperbola branch $ A_1(\bs{\tau}) $ in the $ x$--plane.}
\end{center}
\end{figure}

In order to overcome such difficulties, we manipulate the equations
that define the level sets $A_i(\tau_i)$ (see Def. \ref{HA}), to obtain
an equivalent, partially linear, problem in a 3D space (see
\cite{Bestagini2013} for an introduction on the topic). In order to find the points in
$A_1(\tau_1)\cap A_2(\tau_2)$, we need to solve
the system
$$
\left\{
\begin{array}{l}
\tau_{1}=d_1(\x)-d_0(\x),\\
\tau_{2}=d_2(\x)-d_0(\x).
\end{array}\right.
$$
We introduce a third auxiliary variable $\tau$, and rewrite it as
$$
\left\{
\begin{array}{l}
\tau_{1}-\tau=d_1(\x),\\
\tau_{2}-\tau=d_2(\x),\\
\tau=-d_0(\x).
\end{array}\right.
$$
Again, this is not an algebraic problem, because of the presence of
Euclidean distances. However, by squaring both sides of the
equations, we obtain the polynomial system
$$
\left\{
\begin{array}{l}
(\tau_{1}-\tau)^2=d_1(\x)^2,\\
(\tau_{2}-\tau)^2=d_2(\x)^2,\\
\tau^2=d_0(\x)^2.
\end{array}\right.
$$
In geometric terms, this corresponds to studying the intersection of
three cones in the 3D space described by the triplets $ (x,y,\tau)$.
As described in \cite{Compagnoni2012, Bestagini2013} this problem
representation is given in the space--range reference frame.
For the given TDOA measurements $(\tau_1,\tau_2)$, a solution
$(\bar x,\bar y,\bar \tau) $ of the system gives an admissible position
$(\bar x,\bar y)$ of the source in the $x$--plane and the corresponding
time of emission $\bar \tau$ of the signal, with respect to the time of
arrival at the reference microphone $\m{0}$.
We are actually only interested in the solutions with
$\bar\tau\leq \rm{min}(\tau_1,\tau_2,0)$, i.e. in the points that lie on
the three negative half--cones. Then, we can use the third equation
to simplify the others, to obtain
\begin{equation}\label{eq:coneplanesyst}
\left\{
\begin{array}{l}
\mathbf{d_{10}}\cdot\mathbf{d_0}(\x)-\tau_1\,\tau=\frac{d_{10}^2-\tau_{1}^2}{2},\\
\mathbf{d_{20}}\cdot\mathbf{d_0}(\x)-\tau_2\,\tau=\frac{d_{20}^2-\tau_{2}^2}{2},\\[1mm]
d_0(\x)^2-\tau^2=0,\\[1mm]
\tau\leq \rm{min}(\tau_1,\tau_2,0).
\end{array}\right.
\end{equation}
We conclude that, from a mathematical standpoint, that of TDOA-based
localization is a semi--algebraic and partially linear problem, given by the
intersection of two planes (a line) and a half--cone. This is shown in
Fig. \ref{fig:ConiPianoIperbole}.
Notice that the equations in system (\ref{eq:coneplanesyst}) involve expressions that are
very similar to the standard 3D scalar products and norms, up to a minus
sign in each monomial involving the variable $\tau$ or $(\tau_1,\tau_2)$.
This suggests that, in order to describe and handle all the previous
geometrical objects, an appropriate mathematical framework is the
3D Minkowski space. In the rest of the manuscript, we will explore
this approach and, in particular, we will carry out our analysis using
the exterior algebra formalism (see also \cite{Coll2009,Coll2012} for
a similar analysis). We refer to Appendix A for a concise illustration
of the mathematical tools we are going to use.

Let $ \mb{e_1}$, $\mb{e_2}$ and $\mb{e_3}$ be the unit vectors
of the axes $x$, $y$ and  $\tau$, respectively. Given the pair
$\bs{\tau}=(\tau_1,\tau_2)$ on the $\tau$--plane, we define the
points $\mb{M_i}(\boldsymbol{\tau})=(x_i, y_i, \tau_i),\ i=1, 2,$
and $\mb{M_0} = (x_0, y_0, 0)$.
Given a generic point $\mb{X}=(x,y,\tau)$ in 3D space,
the displacement vectors are defined as $ \mb{D_i}(\mb{X},
\boldsymbol{\tau}) = \mb{X} - \mb{M_i}(\boldsymbol{\tau})$.
Furthermore, we set $ \mb{D_{ji}}(\boldsymbol{\tau}) =
\mb{M_j}(\boldsymbol{\tau}) - \mb{M_i}(\boldsymbol{\tau})$,
for $0 \leq i < j \leq 2$. Notice that, in order to render the
notation more uniform, we left all points and vectors as functions
of $\bt$, although many of them actually depend on a single TDOA.

\begin{definition} \label{cone-plane} For $ i=0, 1, 2,$ we set
\begin{enumerate}
\item $ C_i(\boldsymbol{\tau}) = \{ \mb{X} \in \RR^{2,1} \ \vert \
\parallel \mb{D_i}(\mb{X}, \boldsymbol{\tau}) \parallel^2 = 0 \};$
\item $ C_i(\boldsymbol{\tau})^- = \{ \mb{X} \in
C_i(\boldsymbol{\tau}) \ \vert \ \langle \mb{D_i}(\mb{X},
\boldsymbol{\tau}), \mb{e_3} \rangle \geq 0 \}.$ \end{enumerate}
Moreover, for $ i=1,2,$ we set $$ \Pi_i(\boldsymbol{\tau}) = \{
\mb{X} \in \RR^{2,1} \ \vert \ \langle
\mb{D_{i0}}(\boldsymbol{\tau}), \mb{D_0}(\mb{X},
\boldsymbol{\tau}) \rangle = \frac 12 \parallel
\mb{D_{i0}}(\boldsymbol{\tau}) \parallel^2 \} $$ and $
L_{21}(\boldsymbol{\tau}) = \Pi_1(\boldsymbol{\tau}) \cap
\Pi_2(\boldsymbol{\tau}).$
\end{definition}
$ C_i(\bs{\tau}) $ is a right circular cone with $
\mb{M_i}(\bs{\tau}) $ as vertex, and $ C_i(\bs{\tau})^- $ is a
half--cone, while $ \Pi_i(\bs{\tau}) $ is a plane through
$(\mb{M_0}(\bs{\tau}) + \mb{M_i}(\bs{\tau}))/2$.
Using the exterior algebra formalism (see eq. (\ref{minkowski3})
and the preceding discussion in Appendix \ref{app:A}),
$\Pi_i$ is given by
\begin{equation} \label{eq-piano}
\mb{i}_{\mb{D_0}(\mb{X})}(\mb{D_{i0}}(\boldsymbol{\tau})^\flat) =
\frac 12 \parallel \mb{D_{i0}}(\boldsymbol{\tau}) \parallel^2.
\end{equation}
Finally, if $\mb{D_{10}}(\bs{\tau})$ and $\mb{D_{20}}(\bs{\tau})$
are linearly independent, then $L_{21}(\bs{\tau})$ is the line of
equation
\begin{equation} \label{L21}\fl
\mb{i}_{\mb{D_0}(\mb{X})}
(\mb{D_{10}}(\boldsymbol{\tau})^\flat \wedge
\mb{D_{20}}(\boldsymbol{\tau})^\flat) = \frac 12 \parallel
\mb{D_{10}}(\boldsymbol{\tau}) \parallel^2
\mb{D_{20}}(\boldsymbol{\tau})^\flat - \frac 12 \parallel
\mb{D_{20}}(\boldsymbol{\tau}) \parallel^2
\mb{D_{10}}(\boldsymbol{\tau})^\flat.
\end{equation}
We are now ready to discuss the link that exists between the
geometry of the Minkowski space and the TDOA--based
localization.
As above, we set $A_i(\bs{\tau})=A_i(\tau_i)$.
\begin{theorem} \label{thcopi} Let $ \pi: \RR^{2,1} \to \RR^2 $ be
the projection onto the $x$--plane. Then
\begin{enumerate}
\item
$ \pi(C_0^- \cap C_i(\bs{\tau})^-) = A_i(\bs{\tau}) $
\mbox{if }
$0\leq|\tau_i|\leq d_{i0},$ for $i=1,2;$
\item
$ \pi(C_0^- \cap \Pi_i(\bs{\tau})) = \left\{
\begin{array}{ll}
A_i(\bs{\tau}) & \mbox{if } -d_{i0}<\tau_i \leq d_{i0}\\
A_i(\bs{\tau}) \cup r_{j}^0 & \mbox{if } \tau_i = -d_{i0}
\end{array} \right.$
with $ i \not= j.$
\end{enumerate}
\end{theorem}

\noindent\emph{Proof.} Let $\mb{x} = \pi(\mb{X})$. We
therefore have $\mb{X} = (\mb{x}, \tau)$. According to
Definition \ref{cone-plane}, we obtain $\mb{X} \in C_0^-$
if, and only if, $ \parallel \mb{D_0}(\mb{X},\bs{\tau})
\parallel^2 = 0 $ and $ \langle \mb{D_0}(\mb{X},\bs{\tau}),
\mb{e_3} \rangle > 0$, which means that
$d_0(\mb{x})^2 - \tau^2 =0,\ -\tau > 0 $, therefore we finally
obtain $d_0(\mb{x}) = -\tau,\ \tau < 0$.
Similarly, $ \mb{X} \in C_i(\bs{\tau})^- $ is equivalent to
$d_i(\mb{x}) = -(\tau - \tau_i),\ \tau < \tau_i$. As a
consequence, $\mb{X}\in C_0^- \cap C_i(\bs{\tau})^- $ if,
and only if, $d_i(\mb{x}) - d_0(\mb{x}) = \tau_i,\ \tau < \min(0, \tau_i)$,
i.e. $ \mb{x} \in A_i(\bs{\tau})$, therefore the first claim follows.

Then, we remark that $ \mb{D_i}(\mb{X},
\bs{\tau}) = \mb{D_0}(\mb{X}, \bs{\tau}) + \mb{D_{i0}}(\bs{\tau})
$, that implies $$ \parallel \mb{D_i}(\mb{X}, \bs{\tau}) \parallel^2 =
\parallel \mb{D_0}(\mb{X}, \bs{\tau}) \parallel^2 + 2 \langle
\mb{D_0}(\mb{X}, \bs{\tau}), \mb{D_{i0}}(\bs{\tau}) \rangle +
\parallel \mb{D_{i0}}(\bs{\tau}) \parallel^2.$$ Hence, $ \mb{X}
\in C_0 \cap \Pi(\bs{\tau}) $ if, and only if, $ \mb{X} \in C_0
\cap C_i(\bs{\tau}),$ and, using the first claim, we get $ \pi(C_0^- \cap
\Pi_i(\bs{\tau})) \supseteq A_i(\bs{\tau})$.
$ C_0^- \cap \Pi_i(\bs{\tau}) $ is degenerate, precisely a
half--line, if, and only if, $ \mb{M_0}\in \Pi_i(\bs{\tau}),$ i.e. $ 0
= \langle \mb{D_{i0}}(\bs{\tau}), \mb{0} \rangle = \frac 12
\parallel \mb{D_{i0}}(\bs{\tau}) \parallel^2.$ The last condition
is equivalent to $ \mb{M_i}(\bs{\tau}) \in C_0,$ or $ \tau_i^2 =
d_{i0}.$ Hence, if $ \tau_i^2 \not= d_{i0}^2,$ $ \pi(C_0^- \cap
\Pi_i(\bs{\tau})) $ is a hyperbola branch and the first equality
follows. Otherwise, if $ \tau_i^2 = d_{i0}^2,$ then $ \pi(C_0 \cap
\Pi(\bs{\tau})) = r_j.$ It is easy to check that $ (\mb{m_i},
-d_{i0}) \in C_0^- $ and that $ (\mb{m_i}, d_{i0}) \in C_0
\setminus C_0^-.$ So, $ \pi(C_0 \cap \Pi(\bs{\tau})) =
A_i(\bs{\tau}) \cup r_j^0 $ if $ \tau_i = -d_{i0}.$
\hfill$\square$\vspace{1mm}


\section{First properties of the image of $ \bs{\tau_2} $}\label{sec:image}

We now study the set of admissible pseudoranges, i.e. the image
$\rm{Im}(\bs{\tau_2})$ of the TDOA map, in the $ \tau$--plane. In
particular, in this Section we begin with focusing on the dimension of the image
and then we prove that $ \mbox{Im}(\bs{\tau_2}) $ is contained
within a bounded convex set in the $ \tau$--plane. These preliminary
results are quite similar for both cases of generic and collinear
microphone configurations, which is the reason why we collect
them together in this Section. For the definition and properties of
convex polytopes, see \cite{Matousek2002} among the many
available references.
\begin{theorem} $ \mbox{Im}(\bs{\tau_2}) $ is locally the
$\tau$--plane.
\end{theorem}

\noindent\emph{Proof.} Let us assume that $ \mb{\bar{x}} $ is
a point where $\bs{\tau_2} $ is regular, i.e. where the Jacobian
matrix $J(\mb{\bar{x}}) $ has rank $2$ (see Theorem \ref{n=2}).
The map $\bs{\tau_2} $ can be written as
$$
\left\{ \begin{array}{l} d_1(\mb{x}) - d_0(\mb{x}) = \tau_1 \\ d_2(\mb{x})
- d_0(\mb{x}) = \tau_2 \end{array} \right.
$$
and $\bs{\bar{\tau}} = \bs{\tau_2}(\mb{\bar{x}}) $ is a solution of the
system. The Implicit Function Theorem guarantees that there exist
functions $ x = x(\bs{\tau})$ and $y=y(\bs{\tau})$, which are defined
in a neighborhood of $ \bs{\bar{\tau}} $ and take on values in a
neighborhood of $ \mb{\bar{x}} $ so that the given system will be
equivalent to
$$
\left\{ \begin{array}{l} x = x(\bs{\tau}) \\
y=y(\bs{\tau}) \end{array} \right.
\; ,
$$
therefore the claim follows.
\hfill$\square$\vspace{1mm}

\begin{figure}[htb]
\begin{center}
\resizebox{13cm}{!}{
  \includegraphics
  {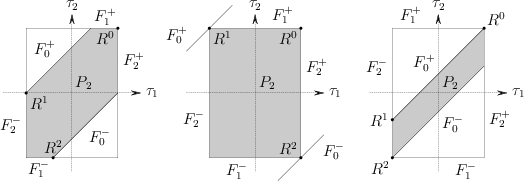}}\\
  \caption{\label{fig:politopi} Left-hand side: polygon $P_2$ (in shaded gray) under
  the assumption that the points $ \m{0}$, $\m{1}$ and $\m{2} $
  are not collinear. Center: polygon $P_2$ (in shaded gray) in the case of three collinear points with $\m{0}$
  between $\m{1}$ and $\m{2}$. Right-hand side: polygon $P_2$ (in shaded gray) when the sensors lie
  on a line, but with $\m{1}$ between $\m{0}$ and $\m{2}$.
  The case with $ \m{2} $ between $\m{0}$ and $\m{1}$ can be
  obtained from the image on the right by swapping the role of $\tau_1$ and
  $\tau_2$.}
\end{center}
\end{figure}

\noindent
In Lemma \ref{LemmaTriang} we showed that the TDOAs are
constrained by the triangular inequalities. In the rest of this
Section we will show that, as a consequence of these inequalities,
$\bs{\tau_2}$ maps the $x$--plane onto a specific bounded
region in the $\tau$--plane.
\begin{definition}\label{def:polytope} Let
$$
P_2 = \{ \bs{\tau} \in \RR^2 |\,
\Vert \mb{D_{ji}}(\bs{\tau}) \Vert^2 \geq 0,\ 0\leq i < j \leq 2
\},
$$
and  $ k \in \{0, 1, 2\} $ be different from $ i, j$, for
$ 0 \leq i < j \leq 2$.
We define
$$
F_{k}^+ = \{ \bs{\tau} \in P_2 |\,
\Vert \mb{D_{ji}}(\bs{\tau}) \Vert^2=0,\
\PM{\mb{D_{ji}}(\bs{\tau}),\mb{e_3}}<0 \},
$$
$$ F_{k}^- =
\{\bt \in P_2 |\, \Vert\mb{D_{ji}}(\bs{\tau})\Vert^2=0,\
\PM{\mb{D_{ji}}(\bs{\tau}),\mb{e_3}}>0 \},$$\\[-7mm]
$$
R^0 =F_{1}^+\cap F_{2}^+,\qquad R^1 =F_{0}^+\cap F_{2}^-,\qquad
R^2 =F_{0}^-\cap F_{1}^-.
$$
\end{definition}
Before we proceed with studying the relation between $P_2$ and
$ \mbox{Im}(\bs{\tau_2})$, let us describe the geometric
properties of this set.
In Fig. \ref{fig:politopi}, we show some examples of $P_2$
(in gray), for different positions of the points $\m{0}$, $\m{1}$
and $\m{2}$.
\begin{theorem}
\label{polytope}
$P_2$ is a polygon (a $ 2$--dimensional convex polytope).
Moreover, if the points $\m{0}$, $\m{1}$ and $\m{2}$ are not collinear,
then $P_2$ has exactly $6$ facets $F_{k}^\pm$, which drop to
$4$ if the points are collinear.
\end{theorem}
\noindent\emph{Proof.}
As a first step we notice that
$$
\Vert \mb{D_{ji}}(\bs{\tau}) \Vert^2=d_{ji}^2-(\tau_j-\tau_i)^2,
$$
therefore
\begin{equation}\label{eq:semipiani}\qquad
\Vert \mb{D_{ji}}(\bs{\tau}) \Vert^2\geq 0
\quad\Leftrightarrow\quad d_{ji}\geq |\tau_j-\tau_i|.
\end{equation}

The set $ P_2 $ is a $2$--dimensional convex polytope because,
according to (\ref{eq:semipiani}), it is the intersection of half--planes
and it contains an open neighborhood of $\mb{0} = (0, 0) \in \RR^2$.
In fact, the coordinates of $ \mb{0} $ satisfy all the finitely many
strict inequalities defining $ P_2$, which implies that also a
sufficiently small open disc centered at $\mb{0}$ belongs to $P_2$.

In order to prove the rest of the statement, we need to show that
the inequalities defining $P_2$ are redundant if, and only if,
$ \mb{m_0}, \mb{m_1}, \mb{m_2} $ are collinear. Let us consider
$$
\left\{ \begin{array}{l} -d_{10} \leq \tau_1 \leq d_{10} \\
-d_{20} \leq \tau_2 \leq d_{20} \\ -d_{21} \leq \tau_2 - \tau_1
\leq d_{21} \end{array} \right. \;.
$$
The first two inequalities define a rectangle whose sides are
parallel to the $ \tau_i$ axes. The lines $\tau_2 - \tau_1 = d_{21}$
and $ \tau_2 = d_{20} $ meet at $(d_{20} - d_{21}, d_{20})$,
which lies between $ (-d_{10}, d_{20})$ and $ (d_{10}, d_{20})$
if, and only if, $ \vert d_{20} - d_{21} \vert < d_{10}$, i.e. when
the points $ \m{0}, \m{1}, \m{2} $ are not collinear.
Through a similar reasoning we can show that, if the three points
$ \m{i}$ are not collinear, the line $ \tau_2 - \tau_1 = d_{21} $
meets $ \tau_1 = -d_{10} $ at $ (-d_{10}, d_{21}-d_{10})$, while
$ \tau_2 -\tau_1 = -d_{21} $ meets $ \tau_1=d_{10} $ at
$ (d_{10}, -d_{21}+d_{10}) $ and $ \tau_2=-d_{20} $ at
$ (d_{21}-d_{20},-d_{20})$.
An easy check proves that $ P_2 $ is a hexagon of vertices
$ (d_{10},d_{20}), (d_{20} - d_{21}, d_{20}),
(-d_{10}, d_{21}-d_{10}), (-d_{10},-d_{20}),
(d_{21}-d_{20},-d_{20}), (d_{10}, -d_{21}+d_{10})$.

On the other hand, if $ \mb{m_0}, \mb{m_1}, \mb{m_2}$ are
collinear, $ P_2 $ ends up having $4$ sides. There are three
possible configurations: $(i)$ $ \mb{m_0} $ lies between
$ \mb{m_1} $ and $ \mb{m_2};$ $(ii)$ $ \mb{m_1} $ lies
between $ \mb{m_0} $ and $ \mb{m_2}$; $(iii)$ $ \mb{m_2} $
lies between $ \mb{m_0} $ and $ \mb{m_1}$. In case $(i)$
we have that $ d_{21} = d_{10} + d_{20}$, therefore
$ -d_{21} \leq \tau_2 - \tau_1 \leq d_{21} $ are redundant.
In case $(ii)$ we have that $ d_{20} = d_{10} + d_{21} $
and $ -d_{20} \leq \tau_2 \leq d_{20} $ give no restrictions
to the others. In case $(iii)$, $ -d_{10} \leq \tau_1 \leq d_{10}$
are redundant as it follows from $ d_{10} = d_{20} + d_{21}$.
\hfill$\square$\vspace{1mm}\\

\noindent
For further reference, we name the vertices of the rectangle
$ -d_{10} \leq \tau_1 \leq d_{10}, -d_{20} \leq \tau_2 \leq d_{20}$,
recalling that $ R^0 = (d_{10}, d_{20}) $
(see Definition \ref{def:polytope}).
\begin{definition}
\label{3-vert}
Let $R^* = (-d_{10}, d_{20})$, $R^0_1 = (-d_{10}, -d_{20})$,
and $R^*_1 = (d_{10}, -d_{20})$.
\end{definition}

\noindent We are now ready to present the main result of this section.
\begin{proposition}
\label{th:preimages}
$ \mbox{Im}(\bs{\tau_2}) \subsetneq P_2$. Moreover,
$\bs{\tau_2}^{-1}(F_{k}^\pm) =r_{k}^\pm$, $k=0,1,2$, and,
if $\m{0},\m{1},\m{2}$ are not collinear, then
$\bs{\tau_2}^{-1}(R^k)=\m{k}$, $k=0,1,2$.
\end{proposition}

\noindent\emph{Proof.}
The first statement is a direct consequence of Definition
\ref{def:polytope}, relation (\ref{eq:semipiani}) and Lemma \ref{LemmaTriang}.
Let us now consider $ \mb{x}$ such that $ \pm d_{ji} = \tau_j(\mb{x}) -
\tau_i(\mb{x}) = d_j(\mb{x}) - d_i(\mb{x})$. Using Lemma
\ref{LemmaTriang} we get $\mb{x} \in r_{k}^\pm$, as
claimed. As the preimage of the intersection of two sets is
equal to the intersection of the respective preimages, the last
statement follows from Definition \ref{rette}. Finally, the
vertices of $P_2$ that are different from $R^0$, $R^1$ and
$R^2 $ are not in $\mbox{Im}(\bs{\tau_2})$, because the
corresponding half--lines do not meet, as it is easy to verify
in all the possible cases. For example, if $\m{0}$, $\m{1}$
and $\m{2} $ are not collinear, then $r_1^+ $ and $ r_0^+ $
do not meet, which implies $ (d_{20}-d_{21}, d_{20}) \in P_2
\setminus \mbox{Im}(\bs{\tau_2})$.
\hfill$\square$\vspace{1mm}

\section{The localization problem in the general case}
\label{sec:loc-gen}
In this Section we offer further insight on the TDOA
map under the assumption that $ \m{0}, \m{1}, \m{2} $
are not collinear.
Subsections \ref{sei-zero} and \ref{sei-uno} contain some
preliminary mathematical results. In Subsection \ref{sei-zero}
we show how the preimages of the $\bs{\tau_2}$ map are
strictly related to the non--positive real roots of a degree-$2$
equation, whose coefficients are polynomials in $ \bt $ (see
eq. (\ref{eq-deg-2}) and the proof of Theorem \ref{th:imtau2}).
In order to use the Descartes' rule of signs for the characterization
of the roots, in Subsection \ref{sei-uno} we give the necessary
background on the zero sets of such coefficients and on the
sign that the polynomials take on in the $\tau$--plane.
The main results of this Section are offered in Subsections
\ref{sei-due} and \ref{sei-tre}. In the former we completely
describe Im$(\bs{\tau_2})$ and the cardinality of each fiber,
while in the latter we derive a visual representation of the
different preimage regions of $\bs{\tau_2}$ in the $x$--plane,
and find the locus where $ \bs{\tau_2} $ is $1$--to--$1$.
The two Subsections \ref{sei-due} and \ref{sei-tre} also offer
an interpretation of such results from the perspective of the
localization problem.

This Section is, in fact, quite central for the manuscript,
and the results included here are mainly proven using
techniques coming from algebraic geometry. A brief
presentation of the tools of algebraic geometry that are
needed for this purpose is included in Appendix \ref{app:B}.
In order to improve the readability of this Section, we collected
some of the proofs in Subsection \ref{sei-quattro}.

\subsection{The quadratic equation}
\label{sei-zero}
As discussed in the previous Sections,
$ \bs{\tau} \in \mbox{Im}(\bs{\tau_2})$ if, and only if,
$ A_1(\bs{\tau}) \cap A_2(\bs{\tau}) \not= \emptyset$.
According to Theorem \ref{thcopi}, we have
$A_1(\bs{\tau}) \cap A_2(\bs{\tau}) \subseteq
\pi(C_0^- \cap L_{21}(\bs{\tau}))$, therefore the analysis
of the intersection $C_0^- \cap L_{21}(\bs{\tau})$ plays
a crucial role in characterizing the TDOA map.
We begin with studying the line $L_{21}(\bt)$ of defining
eq. (\ref{L21}).

Assuming that the microphones are not aligned, we have
\begin{equation}\label{eq:M1wM2}
\eqalign{\mb{D_1}(\mb{X},\bs{\tau}) \wedge \mb{D_2}(\mb{X},\bs{\tau}) =
(\mb{d_1}(\mb{x}) + \tau_1 \mb{e_3}) \wedge (\mb{d_{2}}(\mb{x}) +
\tau_2\mb{e_3}) =\cr
= \mb{d_1}(\mb{x}) \wedge \mb{d_2}(\mb{x}) +
(\tau_2 \mb{d_1}(\mb{x}) - \tau_1 \mb{d_2}(\mb{x}) ) \wedge
\mb{e_3} \neq 0}\label{eq2}
\end{equation}
because $ \mb{d_1}(\mb{x}) $ and $ \mb{d_2}(\mb{x}) $ are linearly
independent. Consequently $ \mb{D_1}(\mb{X},\bs{\tau})$ and
$\mb{D_2}(\mb{X}, \bs{\tau}) $ are linearly independent as well
for every $ \bs{\tau} \in \RR^2$. Let
\begin{equation}\label{eq:M1wM2we3}
\bs{\Omega}=\mb{D_{10}}(\bs{\tau}) \wedge \mb{D_{20}}(\bs{\tau})
\wedge \mb{e_3} = \mb{d_{10}} \wedge \mb{d_{20}} \wedge \mb{e_3}
\neq 0 \;,
\end{equation}
which is a 3--form (see Section \ref{app:AMink} in Appendix
\ref{app:A}). With no loss of generality, we can assume that
$\bs\Omega$ is positively oriented, i.e. $\bs{\Omega}=k\bs{\omega}$
with $k>0$, therefore $\langle\bs{\Omega},\bs{\omega}\rangle=-k <0$.
\begin{lemma}
\label{lm:paramL21}
For any $\bt\in\RR^2$, $L_{21}(\bt)=\Pi_1(\bt) \cap \Pi_2(\bt)$ is
a line. A parametric representation of $L_{21}(\bt)$ is
$\X(\lambda;\bt)=\mb{L_0}(\bs{\tau}) + \lambda \mb{v}(\bs{\tau})$,
where
\begin{equation*}\fl\qquad
\mb{v}(\bs{\tau})=\ast(\mb{D_{10}}(\bs{\tau}) \wedge
\mb{D_{20}}(\bs{\tau})) = \ast( (\mb{d_{10}} \wedge
\mb{d_{20}}) + (\tau_2 \mb{d_{10}} - \tau_1
\mb{d_{20}}) \wedge \mb{e_3})
\end{equation*}
and the displacement vector of $\mb{L_0}(\bs{\tau})$ is
\begin{eqnarray*}\fl\qquad
\mb{D_0}(\mb{L_0}(\bs{\tau}))&=&
\frac{1}{2\ast\bs{\Omega}} \ast\! \left(
\left( \Vert\mb{D_{10}}(\bs{\tau})\Vert^2
\mb{D_{20}}(\bs{\tau})-\Vert\mb{D_{20}}(\bs{\tau})\Vert^2
\mb{D_{10}}(\bs{\tau}) \right) \wedge \mb{e_3} \right)= \\
&=& -\frac{\ast \left( \left(
\Vert\mb{D_{10}}(\bs{\tau})\Vert^2\mb{d_{20}} -
\Vert\mb{D_{20}}(\bs{\tau})\Vert^2 \mb{d_{10}} \right)
\wedge \mb{e_3} \right)} {2\Vert\mb{d_{10}} \wedge
\mb{d_{20}} \Vert}.
\end{eqnarray*}
\end{lemma}
\noindent\emph{Proof.} See Subsection \ref{sei-quattro}.
\hfill$\square$\vspace{1mm}

\begin{remark}\rm
The point $\mb{L_0}(\bs{\tau}) $ is the
intersection between $L_{21}(\bt)$ and the $x$--plane. In fact,
from the properties of the Hodge $ \ast $ operator, we know
that the component of $ \mb{D_0}(\mb{L_0}(\bs{\tau})) $ along
$\mb{e_3} $ is zero.
\end{remark}

We can turn our attention to the study of $C_0^- \cap L_{21}(\bs{\tau})$.
From the definition of $ C_0$ and Lemma \ref{lm:paramL21} follows
that a point $\X(\lambda;\bt)$ of the line $L_{21}(\bs{\tau})$ lies on
$C_0 $ if the vector $ \mb{D_0}(\mb{L_0}(\bs{\tau}))
+\lambda \mb{v}(\bs{\tau})$ is isotropic with respect to the bilinear
form $b$. This means that $ \Vert \mb{D_0}(\mb{L_0}(\bs{\tau}))
+ \lambda \mb{v}(\bs{\tau})\Vert^2 = 0$ or, more explicitly,
\begin{equation} \label{eq-deg-2}
\Vert \mb{v}(\bs{\tau}) \Vert^2 \lambda^2 + 2 \lambda
\langle \mb{D_0}(\mb{L_0}(\bs{\tau})) , \mb{v}(\bs{\tau}) \rangle
+ \Vert \mb{D_0}(\mb{L_0}(\bs{\tau})) \Vert^2 = 0.
\end{equation}
This equation in $ \lambda \in \RR $ has a degree that does not
exceed $2$, and coefficients that depend on $ \bs{\tau}$.
\begin{definition} Let
\begin{enumerate}
\item $ a(\bs{\tau}) = \Vert \mb{v}(\bs{\tau}) \Vert^2 = \Vert
\tau_2 \mb{d_{10}} - \tau_1 \mb{d_{20}} \Vert^2 - \Vert
\mb{d_{10}} \wedge \mb{d_{20}} \Vert^2;$ \item $ b(\bs{\tau}) =
\langle \mb{D_0}(\mb{L_0}(\bs{\tau})) , \mb{v}(\bs{\tau}) \rangle
= \displaystyle \frac{\langle \tau_2 \mb{d_{10}} - \tau_1
\mb{d_{20}}, \Vert \mb{D_{20}}(\bs{\tau})\Vert^2 \mb{d_{10}} -
\Vert \mb{D_{10}}(\bs{\tau}) \Vert^2 \mb{d_{20}} \rangle}{2 \Vert
\mb{d_{10}} \wedge \mb{d_{20}} \Vert};$ \item $ c(\bs{\tau}) =
\Vert \mb{D_0}(\mb{L_0}(\bs{\tau})) \Vert^2 = \displaystyle
\frac{\left\Vert \Vert \mb{D_{10}}(\bs{\tau}) \Vert^2 \mb{d_{20}}
- \Vert \mb{D_{20}}(\bs{\tau}) \Vert^2 \mb{d_{10}}
\right\Vert^2}{4 \Vert \mb{d_{10}} \wedge \mb{d_{20}} \Vert^2}.$
\end{enumerate}
\end{definition}
Eq. (\ref{eq-deg-2}) can be rewritten as
\begin{equation}\label{eq-deg-2b}
a(\bs{\tau}) \lambda^2 + 2 b(\bs{\tau}) \lambda + c(\bs{\tau}) = 0.
\end{equation}

\noindent The explicit solution of eq. (\ref{eq-deg-2}) will be derived
in Subsection \ref{sei-tre}.

\subsection{The study of the coefficients}
\label{sei-uno}

In order to study the solutions of the quadratic equation
(\ref{eq-deg-2}), we need to use Descartes' rule of signs.
To apply it, we first describe the zero set of
the coefficients $a(\bt)$, $b(\bt)$, and $c(\bt)$; then we study
the sign of these coefficients wherever they do not vanish.
As stated above, the main mathematical tools that are used in
this Subsection come from algebraic geometry because
$a(\bt)$, $b(\bt)$ and $c(\bt)$ are polynomials with real
coefficients (see Appendix B for a short introduction or
\cite{Bix1998,clos,hart}).

Let us first describe the vanishing locus of $c(\bt)$, over both
$\RR$, where it is particularly simple, and $ \mathbb{C}$.
\begin{proposition}
\label{ctau}
$ c(\bs{\tau}) \geq 0 $ for every $ \bs{\tau} \in \RR^2$.
Moreover, $ c(\bs{\tau}) = 0$ if and only if
$\bs{\tau} \in \{ R^0, R^*, R^*_1, R^0_1 \}$.
On the complex field $ \mathbb{C}$, $c(\bs{\tau})$
factors as the product of two degree-$2$ polynomials.
\end{proposition}

\noindent\emph{Proof.}  See Subsection \ref{sei-quattro}.
\hfill$\square$\vspace{1mm}

In order to analyze the sign of $a(\bs{\tau})$, we need to
introduce some notation.
\begin{definition}
We define three subsets of the $\tau$--plane, according to the sign of $a(\bt)$:
\begin{itemize}
\item $E=\{\bt\in\RR^2 \ \vert \ a(\bs{\tau}) =0\};$ \item
$E^+=\{\bt\in\RR^2 \ \vert \ a(\bs{\tau}) > 0\};$ \item
$E^-=\{\bt\in\RR^2 \ \vert \ a(\bs{\tau}) < 0\}$.
\end{itemize}
\end{definition}

\begin{proposition} \label{atau} $ E \subset P_2 $ is an ellipse
centered in $\mb{0}= (0,0) $, and it represents the only conic that
is tangent to all sides of the hexagon $P_2$.
\end{proposition}

\noindent\emph{Proof.}  See Subsection \ref{sei-quattro}.
\hfill$\square$\vspace{1mm}

\begin{figure}[htb]
\begin{center}
\resizebox{5cm}{!}{
  \includegraphics
  {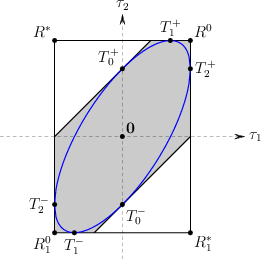}}\\
  \caption{\label{fig:slide18}The ellipse $E$ (in blue) is tangent
  to each side of the  hexagon $P_2$ (in gray).
  We have $11$ distinguished points: the center $ \mb{0}$
  of $E$, the six tangency points $ T_i^{\pm}, i=0,1,2$, and the
  vertices of the rectangle $R_1^0, R^0, R^*$ and $R_1^*$.}
\end{center}
\end{figure}

\noindent The ellipse $E$ and some specific points on the
polytope $P_2$ are shown in Fig. \ref{fig:slide18}.
As the tangency points will eventually show up in the study
of the vanishing locus of $b(\bt)$, we define them here for
further reference.
\begin{definition} \label{tang-points}
Let $0 \leq i,j,k \leq 2 $ with $k < j$ and $k \neq j$. Then
$$
T_i^+ = \left( \langle \mb{d_{10}}, \mb{\tilde{d_{jk}}} \rangle,
\langle \mb{d_{20}}, \mb{\tilde{d_{jk}}} \rangle \right)
\quad
\mbox{ and }
\quad
T_i^- = \left( -\langle \mb{d_{10}}, \mb{\tilde{d_{jk}}}
\rangle, -\langle \mb{d_{20}}, \mb{\tilde{d_{jk}}} \rangle \right),
$$
where, according to our current notation,
$\mb{\tilde{d_{jk}}}=\frac{\mb{d_{jk}}}{d_{jk}}$.
\end{definition}

\begin{remark}\rm
For every non-collinear choice of $ \m{0}$, $\m{1}$, $\m{2}$,
$E$ is smooth. In fact, $ \nabla a(\bs{\tau}) = (0,0)$ is the
homogeneous linear system
$$
\left\{ \begin{array}{l}
d_{20}^2 \tau_1 - \langle \mb{d_{10}}, \mb{d_{20}} \rangle \tau_2
= 0 \\ - \langle \mb{d_{10}}, \mb{d_{20}} \rangle \tau_1 +
d_{10}^2 \tau_2 = 0 \end{array} \right.
$$
whose only solution is $\mb{0} \notin E$, because the matrix of
the coefficients of the variables has determinant
$ \Vert \mb{d_{10}} \wedge \mb{d_{20}} \Vert^2 \not= 0$.
\end{remark}
We conclude the analysis of the sign of $a(\bt)$ by noticing that
the set $E^- $ contains the origin $ \mb{0}$, therefore it is the
bounded connected component of $ \RR^2 \setminus E$.
Similarly, $ R^0 \in E^+ $ therefore $E^+$ is the unbounded
connected component of $ \RR^2 \setminus E$.

\begin{figure}[htb]
\begin{center}
\resizebox{12.5cm}{!}{
  \includegraphics
  {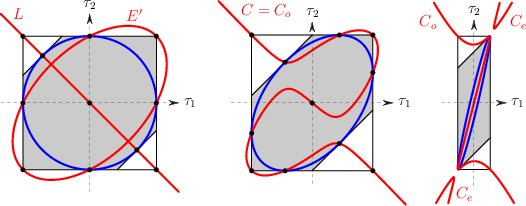}}
  \caption{\label{fig:slide19}Examples of cubics $C$: on the left-hand side it is  singular. In the centre it is an oval and on the right-hand side are two ovals. The curve $ E $ is shown in blue, $ C $ in red, and
  the hexagon $ P_2 $ in shaded gray. The $ 11 $ distinguished points are
  marked in the first two pictures, but not in the last one because
  $ 4 $ of them are very close to each other on the upper--right
  vertex of the rectangle, and similarly for the $ 4 $ ones, which
  are close to the opposite vertex. In all the three cases
  $ P_2 $ is an hexagon, but it exhibits two very short sides in the right-hand picture.}
\end{center}
\end{figure}

The analysis of the sign of the last coefficients $ b(\bs{\tau}) $
is a bit more involved. Let us define the notations as done for
$a(\bs{\tau})$.
\begin{definition}
We define three subsets of the $ \tau$--plane, according to the
sign of $b(\bs{\tau})$:
\begin{itemize}
\item $ C=\{\bt\in\RR^2 \ \vert \ b(\bs{\tau}) = 0 \};$ \item $
C^+=\{\bt\in\RR^2 \ \vert \ b(\bs{\tau}) > 0 \};$ \item $
C^-=\{\bt\in\RR^2 \ \vert \ b(\bs{\tau}) <0\}$.
\end{itemize}
\end{definition}
As our aim is to study the relative position of $ P_2 $ and the
sets $C$, $C^+$, and $C^-$, we need more of an in-depth
understanding of the curve $C$ (see Figure \ref{fig:slide19} for some examples of this curve). We will first analyze the role
of the $11$ distinguished points marked in Fig. \ref{fig:slide18}
for the study of $c(\bt) = 0$ and $a(\bt) = 0$ in connection with
$b(\bt) = 0$. We will then look for special displacement
positions of $\mb{m_0}$, $\mb{m_1}$ and $\mb{m_2}$, which
force $C$ to be non--irreducible. In fact, the irreducibility of $C$
has an impact on the topological properties of $C^+$ and $C^-$,
particularly on their connectedness by arcs. We will finally study
the connected components of $C$.

\begin{proposition}
\label{btau}
$ C $ is a cubic curve with $2$--fold rotational symmetry with
respect to $ \mb{0}$, which contains $T_0^\pm$, $T_1^\pm$,
$T_2^\pm$, $R^0$, $R^0_1$, $R^*$, $R^*_1$ and $\mb{0}$.
The tangent lines to $C$ at $R^0$, $R^0_1$, $R^*$, $R^*_1$
are orthogonal to $F_0^+$, therefore $C$ is smooth at the
above four points. Finally, $C$ transversally intersects
both $E$ and the lines that support the sides of $P_2$.
\end{proposition}

\noindent\emph{Proof.}  See Subsection \ref{sei-quattro}.
\hfill$\square$

\begin{proposition} \label{sing-cubic}
$ C $ is a smooth curve, unless $ d_{10} =d_{20}$.
In this case $C$ is the union of the line $ L: \tau_1 + \tau_2 = 0 $
and the conic
$ E': \tau_1^2 -(\langle \mb{\tilde{d}_{10}}, \mb{\tilde{d}_{20}} \rangle + 1)
\tau_1 \tau_2 + \tau_2^2 + d_{10}^2 (\langle \mb{\tilde{d}_{10}},
\mb{\tilde{d}_{20}} \rangle - 1) = 0$.
\end{proposition}

\noindent\emph{Proof.}  See Subsection \ref{sei-quattro}.
\hfill$\square$\vspace{2mm}

For the sake of completeness, we now investigate the uniqueness of
this cubic curve by showing that $ C $ is completely determined
by the positions of the points $\mb{m_0}$, $\mb{m_1}$, $\mb{m_2}$.

\begin{proposition}\label{th:uniquenessC}
$C$ is the unique cubic curve that contains the points $T_0^\pm$,
$T_1^\pm$, $T_2^\pm$, $\mb{0}$, $R^0$, $R^0_1$, $R^*$, $R^*_1$.
\end{proposition}

\noindent\emph{Proof.}  See Subsection \ref{sei-quattro}.
\hfill$\square$

\begin{remark}\rm Due to the $ 2$--fold rotational symmetry
around $ \mb{0}$, and the fact that $C$ is smooth at $ \mb{0}$,
we can conclude that $\mb{0}$ is an inflectional point for $C$.
\end{remark}

The cubic curve $C$, where smooth, has genus $1$. Therefore,
in the $\tau$--plane, it can have either $1$ or $2$ ovals, in compliance
with Harnack's Theorem \ref{harnack} (see Fig. \ref{fig:slide19}).
Depending on the position of $\m{0}$, $\m{1}$, $\m{2}$, both cases
are possible. Following standard notation, the two ovals are called
$C_o$, the odd oval, and $ C_e$, the even one (this could be
missing), and, at least in the projective plane $\PP_\RR^2$, they
are the connected components of $C$.
The importance of studying the connected components of $C$
rests on the fact that $C$ divides every neighborhood of a point
$P \in C$ in two sets, one in $C^+$, the other in $C^-$. Therefore,
we need to locate $ C_o $ and $ C_e $ with respect to $ P_2.$

\begin{proposition}\label{th:C_o}
The points $ T_0^\pm, T_1^\pm, T_2^\pm, \mb{0}, R^0, R^*,
R^0_1, R^*_1 $ belong to the same connected component
$ C_o $ of $C$, which is the only one that intersects
$ P_2$.
\end{proposition}

\noindent\emph{Proof.}  See Subsection \ref{sei-quattro}.
\hfill$\square$\vspace{1mm}

Now we can complete the study of the sign of $ b(\bt)$
within $P_2$. Let us first assume that $ C $ is smooth.
Due to the rotational symmetry of $C$, the component
$ C_o $ is connected in the affine plane $\RR^2$ as well,
and it divides the $ \tau$--plane in two disjoint sets, which
we name $C_o^+$ and $C_o^-$. Due to Proposition
\ref{th:C_o}, $ b(\bs{\tau})$ does not change sign on
$P_2\cap C_o^+$ and $P_2\cap C_o^-$, therefore we
have $P_2\cap C^+=P_2\cap C_o^+$ and
$P_2\cap C^-=P_2\cap C_o^-$ (possibly with
$C_o^+,C_o^-$ in swapped order). In particular,
evaluating $b(\bt)$ at the vertices of $P_2$, we have
that $C_o^+$ is the connected component of
$\RR^2\setminus C_o$ containing $R^1, R^2$.

Finally, if $ C $ is singular we have $ C = L \cup E'$
(see Fig. \ref{fig:slide19}). There are four disjoint regions
in the $\tau$--plane having different signs. Again by
evaluating $b(\bt)$ at the vertices of $P_2$, we obtain
that $C^+$ is the union of the region outside $E'$ in the
half plane containing $R^1, R^2$ plus the region inside
$E'$ in the complementary half plane.

\subsection{The image of $ \bs{\tau_2}$} \label{sei-due}

\begin{figure}[htb]
\begin{center}
\resizebox{5cm}{!}{
  \includegraphics
  {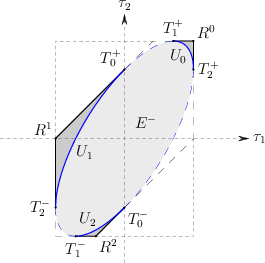}}
  \caption{\label{fig:tauimage}The image of $ \bs{\tau_2}$ is the
  gray subset of $P_2$. In the light gray region marked as $E^-$ the map
  $ \bs{\tau_2} $ is $ 1$--to--$1$, while in the medium gray
  regions $U_0 \cup U_1 \cup U_2$ the map $ \bs{\tau_2} $ is
  $1$--to--$2$. The continuous part of $\partial P_2$ and $E$,
  and the vertices $R^i$, are in the image, where $ \bs{\tau_2}$
  is $ 1$--to--$1$. The dashed part of $\partial P_2$ and $E$,
  and the tangency points $T_i^\pm$, do not belong to
  Im($\bs{\tau_2}$). We remark that the triangles $U_0,U_1,U_2$
  stay on the same connected component of
  $\RR^2\setminus C_o$ (see Fig. \ref{fig:slide19})}
\end{center}
\end{figure}

In this Subsection we achieve one of the main goals of the
manuscript, as we derive the complete and explicit description
of Im$(\bs{\tau_2})$, i.e. the set of admissible TDOAs.
These results are summarized in Fig. \ref{fig:tauimage}.
In the following, we will denote the closure of a set $U$ as $\bar{U}$ and its interior as $ \mathring{U} $.

\begin{definition} The set $ \mathring{P_2} \cap E^+ \cap C^+ $ is the union of $
three $ disjoint connected components that we name $ U_0, U_1, U_2,$ where
$ R^i \in \bar{U}_i $ for $ i = 0, 1, 2$.
\end{definition}
\begin{theorem}\label{th:imtau2}
$\mbox{Im}(\bs{\tau_2}) = E^- \cup \bar{U}_0 \cup \bar{U}_1 \cup
\bar{U}_2 \setminus \{ T_0^\pm, T_1^\pm, T_2^\pm \}.$ Moreover,
$$
\vert\bs{\tau_2}^{-1}(\bt)\vert=
\cases{
2 & if $\bt \in U_0 \cup U_1 \cup U_2$,\\
1 & if $\bt \in \mbox{Im}(\bs{\tau_2}) \setminus U_0 \cup U_1 \cup U_2$.}
$$
\end{theorem}

\noindent\emph{Proof.} Consider the equation (\ref{eq-deg-2b})
$$ a(\bs{\tau}) \lambda^2 + 2 b(\bs{\tau}) \lambda + c(\bs{\tau})
= 0,$$ with $ \bs{\tau} \in P_2$. The reduced discriminant
$ \Delta(\bs{\tau})/4 = b(\bs{\tau})^2 - a(\bs{\tau}) c(\bs{\tau})$
is a degree-$6$ polynomial that vanishes if $ L_{21}(\bs{\tau})$
is tangent to the cone $ C_0$.
According to Theorem \ref{thcopi}, this condition is equivalent to
$A_1(\bs{\tau})$ and $A_2(\bs{\tau})$ intersecting tangentially.
According to Proposition \ref{n=2-geom}, this happens exactly
if $ \mb{x} \in r_0^\pm \cup r_1^\pm \cup r_2^\pm$. Hence,
$ \bs{\tau} \in \bs{\tau_2}(r_0^\pm \cup r_1^\pm \cup
r_2^\pm) = F_0^\pm \cup F_1^\pm \cup F_2^\pm$, which implies
$ \Delta(\bs{\tau}) = 0 $ if, and only if, $ \bs{\tau} \in
\partial P_2$. On the other hand, $ \Delta(\mb{0}) = -a(\mb{0})
c(\mb{0}) > 0 $ because $ \mb{0} \in E^-$, therefore
$\Delta(\bs{\tau}) > 0 $ for $ \bs{\tau} \in \mathring{P}_2$.
As a consequence, equation (\ref{eq-deg-2b}) has real
solutions for any $ \bs{\tau} \in P_2$.

According to Theorem \ref{thcopi}, we are looking for $\bs{\tau}$
that satisfies  $ C_0^- \cap L_{21}(\bs{\tau}) \not= \emptyset$:
\begin{equation*}\fl\qquad\qquad\eqalign{
0 \leq \langle \mb{D_0}(\mb{L_0}(\bs{\tau})) + \lambda
\mb{v}(\bs{\tau}), \mb{e_3} \rangle = \langle
\mb{D_0}(\mb{L_0}(\bs{\tau})), \mb{e_3} \rangle + \lambda \langle
\mb{v}(\bs{\tau}), \mb{e_3} \rangle =\cr
\quad = \lambda \langle
*(\mb{d_{10}} \wedge \mb{d_{20}}) , \mb{e_3} \rangle =
\lambda \langle \ast(\mb{d_{10}} \wedge \mb{d_{20}}), \ast(
\mb{e_1} \wedge \mb{e_2}) \rangle =\cr
\quad = - \lambda
\langle \mb{d_{10}} \wedge \mb{d_{20}}, \mb{e_1} \wedge \mb{e_2}
\rangle = \lambda \langle \mb{d_{10}} \wedge \mb{d_{20}} \wedge
\mb{e_3}, \bs{\omega} \rangle = \lambda \langle \bs{\Omega},
\bs{\omega} \rangle,}
\end{equation*}
which narrows down to $ \lambda \leq 0$, as
$\langle \bs{\Omega}, \bs{\omega} \rangle < 0$.

Let us first consider the case $ \lambda = 0$, which is equivalent
to $c(\bs{\tau}) = 0$.
From Proposition \ref{ctau}, we know that $ c(\bs{\tau}) \geq 0 $
for any $ \bs{\tau} \in P_2,$ and $ c(\bs{\tau}) = 0 $ if, and
only if, $ \bs{\tau} \in \{ R^0, R^*, R^0_1, R^*_1 \}$. At the
four considered points, also $ b(\bs{\tau}) = 0 $ and $
\Delta(\bs{\tau}) = 0.$ Hence, $ \lambda = 0 $ is the only
solution with multiplicity $ 2,$ if $ \bs{\tau} = R^0 $ or $
\bs{\tau} = R^0_1,$ the other two points not being in $ P_2.$
However, the half--lines $ r_1^+ $ and $ r_2^+ $ meet at $ \mb{x}
= \m{0},$ while $ r_1^- \cap r_2^- = \emptyset$. Consequently,
$\bs{\tau_2}^{-1}(R^0) = \m{0}$, while $ R^0_1 \notin
\mbox{Im}(\bs{\tau_2})$.

Let us now assume $ \lambda \not= 0,$ i.e. $ c(\bs{\tau}) > 0$,
and consider all the possible cases, one at a time.
The main (and essentially unique) tool is Descartes' rule of
signs for determining the number of positive roots of a
polynomial equation, with real coefficients and real roots.
\begin{description}
\item[Case (i):] \rm $ a(\bs{\tau}) = b(\bs{\tau}) =
0$.\\
Eq. (\ref{eq-deg-2b}) has no solution, therefore
$ E \cap C=\{ T_0^\pm, T_1^\pm, T_2^\pm\}$ is not
in $ \mbox{Im}(\bs{\tau_2})$.

\item[Case (ii):]  \rm $ a(\bs{\tau}) = 0, b(\bs{\tau}) \not= 0$.\\
Eq. (\ref{eq-deg-2b}) has the only solution $
\lambda = - c(\bs{\tau}) / 2 b(\bs{\tau}) $ for each $ \bs{\tau}
\in E \setminus \{ T_0^\pm, T_1^\pm, T_2^\pm \}.$ Moreover, $
\lambda < 0 $ if, and only if, $ b(\bs{\tau}) > 0 $ i.e. $
\bs{\tau} \in E \cap C^+ = (\partial U_0 \cup \partial U_1 \cup
\partial U_2) \cap E \setminus \{ T_0^\pm, T_1^\pm, T_2^\pm \}.$

\item[Case (iii):]  \rm $ a(\bs{\tau}) < 0$.\\
Equation (\ref{eq-deg-2b}) has one negative root and one
positive root, thus $ E^- \subset \mbox{Im}(\bs{\tau_2}) $ and
$ \vert \bs{\tau_2}^{-1}(\bs{\tau}) \vert = 1 $ for each
$ \bs{\tau} \in E^-.$

\item[Case (iv):]  \rm $ a(\bs{\tau}) > 0, b(\bs{\tau}) < 0$.\\
Eq. (\ref{eq-deg-2b}) has two positive roots, thus
$ E^+ \cap C^- \cap \mbox{Im}(\bs{\tau_2}) = \emptyset.$

\item[Case (v):]  \rm $ a(\bs{\tau}) > 0, b(\bs{\tau}) > 0,
\Delta(\bs{\tau}) = 0$.\\
Eq. (\ref{eq-deg-2b}) has one negative root with
multiplicity $ 2,$ thus $ \vert \bs{\tau_2}^{-1}(\bs{\tau})
\vert = 1 $ for each $ \bs{\tau} \in E^+ \cap C^+ \cap \partial
P_2.$ In particular, $ \bs{\tau_2}^{-1}(R^j) = \m{j} $ for $ j =
1, 2.$

\item[Case (vi):] \rm $ a(\bs{\tau}) > 0, b(\bs{\tau})
> 0, \Delta(\bs{\tau}) > 0$.\\
Eq. (\ref{eq-deg-2b}) has two distinct negative
roots, therefore $ \vert \bs{\tau_2}^{-1}(\bs{\tau}) \vert = 2 $ for
any $ \bs{\tau} \in U_0 \cup U_1 \cup U_2$.
\hfill$\square$
\end{description}

\begin{remark}\rm
Theorem \ref{th:imtau2} agrees with Theorem \ref{thcopi}.
The exact relationship between $ A_1(\bs{\tau}) \cap A_2(\bs{\tau})$
and $ \pi(C_0^-\cap L_{21}(\bs{\tau}))$ is the following:
\begin{description}
\item[(1)]
If $\bt \notin F_{2}^-\cup F_{1}^-$, then $\pi(C_0^-\cap
L_{21}(\bt))=A_1(\bt)\cap A_2(\bt)$.
\item[(2)]
If $\bt\in F_{2}^-\setminus \{ R_{1}^0 \},$ then
$\tau_1=-d_{10}$ and $-d_{20}<\tau_2\leq d_{21}-d_{10}$.
Thus
$$
\pi(C_0^-\cap L_{21}(\bt))= (A_1(\bt)\cap A_2(\bt))\cup
(r_{2}^0\cap A_2(\bt)),
$$
where $A_1(\bt)=r_2^-$. If $\x\in r_{2}^0\cap A_2(\bt)$,
then $d_0(\x)=d_{10}-d_1(\x)$ and
$$
\tau_2(\x)=d_2(\x)-d_0(\x)=d_2(\x)+d_1(\x)-d_{10}\geq d_{21}-d_{10},
$$
where we used the triangular inequality. It follows that
$\tau_2(\x)=d_{21}-d_{10}$ and
$\x=\m{1}$, so $r_{2}^0\cap A_2(\bt)=A_1(\bt)\cap A_2(\bt)$ and,
again, $\pi(C_0^-\cap L_{21}(\bt))=A_1(\bt)\cap A_2(\bt)$.
The case $\bt\in F_{1}^-\setminus \{ R_{1}^0 \} $ is similar.
\item[(3)]
If $\bt= R_{1}^0 \notin \mbox{Im}(\bs{\tau_2}),$ then
$A_1(\bt)\cap A_2(\bt)=\emptyset$ while
$\pi(C_0^-\cap L_{21}(\bt))=r_{1}^0\cup r_2^0=\m{0}.$
\end{description}
\end{remark}\vspace{3mm}

Theorem \ref{th:imtau2} can be nicely interpreted in terms of the
two-dimensional and the three-dimensional intersection problems.
Here we use some standard Minkowski and relativistic conventions
used, for example, in \cite{Abraham1988,Penrose2007}.
\begin{enumerate}
\item
$\bt \in E$ if, and only if, $\mb{v}(\bs{\tau}) $ is isotropic, or light--like.
In this case, the line $L_{21}(\bt)$ is parallel to a generatrix of the
cone, therefore it meets $C_0$ at an ideal point. On the $x$--plane
this means that the level sets $A_1(\bt)$ and $A_2(\bt)$ have one
parallel asymptote. With respect to the localization problem, $\bt\in E$
means that there could exist a source whose distance from the
microphones is large compared to $d_{10}$ and $d_{20}$.
Along $E$, the two TDOAs are not independent and we are able
to recover information only about the direction of arrival of the signal,
and not on the source location.
Things complicate further if $ \bs{\tau} \in E\cap\mbox{Im}(\bs{\tau_2})$,
as the level sets $A_1(\bt)$ and $A_2(\bt)$ also meet at a point at
finite distance, corresponding to another admissible source location.
\item
$\bt\in E^-$ if, and only if $ \mb{v}(\bs{\tau}) $ is time--like, pointing
to the interior of the cone $C_0$. In this case, the line $L_{21}(\bt)$
intersects both half--cones and, on the $x$--plane, the level sets
$A_1(\bt)$ and $A_2(\bt)$ meet at a single point. This is the most
desirable case for localization purposes: a $\bt$ corresponds to a
unique source position $\x$.
\item
$\bt\in E^+$ if, and only if, $\mb{v}(\bs{\tau})$ is space--like,
pointing to the exterior of the cone $C_0$. In this case, the line
$L_{21}(\bt)$ intersects only one half--cone, depending on the
position of the point $\mb{L_0}(\bs{\tau})$ and the direction of
$\mb{v}(\bs{\tau})$. On the $x$--plane, the level sets
$A_1(\bt), A_2(\bt)$ either do not intersect or intersect at two
distinct points. In the last case, for a given $\bt$ there are two
admissible source positions. Following the discussion at point
(i), a source runs away to infinity as $\bt$ gets close to $E$,
while the other remains at a finite position, which suggests a
possible way to distinguish between them if one has some
a-priori knowledge on the source location.
Finally, we observe that the two solutions overlap if
$\bt\in\partial P_2$, which corresponds to $\x$ in the
degeneracy locus.
\end{enumerate}
If $\bt\in E^-$, the localization is still possible even in a noisy
scenario, but we experience a loss in precision and stability
as $ \bt $ approach $E$ (see also the discussion in Section \ref{sec:Impact}).

\subsection{The inverse image}
\label{sei-tre}
We are now ready to reverse the analysis. In fact, the
description of Im($\bs{\tau_2}$) allows us to analyze the
dual situation in the physical $x$--plane. For any given
$ \bs{\tau} \in \mbox{Im}(\bs{\tau_2}) $ and a negative solution
$ \lambda $ of eq. (\ref{eq-deg-2}), we have the corresponding
preimage in the $x$--plane
\begin{equation}
\label{eq:inv-image} \mb{x}(\bs{\tau}) = \mb{L_0}(\bs{\tau}) +
\lambda \ast((\tau_2 \mb{d_{10}} - \tau_1 \mb{d_{20}})
\wedge \mb{e_3}), \end{equation} where $ \ast((\tau_2
\mb{d_{10}} - \tau_1 \mb{d_{20}}) \wedge \mb{e_3}) $
is the projection of $ \mb{v}(\bs{\tau}) $ on the subspace
spanned by $ \mb{e_1}, \mb{e_2}$.  Roughly speaking, we
can identify two distinct regions: the preimage of the interior
of the ellipse, where the TDOA map is $1$--to--$1$ and the
source localization is possible, and the preimage of the three
triangles $U_i,\ i=0,1,2$, where the map is $2$--to--$1$
and there is no way to locate the source. The region of
transition is also known in the literature as the bifurcation
region \cite{Coll2012}. In this subsection we offer a complete
geometric description of the above sets.

Notice that that formula (\ref{eq:inv-image}) gives the exact
solutions $\x$ to the localization problem for any given
measurements $\bt$, and it can be used as the starting point
and building block for a local error propagation analysis in
the case of noisy measurements or even with sensor
calibration uncertainty.

\begin{definition} \label{bif-curve}
Let $ E $ be the ellipse in the $ \tau$--plane defined by
$a(\bs{\tau}) = 0$. We call $ \tilde{E} $ its inverse image
contained in the $ x$--plane, and we refer to it as the
bifurcation curve.
\end{definition}

As we said in the discussion at the end of
Subsection \ref{sei-due}, for $\bt\in E$ we have an
admissible source position at an ideal point of the
$x$--plane and, possibly, one more at a finite distance
from the sensors. In the affine plane, the curve
$\tilde{E}$ is exactly the set of these last points.
According to Definition \ref{bif-curve}, $\tilde{E}$
is the preimage of $E$, therefore it can be studied
using formula (\ref{eq:inv-image}). We recall that for
$\bt\in E$ we have $ a(\bs{\tau}) = 0$, therefore
eq. (\ref{eq-deg-2}) has a unique solution in
$\lambda(\bt)=-c(\bs{\tau})/2 b(\bs{\tau})$,
which corresponds to the unique preimage
\begin{equation}\label{eq:paramquintica}
\mb{x}(\bs{\tau}) = \mb{L_0}(\bs{\tau}) -
\frac{c(\bs{\tau})}{2 b(\bs{\tau})} \ast((\tau_2 \mb{d_{10}} -
\tau_1 \mb{d_{20}}) \wedge \mb{e_3}).
\end{equation}
In the next Theorem, we show that the function
(\ref{eq:paramquintica}) restricted on $E$ is a
rational parametrization of degree $5$ of the
bifurcation curve $\tilde{E}$. This means that
$\tilde{E}$ admits a characterization as an
algebraic curve.
\begin{theorem} \label{bif-par}
$\tilde{E} $ is a rational degree--$5$ curve, whose
ideal points are the ones of the lines $ r_0, r_1, r_2$
and the cyclic points of $\PP_\CC^2.$
\end{theorem}
\noindent\emph{Proof.}  See Subsection \ref{sei-quattro}.
\hfill$\square$\vspace{1mm}

\begin{figure}[htb]
\begin{center}
\resizebox{11cm}{!}{
  \includegraphics
  {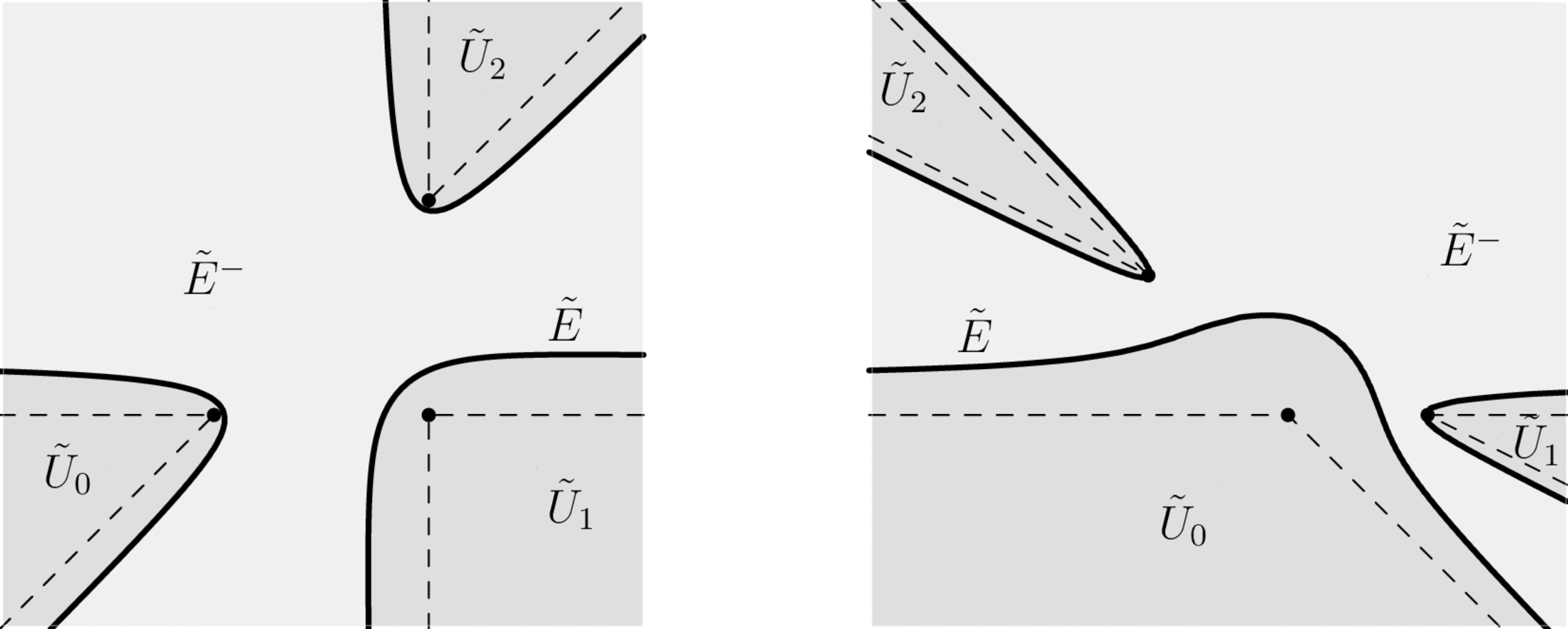}}
\caption{\label{fig:x-plane}Two examples of the different localization
       regions and the quintic $ \tilde{E} $ in the $x$--plane.
       The microphones $\m{0},\m{1}$ and $\m{2}$ are the in the points marked with black dots. Locations of the microphones are $\m{0}=(0,0),\ \m{1}=(2,0)$,
       $ \m{2}=(2,2) $ and $ \m{2}=(-2,2) $ on the left and the right, respectively.
       Each quintic $\tilde{E}$ separates the light gray region $\tilde{E}^-$,
       where the map $\bs{\tau_2}$ is 1--1 and it is possible to localize the
       source, and the dark gray region
       $\tilde{U}_0\cup\tilde{U}_1\cup\tilde{U}_2$, where $\bs{\tau_2}$
       is 2--1 and the localization is not unique.
       The dashed lines represent the degeneracy locus of $\bs{\tau_2}$.}
\end{center}
\end{figure}

In Fig. \ref{fig:x-plane} we show two examples of the
quintic $\tilde{E}$. The real part of $ \tilde{E} $ consists
of three disjoint arcs, one for each arc of $E$ contained
in $ \mbox{Im}(\bs{\tau_2})$.
The points $ \m{0}, \m{1}, \m{2} $ do not belong to
$\tilde{E}$, as their images via $\bs{\tau_2}$ are not
on $E$.
Notice that no arc is bounded, as $ \tilde{E} $ has genus $0$.
In particular, when $\bt $ approaches a point $T_i^\pm$
in $ E \cap C $ the denominator of $\x(\bt)$ approaches to
zero and $ \tilde{E} $ goes to infinity.
As for the smoothness of $\tilde{E}$, the curve has no
self--intersection because each point of the $ x$--plane
has one image in the $\tau$--plane. Furthermore, it is quite
easy to show that cusps are not allowed on $ \tilde{E}$
either. In fact, $E$ is regularly parameterized and the
Jacobian matrix of $\bs{\tau_2} $ is invertible on $\tilde{E}$,
which implies the regularity of $\x(\bt)$.
Quite clearly, on the complex plane $ \tilde{E} $ is bound
to have singular points, as $ \tilde{E} $ is an algebraic
rational quintic curve.
In Appendix C we include the source code in
Singular \cite{DGPS} language for computing the
Cartesian equation (further analysis of the properties of the bifurcation curve is contained in \cite{Compagnoni2013b}).

From Fig. \ref{fig:tauimage} and Fig. \ref{fig:x-plane}, we immediately recognize
what was assessed through simulations and for a specific
sensor configuration in \cite{Spencer2007}. These results,
however, have been here derived in closed form and for
arbitrary sensor geometries, which allows us to characterise
the pre--image in an exhaustive fashion.

The curve $\tilde{E}$ separates the regions of the $x$--plane
where the map $\bs{\tau_2}$ is $1$--to--$1$ or $2$--to--$1$.
We complete the analysis in terms of TDOA--based
localization after introducing and analyzing the preimage of
the open subsets $E^-,U_0,U_1,U_2$ of Im$(\bs{\tau_2})$.

\begin{definition} Let $ \tilde{U}_i $ be the inverse image of
$U_i$ via $ \bs{\tau_2}$, for $ i=0,1,2$, and $ \tilde{E}^- $
be the inverse image of $ E^-$.
\end{definition}
The continuity of $\bs{\tau_2}$ implies that $\tilde{E}^-$,
$\tilde{U}_0$, $\tilde{U}_1$, $\tilde{U}_2$ are open subsets
of the $x$--plane, which are separated by the three arcs of
$\tilde{E}$. Let $F(x,y) = 0$ be a Cartesian equation of
$\tilde{E}$: a point $ \mb{x} \in \tilde{E}^-$ if
$F(\mb{x}) F(\m{0}) < 0$, while
$\x\in\tilde U_0\cup \tilde U_1\cup \tilde U_2$ if
$F(\mb{x}) F(\m{0}) > 0$.
Now, let us focus on the open sets $\tilde{U}_i$.
In this case, without loss of generality, we consider
$i=0$, the other two ones having the same properties.
\begin{proposition}\label{U0Top}
$ \tilde{U}_0 $ has two connected components separated
by $r_1^+\cup r_2^+$, and $ \bs{\tau_2} $ is $ 1$--to--$1 $
on each of them.
\end{proposition}

\noindent\emph{Proof.} See Subsection \ref{sei-quattro}.
\hfill$\square$

\begin{remark} \rm
The previous Proposition can be restated by saying that
$\boldsymbol{\tau_2} : \tilde{U_i} \to U_i $ is a double cover,
for every $ i=0,1,2$. The ramification locus is the union of the
two half--lines through $ \mb{m_i},$ while the branching locus
is union of the two facets of $ P_2 $ through $ R^i$.
\end{remark}

The source localization is possible if $\bt\in E^-$ and,
consequently, $\x\in \tilde E^-$. Otherwise, assume
$\bs{\tau} \in U_0$. According to Proposition \ref{U0Top},
there are two admissible sources in the two disjoint
components of $\tilde{U}_0$. As $\bs{\tau}$ comes
close to $E$, one of its inverse images approaches
a point on $\tilde{E}$, while the other one goes to
infinity. Conversely, if $\bs{\tau}$ approaches
$\partial P_2$, the inverse images of $\bs{\tau}$
come closer to each other and converge to a point
on the degeneracy locus $ r_1^+ \cup r_2^+$.
As we said above, in a realistic noisy scenario, we
end up with poor localization in the proximity of $\tilde{E}$.

\subsection{Proofs of the results} \label{sei-quattro}

\noindent\emph{Proof of Lemma \ref{lm:paramL21}.} As remarked before eq. (\ref{L21}), $L_{21}(\bs{\tau}) $ is a line because $ \mb{D_{10}}(\bs{\tau}) $ and $ \mb{D_{20}}(\bs{\tau}) $ are linearly
independent. Thus, the equation of $L_{21}(\bs{\tau}) $ is (\ref{L21}):
$$
\mb{i}_{\mb{D_0}(\mb{X})}
(\mb{D_{10}}(\boldsymbol{\tau})^\flat \wedge
\mb{D_{20}}(\boldsymbol{\tau})^\flat) = \frac 12 \parallel
\mb{D_{10}}(\boldsymbol{\tau}) \parallel^2
\mb{D_{20}}(\boldsymbol{\tau})^\flat - \frac 12 \parallel
\mb{D_{20}}(\boldsymbol{\tau}) \parallel^2
\mb{D_{10}}(\boldsymbol{\tau})^\flat.
$$

A vector $\mb{v}(\bs{\tau})$ is parallel to $L_{21}(\bs{\tau}) $ if it is a solution of
 $$
\mb{i}_{\mb{v}(\bs{\tau})}(\mb{D_{10}}(\bs{\tau})^\flat \wedge
\mb{D_{20}}(\bs{\tau})^\flat) = 0
$$
From Corollary \ref{2-forme}, this is equivalent to
$$
\mb{v}(\bs{\tau}) = t \ast( \mb{D_{10}}(\bs{\tau})^\flat
\wedge \mb{D_{20}}(\bs{\tau})^\flat)^\sharp = t \ast( \mb{D_{10}}(\bs{\tau}) \wedge \mb{D_{20}}(\bs{\tau}))
$$
for $ t \in \RR.$ We prove the first claim of the Lemma by setting $ t = 1.$

Then, let $\mb{L_0}(\bs{\tau})$ be the intersection point between $L_{21}(\bs{\tau})$ and the $x$--plane. This implies that
$$
\mb{i}_{\mb{D_0}(\mb{L_0}(\bs{\tau}))}\,\bs{\Omega}^\flat=
\frac 12\left(\parallel
\mb{D_{10}}(\boldsymbol{\tau}) \parallel^2
\mb{D_{20}}(\boldsymbol{\tau})^\flat - \parallel
\mb{D_{20}}(\boldsymbol{\tau}) \parallel^2
\mb{D_{10}}(\boldsymbol{\tau})^\flat\right)\wedge\mb{e_3}^\flat.
$$
Therefore, the second claim follows from Lemma \ref{lemma:3forma}.
\hfill$\square$\vspace{3mm}

\noindent\emph{Proof of Proposition \ref{ctau}.} As a real function, $ c(\bs{\tau}) \geq 0 $ because $ \Vert
\mb{D_{10}}(\bs{\tau}) \Vert^2 \mb{d_{20}} - \Vert
\mb{D_{20}}(\bs{\tau}) \Vert^2 \mb{d_{10}} $ is in the subspace
spanned by $ \mb{e_1}, \mb{e_2},$ where $ b $ is positive-defined,
and $ \mb{d_{10}} \wedge \mb{d_{20}} $ is parallel to
$ \mb{e_1} \wedge \mb{e_2} $, whose module is equal to $1$.
Furthermore, $ \mb{d_{10}}, \mb{d_{20}} $ are
linearly independent, thus $ c(\bs{\tau}) = 0 $ if, and only if, $ \Vert
\mb{D_{10}}(\bs{\tau}) \Vert^2 = \Vert \mb{D_{20}}(\bs{\tau})
\Vert^2 = 0,$ i.e. $ \tau_1^2 = d_{10}^2,\
\tau_2^2 = d_{20}^2 $, and the claim follows.

The gradient of $ c(\bt) $ is
$$
\nabla c(\bt) =  \frac 1{\Vert \mb{d_{10}} \wedge \mb{d_{20}} \Vert^2}  \left(
\langle -\tau_1 \mb{d_{20}}, \Vert \mb{D_{10}}(\bs{\tau}) \Vert^2
\mb{d_{20}} - \Vert \mb{D_{20}}(\bs{\tau}) \Vert^2 \mb{d_{10}}
\rangle, \right.
$$
$$
\left. \langle \tau_2 \mb{d_{10}}, \Vert
\mb{D_{10}}(\bs{\tau}) \Vert^2 \mb{d_{20}} - \Vert
\mb{D_{20}}(\bs{\tau}) \Vert^2 \mb{d_{10}} \rangle \right),
$$
therefore it vanishes if $ \Vert\mb{D_{10}}(\bs{\tau}) \Vert^2 =
\Vert \mb{D_{20}}(\bs{\tau})\Vert^2 = 0.$
Hence, in $ \mathbb{A}^2_{\mathbb{C}},$ $ c(\bs{\tau}) =
0 $ is a quartic algebraic curve with four singular points, thus
it cannot be irreducible (see Theorem \ref{bound-sing}). After some simple computations, we
obtain
$$
c(\bs{\tau}) = \left( d_{20} e^{-i\theta} (\tau_1^2 - d_{10}^2) -
d_{10} e^{i\theta}(\tau_2^2 - d_{20}^2) \right) \left( d_{20}
e^{i\theta} (\tau_1^2 - d_{10}^2) - d_{10} e^{-i\theta}(\tau_2^2 -
d_{20}^2) \right)
$$
where $ 2 \theta \in (0, \pi) $ is the angle
between $ \mb{d_{10}} $ and $ \mb{d_{20}}.$
\hfill$\square$\vspace{3mm}

\noindent\emph{Proof of Proposition \ref{atau}.} The equation that defines $E$,
i.e.
$$
a(\bt)=-\Vert\mb{d_{10}} \wedge \mb{d_{20}} \Vert^2 + \Vert \tau_2
\mb{d_{10}} - \tau_1 \mb{d_{20}} \Vert^2 = 0 \; ,
$$
has degree $2$, therefore $E$ is a conic in the $\tau$--space.
Considering the assumption of non-collinearity, $\Vert \tau_2 \mb{d_{10}} -
\tau_1 \mb{d_{20}} \Vert^2 $ is a positively-defined quadratic form
and $\Vert\mb{d_{10}}\wedge\mb{d_{20}}\Vert^2>0$.
$E$ is therefore a non-degenerate ellipse containing real points,
whose center is at $\mb{0}$.
Moreover, it is a simple matter of computation to verify that the intersection between $E$ and $F_i^+$ is the point
$ T_i^+ $ with multiplicity $2$, for $i=0,1,2$, and analogously for $F_i^-$ and
$ T_i^- $. $ E $ is therefore tangent to each side of $P_2$.
This implies also that $ E
\subset P_2$. In order to prove the uniqueness of $ E $, we
embed the $ \tau$--plane $ \RR^2 $ into a projective
plane $ \mathbb{P}^2_{\RR},$ and take
the dual projective plane $ \check{\PP}^2_{\RR} $ (see Definition \ref{dual-plane}). In
$ \check{\PP}^2_{\RR} $ there exists one conic through the $ 6 $
points corresponding to the sides of $ P_2 $ and it is the dual conic $ \check{E} $ (see Definition \ref{dual-curve} and Proposition \ref{deg-dual-conic}). Moreover, $ \check{E} $ is unique by Corollary \ref{uni-conic}. We conclude that the uniqueness of $\check{E} $ is equivalent to the uniqueness of $E$.
\hfill$\square$\vspace{3mm}

\noindent\emph{Proof of Proposition \ref{btau}.} $ C $ is defined by the degree--$3$ polynomial
equation $$\bar{b}(\bs{\tau}) = \langle \tau_2 \mb{d_{10}} - \tau_1 \mb{d_{20}},
\Vert \mb{D_{20}}(\bs{\tau}) \Vert^2 \mb{d_{10}} - \Vert
\mb{D_{10}}(\bs{\tau}) \Vert^2 \mb{d_{20}} \rangle = 0,$$ therefore
it is a cubic curve. It is easy to verify that
\begin{itemize}
\item the equation does not change if we replace $ \tau_i $ with
$ -\tau_i, i=1,2$, therefore $ C $ has a $ 2$--fold rotational
symmetry with respect to $ \mb{0}$;
\item $ C $ contains all the $11$ points of the statement.
\end{itemize}

The partial derivatives of $ \bar{b} $ are $$ \frac{\partial
\bar{b}}{\partial \tau_1} = -\langle \mb{d_{20}}, \Vert
\mb{D_{20}}(\bs{\tau}) \Vert^2 \mb{d_{10}} - \Vert
\mb{D_{10}}(\bs{\tau}) \Vert^2 \mb{d_{20}} \rangle + \langle
\tau_2 \mb{d_{10}} - \tau_1 \mb{d_{20}}, 2 \tau_1 \mb{d_{20}}
\rangle, $$ $$ \frac{\partial \bar{b}}{\partial \tau_2} =
\langle \mb{d_{10}}, \Vert \mb{D_{20}}(\bs{\tau}) \Vert^2
\mb{d_{10}} - \Vert \mb{D_{10}}(\bs{\tau}) \Vert^2 \mb{d_{20}}
\rangle - \langle \tau_2 \mb{d_{10}} - \tau_1 \mb{d_{20}}, 2
\tau_2 \mb{d_{10}} \rangle.$$
After simple calculations, we obtain
$ \nabla \bar{b} (R^0) = 2 d_{10} d_{20} \left( d_{10} d_{20}
- \langle \mb{d_{10}}, \mb{d_{20}} \rangle \right) (1,1) $ and $ \nabla \bar{b} (R^*) = -2 d_{10} d_{20} \left( d_{10}
d_{20} + \langle \mb{d_{10}}, \mb{d_{20}} \rangle \right) (1,1).$
The gradient of $ \bar{b} $ is therefore non-zero at both $ R^0 $ and
$ R^*,$ i.e. $ R^0 $ and $ R^* $ are smooth on $ C.$ Moreover, the
tangent lines to $ C $ at $ R^0 $ and $ R^* $ are orthogonal to $
(1,1) $ therefore they are orthogonal to $ F_0^+.$ For symmetry, the
same holds at $ R^0_1 $ and $ R^*_1.$

In compliance with B\'{e}zout's Theorem \ref{bezout}, $ C $ and $ E $ meet
at $ 6 $ points after embedding the $ \tau$--plane into
$ \mathbb{P}^2_{\mathbb{C}}$, but $ C \cap E =
\{ T_0^\pm, T_1^\pm, T_2^\pm \} $, thus $ C $
and $ E $ intersect transversally.
Moreover, we use B\'{e}zout's Theorem also to prove that $ C $ is not
tangent to any line among $ F_2^\pm,$ and $ F_1^\pm,$ because the
points where the curve $ C $ meets each line are known.

Finally, the line containing $ F_0^+ $ meets $ C $ at $ T_0^+ $
plus two other points whose coordinates solve $ \tau_2
= \tau_1 + d_{21}, (d_{21} \tau_1 - \langle \mb{d_{21}},
\mb{d_{10}} \rangle)^2 + \Vert \mb{d_{10}} \wedge \mb{d_{20}}
\Vert^2 = 0,$ therefore they cannot be real. As a
consequence, according to B\'{e}zout's Theorem, we obtain that
$ C $ and $ F_0^+ $ are not tangent.
By symmetry, $ F_0^- $ is not tangent to $ C$ either.
\hfill$\square$\vspace{3mm}

\noindent\emph{Proof of Proposition \ref{sing-cubic}.}
The gradient at $ \mb{0} $
is $ \nabla \bar{b} (\mb{0}) = (-d_{20}^2 \langle \mb{d_{21}},
\mb{d_{10}} \rangle,  d_{10}^2 \langle \mb{d_{21}}, \mb{d_{20}}
\rangle) \not= (0,0).$ Hence, if $ C $ is not smooth, there are
at least two singular points, because of the $ 2$--fold rotational
symmetry, and so $ C $ is reducible.
As $ C $ contains $ T_0^\pm, T_1^\pm, T_2^\pm, R^0, R^*, R^0_1,
R^*_1, \mb{0},$ the only possible splitting of $ C $ is $
E' \cup L,$ with $ L $ the line through $ R^*, R^*_1, \mb{0},
T_0^\pm,$ and $ E' $ the conic through $ T_2^\pm, T_1^\pm, R^0,
R^0_1.$ The point $ T_0^+ $ is collinear with $ \mb{0} $ and $ R^*
$ if, and only if, there exists $ t \in \RR $ such that $$ \left(
\langle \mb{d_{10}}, \mb{\tilde{d}_{21}} \rangle,
\langle \mb{d_{20}}, \mb{\tilde{d}_{21}} \rangle \right) =
t (-d_{10}, d_{20}),$$ therefore $ t = \langle \mb{\tilde{d}_{20}},
\mb{\tilde{d}_{21}} \rangle $ and $ (d_{20} - d_{10})(
\langle \mb{d_{10}}, \mb{d_{20}} \rangle + d_{10} d_{20} ) = 0.$
Since $ \m{0}, \m{1}, \m{2} $ are not collinear, the second
factor is non--zero and $C$ is singular if, and only if,
$d_{10} = d_{20}.$ The equations of $ L $ and $ E' $ are then straightforward.
\hfill$\square$\vspace{3mm}

\noindent\emph{Proof of Proposition \ref{th:uniquenessC}.} If $ C $ is not smooth, any cubic curve containing
the given points contains the line $ L,$ according to B\'{e}zout's
Theorem. The remaining points lie on a unique conic
$ E',$ therefore $ C = E' \cup L $ is unique.

Let us now assume that $ C $ is smooth. We embed the
$ \tau$--plane into $\mathbb{P}^2_{\mathbb{C}},$ and let
$ X = \{ T_0^\pm, T_1^\pm, T_2^\pm \} $ and
$ Y = \{ \mb{0}, R^0, R^0_1, R^*, R^*_1 \}.$
The defining ideal $ I_X $ of $ X $ is generated by $ a_h(\bs{\tau}),
\bar{b}_h(\bs{\tau}) $ obtained by homogenizing $ a(\bs{\tau}) $
and $ \bar{b}(\bs{\tau}),$ because $ X = E \cap C $, as proven
earlier (see Theorem \ref{ideal-of-points}). The ideal $ I_Y $ of $ Y $ is generated by a degree $ 2 $ and two degree $ 3 $ homogeneous polynomials (see Theorem \ref{ideal-of-points}).
Let $ L_1 $ be the line through $ R^0, R^0_1, \mb{0},$
$ L_2 $ be the line through $ R^*, R^*_1,$ and let $ Q =
L_1 \cup L_2$. $Q$ is a reducible conic that is singular at $ \mb{0}.$
With abuse of notation, we also call $ Q $ the defining polynomial
of $Q$ and so $ Q \in I_Y $ by Definition \ref{coord-ring}. Moreover, $ Y \subset C,$ and so $ \bar{b}_h \in I_Y,$ as well. Finally, let $ C' $ be a further cubic curve, whose
defining polynomial is equal to $ C',$ with abuse of notation, so that $ I_Y = \langle Q, \bar{b}_h, C' \rangle.$

\noindent {\bf Claim:} $ C' $ is not a combination of $Q$, $a_h$, and
$\bar{b}_h.$

\noindent If we assume the contrary, we have
$ C' = q_1 a_h + q_2 Q + q_3 \bar{b}_h $ with $ \deg(q_1)
= \deg(q_2) = 1$, and $\deg(q_3) = 0$.
Consequently, we have $ q_1(p) = 0 $ for every
$ p \in Y,$ because $ a_h(p) \not= 0 $ for each $ p \in Y$. So, $ q_1 \in I_Y $ therefore $ q_1 = 0 $
because $ I_Y $ does not contain linear forms. Then, $ C' $ is a
combination of $ Q $ and $ \bar{b}_h,$ but this is not possible
because $ C' $ is a minimal generator of $ I_Y,$ therefore the claim
holds true.

\noindent Hence, $ I_X + I_Y $ is minimally generated by $ a_h, Q,
\bar{b}_h, C'.$ Moreover, since two conics meet at four
points, $(\mathbb{C}[\tau_0, \tau_1, \tau_2] / (a_h, Q) )_3 $ has dimension
$ 4,$ and we obtain
$$ \dim_{\mathbb{C}} \left( \frac{\mathbb{C}[\tau_0, \tau_1,
\tau_2]}{I_X + I_Y} \right)_3 = 10 - 8 = 2. $$
From the exactness of the short sequence of vector spaces (\ref{M-V-seq})
$$
0 \to \left( \frac{\mathbb{C}[\tau_0, \tau_1, \tau_2]}{I_X \cap
I_Y} \right)_3 \to \left( \frac{\mathbb{C}[\tau_0, \tau_1,
\tau_2]}{I_X} \right)_3 \oplus \left( \frac{\mathbb{C}[\tau_0,
\tau_1, \tau_2]}{I_Y} \right)_3 \to \left( \frac{\mathbb{C}[\tau_0,
\tau_1, \tau_2]}{I_X + I_Y} \right)_3 \to 0
$$
we conclude that the dimension of the first item is $6+5-2=9$
and, finally, $\dim_{\mathbb{C}} \left( I_X \cap I_Y \right)_3
= \dim_{\mathbb{C}} \left( \mathbb{C}[\tau_0,
\tau_1, \tau_2] \right)_3 - \dim_{\mathbb{C}} \left(
\frac{\mathbb{C}[\tau_0, \tau_1, \tau_2]}{I_X \cap I_Y} \right)_3
= 10 - 9 = 1.$
\hfill$\square$\vspace{3mm}

\noindent\emph{Proof of Proposition \ref{th:C_o}.} We embed the $ \tau$--plane into $
\mathbb{P}^2_{\mathbb R}.$ The oval $ C_o $
meets all the lines of the projective plane either in $ 1 $
or in $ 3 $ points, up to count the points with their intersection
multiplicity, as discussed after Harnack's Theorem \ref{harnack} in Appendix B. This implies that $C_o$ contains the inflectional
point $\mb{0}$ of $C$. Moreover, from the proof of Proposition
\ref{btau}, the lines supporting $F_0^\pm $ meet $ C $ at $1$
point each, thus $T_0^\pm \in C_o$.

The possible second oval $C_e$ meets every line at an even number of points ($0$ is allowed) and it cannot meet $C_o$. By contradiction, let us assume that $R^* \in C_e$. Hence, the line $F_2^-$ meets $C_e$ either at $ T_2^-$ or $ R^0_1$. This implies that $C_e$ meets the line $F_0^+$, which is a contradiction. We conclude that $R^*$ and, symmetrically, $R^*_1$ lie on $C_o$.

Again by contradiction, we assume that $ R^0 \in C_e$. By looking
at the intersection points of $ C_e $ with the sides of $P_2$, we
obtain $T_1^+,T_2^+ \in C_e$ and, symmetrically,
$R^0_1, T_1^-, T_2^- \in C_e$.
We also observe that $C_e$ meets the tangent line to $C$ at
$R^0$ exclusively at the point $R^0$ itself, and the same holds
true at $ R^0_1$. As $ C_e $ does not meet $ F_0^\pm$, $C_e$
is constrained into the quadrangle formed by $F_0^\pm $ and
the tangent lines to $ C $ at $ R^0, R^0_1$. This quadrangle
contains $ \mb{0}$, therefore either $ C_e $ is the union of two
disjoint ovals, or $C_e$ meets $C_o$. Both cases are not
allowed, thus $ R^0 \in C_o$. This implies that all the
remaining points lie on $C_o$ and the first claim is proven.

We finish the proof by noting that $C_e$ does not meet
any side of $P_2$ and, on the other hand, $C_e$ cannot be
contained in $ P_2.$
\hfill$\square$\vspace{3mm}

\noindent\emph{Proof of Theorem \ref{bif-par}.} If $\bt\in E$, its preimage is given by (\ref{eq:paramquintica}):
$$ \mb{x}(\bs{\tau}) = \mb{L_0}(\bs{\tau}) -
\frac{c(\bs{\tau})}{2 b(\bs{\tau})} \ast((\tau_2 \mb{d_{10}} -
\tau_1 \mb{d_{20}}) \wedge \mb{e_3}).$$
Hence, $\x(\bt)$ gives a point in
the $ x$--plane both if $ \bs{\tau} \in E \cap
\mbox{Im}(\bs{\tau_2}) $ and if $ \bs{\tau} \in E \setminus
(\mbox{Im}(\bs{\tau_2}) \cup C).$ Moreover, because of the symmetry
properties of the polynomials and vectors involved, we have
$ \x(-\bt)=\x(\bt)$, which means that $\x(\bt)$ is a $ 2$--to--$1 $
map from $ E $ to $ \tilde{E}.$

In order to obtain a parametrization of $ \tilde{E}$, we
consider a parametrization of $ E $ via the pencil of lines
through $ \mb{0}.$
Let $ \tau_1 = \mu_1 t, \tau_2 = \mu_2 t $ be a line through $
\mb{0} $ in the $ \tau$--plane, with $ \bs{\mu} = (\mu_1 , \mu_2)
\in \RR^2 \setminus \{ (0,0) \}.$ The line intersects the ellipse
$ E $ at the two points $t = \pm \Vert \mb{d_{10}} \wedge
\mb{d_{20}} \Vert / \Vert \mu_2 \mb{d_{10}} - \mu_1
\mb{d_{20}} \Vert$, which are symmetrical with respect to
$\mb{0}$. Let $ P_0(\bs{\mu}) = \Vert \mu_2 \mb{d_{10}} -
\mu_1 \mb{d_{20}} \Vert^2$. This is a degree--$2$
homogeneous polynomial that vanishes at the ideal points
of $ E$, therefore it is irreducible over $ \RR$.
By substituting $ \bs{\tau} = \frac{\Vert \mb{d_{10}}
\wedge \mb{d_{20}} \Vert}{\sqrt{P_0(\bs{\mu})}} \bs{\mu},$ all the
functions depend on $ \bs{\mu},$ therefore we obtain $$ \Vert
\mb{D_{10}}(\bs{\mu}) \Vert^2 = \frac 1{P_0(\bs{\mu})} \langle
\mu_1 \mb{d_{20}} - \mu_2 \mb{d_{10}} , \mb{d_{10}} \rangle^2,$$
$$ \Vert \mb{D_{20}}(\bs{\mu}) \Vert^2 = \frac 1{P_0(\bs{\mu})}
\langle \mu_1 \mb{d_{20}} - \mu_2 \mb{d_{10}} , \mb{d_{20}}
\rangle^2 $$ which are both ratios of degree--$2$ homogeneous
polynomials. For our convenience, we set $ P_1(\bs{\mu}) = \langle
\mu_1 \mb{d_{20}} - \mu_2 \mb{d_{10}} , \mb{d_{10}} \rangle^2 $
and $ P_2(\bs{\mu}) = \langle \mu_1 \mb{d_{20}} - \mu_2
\mb{d_{10}} , \mb{d_{20}} \rangle^2.$
As $\bt$ depends on $\bs{\mu}$, the polynomials $b(\bt),c(\bt)$ can be computed as depending on $\bs{\mu}$, obtaining
$$ c(\bs{\mu}) =
\frac 1{4 \Vert \mb{d_{10}} \wedge \mb{d_{20}} \Vert^2
P_0(\bs{\mu})^2} \Vert P_1(\bs{\mu}) \mb{d_{20}} - P_2(\bs{\mu})
\mb{d_{10}} \Vert^2,$$
$$ b(\bs{\mu}) = \frac 1{2 \sqrt{P_0(\bs{\mu})^3} } \langle \mu_2
\mb{d_{10}} - \mu_1 \mb{d_{20}}, P_2(\bs{\mu}) \mb{d_{10}} -
P_1(\bs{\mu}) \mb{d_{20}} \rangle = $$
$$ =\frac 1{2
\sqrt{P_0(\bs{\mu})^3}} \langle \mu_2 \mb{d_{10}} - \mu_1
\mb{d_{20}}, \mb{d_{10}} \rangle \langle \mu_2 \mb{d_{10}} - \mu_1
\mb{d_{20}}, \mb{d_{20}} \rangle \langle \mu_2 \mb{d_{10}} - \mu_1
\mb{d_{20}}, \mb{d_{21}} \rangle.$$
Moreover, $$ \ast((\tau_2 \mb{d_{10}} - \tau_1 \mb{d_{20}}) \wedge
\mb{e_3} ) = \frac{\Vert \mb{d_{10}} \wedge \mb{d_{20}}
\Vert}{\sqrt{P_0(\bs{\mu})}} \ast((\mu_2 \mb{d_{10}} - \mu_1
\mb{d_{20}}) \wedge \mb{e_3} ) $$ and $$
\mb{D_0}(\mb{L_0}(\bs{\mu})) = -\frac 1{2 \Vert \mb{d_{10}} \wedge
\mb{d_{20}} \Vert P_0(\bs{\mu})} \ast \left( (P_1(\bs{\mu})
\mb{d_{20}} - P_2(\bs{\mu}) \mb{d_{10}}) \wedge \mb{e_3}
\right).$$
It follows that $ \mb{D_0}(\mb{x(\bs{\mu})}) $ is a ratio of two degree--$5$
homogeneous polynomials.

The denominator is, up to a non zero
scalar, $ P_0(\bs{\mu}) \langle \mu_2 \mb{d_{10}} - \mu_1
\mb{d_{20}}, \mb{d_{10}} \rangle \langle \mu_2 \mb{d_{10}} - \mu_1
\mb{d_{20}}, \mb{d_{20}} \rangle \langle \mu_2 \mb{d_{10}} - \mu_1
\mb{d_{20}}, \mb{d_{21}} \rangle.$ It is easy to check that, if $
\bs{\mu} $ is such that $ \langle \mu_2 \mb{d_{10}} - \mu_1
\mb{d_{20}}, \mb{d_{ij}} \rangle = 0,$ for $ 0 \leq j < i \leq 2,$
then $ c(\bs{\mu}) \not= 0 $ because $ c $ is non--zero
on $ E,$ and $ \ast \left( (\mu_2 \mb{d_{10}} - \mu_1 \mb{d_{20}})
\wedge \mb{e_3} \right) $ does not vanish. Hence, the numerator
does not vanish at the given $ \bs{\mu}.$ We remark that they give
exactly the ideal points of the lines $ r_0, r_1, r_2.$ The ideal
points of $ E $ are the roots of $ P_0(\bs{\mu}),$ i.e.
$ \bs{\mu_1} = (d_{10} e^{-i\theta}, d_{20} e^{i\theta}) $ and $
\bs{\mu_2} = (d_{10} e^{i\theta}, d_{20} e^{-i\theta}) $ (same
notation of Theorem \ref{ctau}). Here we analyze $ \mb{D_0}(\mb{x(\bs{\mu_1})}), $ being the other case analogous. After tedious, though fairly
straightforward computations, the numerators of the coefficients
of $\ast(\mb{d_{10}} \wedge \mb{e_3})$ and $\ast(\mb{d_{20}}
\wedge \mb{e_3}) $ turn out to be
$$
-d_{10}^6 d_{20}^7 e^{i\theta} (e^{i\theta}d_{10} - e^{-i\theta}d_{20})^2 \sin^4(2\theta),
$$
$$
d_{10}^7 d_{20}^6 e^{-i\theta} (e^{i\theta}d_{10} - e^{-i\theta}d_{20})^2 \sin^4(2\theta).
$$
Without loss of generality, we choose a reference system where
$$
\left\{\begin{array}{l}
\mb{d_{10}}=d_{10}(\cos\theta\,\mb{e_1}+\sin\theta\,\mb{e_2})\\
\mb{d_{20}}=d_{20}(\cos\theta\,\mb{e_1}-\sin\theta\,\mb{e_2})
\end{array} \right.
\ \Rightarrow\
\left\{\begin{array}{l}
\ast(\mb{d_{10}} \wedge \mb{e_3})=d_{10}(-\sin\theta\,\mb{e_1}+\cos\theta\,\mb{e_2})\\
\ast(\mb{d_{20}} \wedge \mb{e_3})=d_{20}(\sin\theta\,\mb{e_1}+\cos\theta\,\mb{e_2})
\end{array} \right..
$$
Therefore, $ \mb{x(\bs{\mu_1})} $ is the ideal point
$$
d_{10}^7 d_{20}^7 (e^{i\theta}d_{10} - e^{-i\theta}d_{20})^2 \sin^5(2\theta)(1:i:0).
$$
It is simple to prove that the coefficient cannot vanish for a value of
$ \theta \in (0, \pi/2).$ We conclude that $ \mb{x(\bs{\mu_1})} $ (and similarly $ \mb{x(\bs{\mu_2})} $) is a cyclic point of $\PP_\CC^2.$ Furthermore, the parametric
representation of $ \tilde{E} $ is given by ratios of degree--$5$
polynomials without common factors, and the claim follows.
\hfill$\square$\vspace{3mm}

\noindent\emph{Proof of Proposition \ref{U0Top}.}
The closure $ \overline{\tilde{U}_0} $ of $ \tilde{U}_0 $ contains
$ \m{0}, r_1^+ \cup r_2^+,$ and the arc of $ \tilde{E} $ inverse image of
the arc of $ E \cap \mbox{Im}(\bs{\tau_2}) $ with endpoints $
T_1^+, T_2^+.$ Furthermore, $ \tilde{U}_0 \cup r_1^+ \cup
r_2^+ \cup \m{0} $ is connected because equal to an oval of $ \tilde{E} $
intersected with the Euclidean $ x$--plane, but $ \tilde{E} \cap (r_1^+
\cup r_2^+ \cup \m{0}) $ is the empty set, because their images in
the $ \tau$--plane do not meet. Hence, $ \tilde{U}_0 $ has two
connected components, and $ \bs{\tau_2}: \tilde{U}_0 \to U_0 $ is
a cover.

Let us now assume that the two inverse images of
$ \bs{\tau_0} \in U_0 $ belong to the same connected
component of $ \tilde{U}_0.$ As $ U_0 $ is path--connected,
from the Path Lifting Theorem (see \cite{kosniowski}), it follows
that the inverse images of any other point
$\bs{\tau} \in U_0 $ belong to the same connected component
of $ \tilde{U}_0$ as well. Let $ \x' $ be a point in the other
connected component of $ \tilde{U}_0,$ with $ \bs{\tau'} =
\bs{\tau_2}(\x').$ Hence, $ \bs{\tau'} $ has three inverse
images, contradicting Theorem \ref{th:imtau2}. Thus,
$\bs{\tau_2} $ is $ 1$--to--$1 $ on each connected
component of $\tilde{U}_0$, as claimed.
\hfill$\square$


\section{The localization problem for special configurations}
\label{sec:special-conf}

In this Section we study the behaviour of the TDOA map
$\bs{\tau_2}$, particularly of its image, under the hypothesis
that $\m{0}$, $\m{1}$, and $\m{2}$ lie on a line $r$.
This is equivalent to assuming that
$\mb{d_{20}} = k \mb{d_{10}}$ for some $k\in \RR$,
$k \not= 0,1$. If $ k < 0,$ then $ \m{0} $ lies between
$ \m{1} $ and $ \m{2},$ if $ 0 < k < 1,$ then $ \m{2}
$ lies between $ \m{0} $ and $ \m{1},$ and finally, if $ k > 1,$
then $ \m{1} $ lies between $ \m{0} $ and $ \m{2}.$
As discussed in Section \ref{sec:image}, in this
configuration, the polygon $P_2$ has only four sides.
\\[2mm]
\indent
Let us first consider the case in which $ \mb{D_{10}}(\bs{\tau})$
and $\mb{D_{20}}(\bs{\tau})$ are linearly dependent.
\begin{lemma} The vectors $ \mb{D_{10}}(\bs{\tau})
$ and $ \mb{D_{20}}(\bs{\tau}) $ are linearly dependent if, and
only if, $ d_{10} \tau_2 - \sgn(k) d_{20} \tau_1 = \tau_2 - k \tau_1 = 0.$
\end{lemma}

\noindent\emph{Proof.}
By definition we have $ \mb{D_{i0}}(\bs{\tau}) = \mb{d_{i0}} +
\tau_i \mb{e_3}$. Under the assumption $\mb{d_{20}} = k \mb{d_{10}}$ of this Section,
$ \mb{D_{10}}(\bs{\tau}) $ and $\mb{D_{20}}(\bs{\tau}) $ are
linearly dependent if, and only if, $\tau_2 = k \tau_1 $ or,
equivalently, $ d_{10} \tau_2 - \sgn(k) d_{20} \tau_1 = 0$,
as claimed.
\hfill$\square$\vspace{1mm}

\noindent The line $ d_{10} \tau_2 - \sgn(k) d_{20} \tau_1 = \tau_2 - k
\tau_1 = 0 $ contains the origin $ \mb{0} $ of the $ \tau$--plane,
and two opposite vertices of $ P_2:$ if $ k > 0,$ then it contains
$ (d_{10}, d_{20}),$ while, if $ k < 0,$ it contains $ (-d_{10},
d_{20}).$

\begin{proposition}\label{prop:ind}
Assume $ \mb{D_{10}}(\bs{\tau}) $ and $ \mb{D_{20}}(\bs{\tau}) $
are linearly dependent. Then, either $ d_{10} \not= \pm \tau_1,$
and the intersection of the planes $ \Pi_1(\bs{\tau}) $ and $
\Pi_2(\bs{\tau}) $ is empty, or $ d_{10} = \pm \tau_1,$ and $
\Pi_1(\bs{\tau}) = \Pi_2(\bs{\tau}) \ni \mb{M_0}.$
\end{proposition}

\noindent\emph{Proof.}
By assumption, we have $ \tau_2 = k \tau_1,$
with $ \mb{d_{20}} = k \mb{d_{10}}, k \not= 0,1.$ As a consequence
$\mb{D_{20}}(\bs{\tau}) = k \mb{D_{10}}(\bs{\tau})$, therefore both
$\mb{D_{20}}(\bs{\tau})^\flat = k \mb{D_{10}}(\bs{\tau})^\flat $
and $ \Vert \mb{D_{20}}(\bs{\tau}) \Vert^2 $ $ = k^2 \Vert
\mb{D_{10}}(\bs{\tau}) \Vert^2.$
Let $ \mb{X} \in \Pi_1(\bs{\tau}) \cap \Pi_2(\bs{\tau}).$ From
equation (\ref{eq-piano}) it follows that
$$
\frac 12 k^2 \Vert \mb{D_{10}}(\bs{\tau}) \Vert^2
= \frac 12 \Vert \mb{D_{20}}(\bs{\tau}) \Vert^2 =
\mb{i}_{\mb{D_0}(\mb{X})}(\mb{D_{20}}(\bs{\tau})^\flat) = \langle
\mb{D_0}(\mb{X}) , \mb{D_{20}}(\bs{\tau}) \rangle =
$$
$$
=  k \ \langle \mb{D_0}(\mb{X}) , \mb{D_{10}}(\bs{\tau}) \rangle = k \
\mb{i}_{\mb{D_0}(\mb{X})}(\mb{D_{10}}(\bs{\tau})^\flat) = \frac 12
k \ \Vert \mb{D_{10}}(\bs{\tau}) \Vert^2
$$
Hence, either $ k^2 = k,$ which is not allowed because
$ k \not= 0, 1,$ or $ \Vert \mb{D_{10}}(\bs{\tau}) \Vert^2 = 0.$
The second condition implies $ d_{10} = \pm \tau_1$ and
$ \Pi_1(\bs{\tau}) \ni \mb{M_0},$ which completes the proof.
\hfill$\square$\vspace{1mm}

\noindent Proposition \ref{prop:ind} implies that the points $
\bs{\tau} $ on the line $ \tau_2 - k \tau_1 = 0 $, with $ \tau_1
\not= \pm d_{10},$ are not in $ \mbox{Im}(\bs{\tau_2}).$
Furthermore, with notation of Definition \ref{rette}, we have
\begin{proposition}
$ \bs{\tau_2}(\mb{x}) = (d_{10}, \sgn(k) d_{20}) $ if, and only
if, $ \mb{x} \in r^c,$ and $ \langle \mb{d_0}(\mb{x}) ,
\mb{d_{10}} \rangle < 0,$ while $ \bs{\tau_2}(\mb{x}) = (-d_{10},
-\sgn(k) d_{20}) $ if, and only if, $ \mb{x} \in r^c,$ and $
\langle \mb{d_0}(\mb{x}) , \mb{d_{10}} \rangle > 0.$
\end{proposition}

\noindent\emph{Proof.}
$ \bs{\tau_2}(\mb{x}) = \pm(d_{10}, \sgn(k) d_{20}) $
if, and only if, $ \mb{x} \in r^c.$ Moreover, given $ \mb{x} \in
r^c,$ $ \tau_1(\mb{x}) = d_{10} $ is equivalent to $ \m{0} $ lying
between $ \m{1} $ ad $ \mb{x}.$
\hfill$\square$\vspace{1mm}

Now, we assume that $ \bs{\tau} $ does not belong to the line $
\tau_2 - k \tau_1 = 0.$
\begin{lemma} \label{lm:paramL21-2}
Assume that $ \mb{D_{10}}(\bs{\tau}) $ and $
\mb{D_{20}}(\bs{\tau}) $ are linearly independent. Then, the
parametric equation of the line $ L_{21}(\bs{\tau}) =
\Pi_1(\bs{\tau}) \cap \Pi_2(\bs{\tau}) $ is $ \mb{L_0}(\bs{\tau})
+ \lambda \mb{v}(\bs{\tau}),$ where $$ \mb{v}(\bs{\tau}) =
\ast(\mb{d_{10}} \wedge \mb{e_3}), $$
$$
\mb{D_0}(\mb{L_0}(\bs{\tau})) = -\frac 1{2 d_{10}^2 (k \tau_1 -
\tau_2)} \ast \left( \mb{v}(\bs{\tau}) \wedge \left( \Vert
\mb{D_{20}}(\bs{\tau}) \Vert^2 \mb{D_{10}}(\bs{\tau}) - \Vert
\mb{D_{10}}(\bs{\tau}) \Vert^2 \mb{D_{20}}(\bs{\tau}) \right)
\right).$$
\end{lemma}

\noindent\emph{Proof.} We use the same reasoning as in Lemma \ref{lm:paramL21}.
\hfill$\square$\vspace{1mm}

\noindent The line $ L_{21}(\bs{\tau}) $ is parallel to the $ x$--plane,
because $ \langle \mb{v}(\bs{\tau}) , \mb{e_3} \rangle = 0,$ thus it is not possible for it to intersect both half--cones
$ C_0^+, C_0^-.$ As for the general case, the line $
L_{21}(\bs{\tau}) $ intersects the cone $ C_0 $ if and only if
\begin{equation} \label{eq-deg-2-2} \Vert \mb{v}(\bs{\tau})
\Vert^2 \lambda^2 + 2 \langle \mb{v}(\bs{\tau}) ,
\mb{D_0}(\mb{L_0}(\bs{\tau})) \rangle \lambda + \Vert
\mb{D_0}(\mb{L_0}(\bs{\tau})) \Vert^2 = 0. \end{equation}
In this case, $ \Vert \mb{v}(\bs{\tau}) \Vert^2 = -\langle
\mb{d_{10}} \wedge \mb{e_3}, \mb{d_{10}} \wedge \mb{e_3} \rangle =
d_{10}^2 > 0,$ $ \langle \mb{v}(\bs{\tau}),
\mb{D_0}(\mb{L_0}(\bs{\tau})) \rangle = 0,$ and
$$
\Vert
\mb{D_0}(\mb{L_0}(\bs{\tau})) \Vert^2 = -\frac{(d_{10}^2 -
\tau_1^2)(d_{20}^2 - \tau_2^2)(d_{21}^2 - (\tau_1 - \tau_2)^2)}{4
d_{10}^2 (k \tau_1 - \tau_2)^2} =
$$
$$
= -\frac{\Vert
\mb{D_{10}}(\bs{\tau}) \Vert^2 \ \Vert \mb{D_{20}}(\bs{\tau})
\Vert^2 \ \Vert \mb{D_{21}}(\bs{\tau}) \Vert^2}{4 d_{10}^2
(k\tau_1 -\tau_2)^2}.
$$
As a consequence, the line $ L_{21}(\bs{\tau}) $ intersects the
cone $ C_0 $ if, and only if, $ c(\bs{\tau}) \leq 0.$
Moreover, the two intersections belong to $ C_0^- $ if, and
only if, $ \langle \mb{D_0}(\mb{L_0}(\bs{\tau})) , \mb{e_3}
\rangle > 0,$ which means that $$ \frac{k \tau_1^2 -
\tau_2^2 + d_{10}^2 (k^2-k)}{2 (\tau_2 - k \tau_1)} > 0.$$

\begin{figure}[htb]
\begin{center}
\resizebox{5.3cm}{!}{
  \includegraphics
  {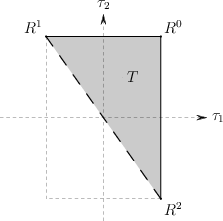}}
  \caption{\label{fig:tauimagespecial}The image of $ \bs{\tau_2}$ under the assumption that $\m{0}$ lies on the segment between $\m{1}$ and $\m{2}$. In the gray region $T$ the map $ \bs{\tau_2} $ is $ 2$--to--$1$. Along the horizontal and vertical sides of $T$ the map is $ 1$--to--$1,$ with the exception of the vertices $R^1,R^2$, where the fibers of $\bs{\tau_2}$ are not finite. Finally, the dashed side of $T$ is not in Im($\bs{\tau_2}$).  }
\end{center}
\end{figure}

Now, we are able to describe the image of $
\bs{\tau_2}.$ The results of the next theorem are illustrated in Fig. \ref{fig:tauimagespecial}, in the subcase with $k<0$, i.e. $\m{0}$ between $\m{1}$ and $\m{2}$ (the other two subcases are similar).

\begin{theorem}\label{th:imspecial}
Let us assume that $\mb{d_{20}} = k \mb{d_{10}}, k \not= 0,1$
and let $ R^i $ be the image of the point $ \m{i} $ in the interior
of $ r^0.$ Then, the image of $ \bs{\tau_2} $ is the triangle
$ T $ with vertices $(d_{10}, \sgn(k) d_{20}),
(-d_{10}, -\sgn(k) d_{20}), R^i$ minus the open segment with
endpoints $(d_{10}, \sgn(k) d_{20}),$ $ (-d_{10}, -\sgn(k) d_{20}).$
Moreover, given $ \bs{\tau} \in \mbox{Im}(\bs{\tau_2}),$ we have
$$ \vert \bs{\tau_2}^{-1}(\bs{\tau}) \vert =
\left\{
\begin{array}{cl}
\infty & \mbox{ if } \bs{\tau} = \pm(d_{10},\sgn(k) d_{20}), \\
2 & \mbox{ if } \bs{\tau} \in \mathring{T}, \\
1 & \mbox{ otherwise.} \end{array} \right.$$
\end{theorem}

\noindent\emph{Proof.}
The case $ \bs{\tau} = \pm(d_{10}, \sgn(k) d_{20}) $ has already
been studied, as well as the case $ \bs{\tau} $ on the line
through them. Let us assume that $ \bs{\tau} $ does not lie on the
line $d_{10} \tau_2 - \sgn(k) d_{20} \tau_1=0.$ Eq.
(\ref{eq-deg-2-2}) has two real distinct roots if, and only if, $
c(\bs{\tau}) < 0.$ Quite clearly this happens if, and only if,
$ \bs{\tau} \in \mathring{P_2}.$ Furthermore, the same equation
has a multiplicity--two root if, and only if, $ c(\bs{\tau}) = 0,$
i.e. $ \bs{\tau} \in \partial P_2.$ Finally, the intersection points of $
L_{21}(\bs{\tau}) $ and $ C_0 $ are in $ C_0^- $ if, and only if,
$ \frac{k \tau_1^2 - \tau_2^2 + d_{10}^2 (k^2-k)}{2 (\tau_2 - k
\tau_1)} > 0.$

The equation $ e(\bs{\tau}) = k \tau_1^2 - \tau_2^2
+ d_{10}^2 (k^2-k)=0 $ defines a conic $ C' $ through the four
points $ (\pm d_{10}, \pm d_{20}).$ If $ k < 0,$ $ C' $ is an
ellipse with real points, and so $ P_2 $ is inscribed into $ C'.$
Moreover, $ e(\mb{0}) > 0,$ and so $ e(\bs{\tau}) > 0 $ for each $
\bs{\tau} \in P_2 $ except the four points $ (\pm d_{10}, \pm
d_{20}).$ If $ k > 0,$ $ C' $ is a hyperbola. The tangent line to
$ C' $ at $ R^0 = (d_{10}, d_{20}) $ is $ F_0^+ $ if $ k > 1,$ ($
F_0^- $ if $ 0 < k < 1,$ respectively), while the tangent line to
$ C'$ at $ R^2 $ is $ F_0^- $ if $ k > 1 $ ($ F_0^+ $ if $ 0 < k <
1,$ respectively). Finally, if $0 < k < 1 $ then $R^0 $ and
$ (d_{10}, -d_{20}) $ belong to the same branch of $ C'$
($ (-d_{10}, d_{20}) $ if $ k > 1,$ respectively).
As a consequence, $ e(\bs{\tau}) $ does not change sign in
$ P_2.$ More precisely, $ e(\bs{\tau}) $ has the same sign
as $ k^2-k $ for each $ \bs{\tau} \in P_2,$ except for
$\bt= \pm (d_{10}, d_{20}) $, where it vanishes.

On the other hand, after a rather strightforward computation we
find that the linear polynomial $ \tau_2 - k \tau_1 $ has the same
sign as $ k^2-k$ at the vertex $R^i$, therefore the ratio
$ \frac{k \tau_1^2 - \tau_2^2 + d_{10}^2 (k^2-k)}{2 (\tau_2 - k \tau_1)}$
is positive at any point in the interior of the triangle $T$. This
proves that each point $ \bs{\tau} $ in $ \mathring{T} $ has
two distinct preimages.

Finally, for $\bt$ on the two remaining sides of $T$, eq.
(\ref{eq-deg-2-2}) has only one root of multiplicity $ 2$, which
implies $|\bs{\tau_2}^{-1}(\bt)|=1$.
\hfill$\square$\vspace{1mm}

The preimages of
$ \bs{\tau} \in \mbox{Im}(\bs{\tau_2})$ in the $x$--plane are
$$ \mb{x}(\bt)
= \pi(\mb{L_0}(\bs{\tau})) + \lambda\, \mb{v}(\bs{\tau}),$$
where $
\lambda = \pm \Vert \mb{D_{10}}(\bs{\tau}) \Vert \ \Vert
\mb{D_{20}}(\bs{\tau}) \Vert \ \Vert \mb{D_{21}}(\bs{\tau}) \Vert
/ 2 d_{10}^2 \vert k \tau_1 - \tau_2 \vert$ and $ \pi $ is the
projection onto the $ x$--plane. Moreover, we have
$$
\mb{D_0}(\pi(\mb{L_0}(\bs{\tau}))) = \frac{\Vert
\mb{D_{20}}(\bs{\tau}) \Vert^2 \tau_1 - \Vert
\mb{D_{10}}(\bs{\tau}) \Vert^2 k \tau_2}{2 d_{10}^2 (k \tau_1 -
\tau_2)}\, \mb{d_{10}}.
$$

In order to interpret the results, we notice that in the aligned
configuration, the foci of $ A_1(\bs{\tau}), A_2(\bs{\tau}) $
belong to the line $r$, therefore the two level sets
$A_1(\bs{\tau}), A_2(\bt) $ are both symmetrical with respect
to $r$.  We are in the $1$--to--$1$ situation if, and only if, the source
$\x$ belongs to $r^0$, corresponding to $ A_1(\bt), A_2(\bt) $ tangentially
intersecting at $\x$.
In the general case, when $\bt\in\mathring{T}$, the level sets meet
at two distinct symmetrical points. This agrees with the classical
statement that it is not possible to distinguish between symmetric
configuration of the source, with respect to $r$, using a linear array of
receivers.

The degenerate situation occurs for $\x\in r^c$, dual to $\bt$ equal both to $ (d_{10}, \sgn(k) d_{20}) $ and $(-d_{10}, -\sgn(k) d_{20}) $. In this case the localization of the source is totally unavailable, because the preimages contain infinitely many points of the $x$--plane. Finally, the points on the interior of the side $ \tau_2 - k \tau_1 = 0 $ correspond to $ A_1(\bs{\tau}), A_2(\bs{\tau}) $ with
parallel asymptotic lines and empty intersection.


\section{The image of the complete TDOA map}
\label{sec:symmetry}

\begin{figure}[htb]
\begin{center}
\resizebox{8cm}{!}{
  \includegraphics
  {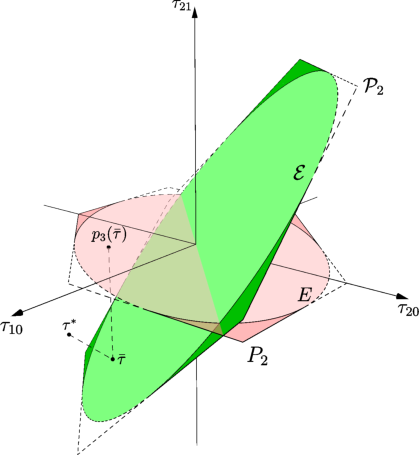}}
  \caption{\label{fig:taucomimage}The image of
  $\bs{\tau_2^*}$ is the subset of $\mathcal{P}_2$ in
  green, while the image of $ \bs{\tau_2}$ is the subset
  of $P_2$ in red. There is a 1--to--1 correspondence
  between Im($\bs{\tau_2^*}$) and Im($\bs{\tau_2}$)
  via the projection map $p_3$.
  In the lightly shaded regions, the TDOA maps are
  $1$--to--$1$, while in the more darkly shaded regions
  the maps are $2$--to--$1$.
  As explained in Section \ref{sec:Conclusions}, three
  noisy TDOAs define a point $\bs{\tau^*}$ outside
  $\mathcal{P}_2$. The Maximum Likelihood Estimator
  computes the projection
  $\bar{\bt}=p_{\mathcal{H}}(\bs{\tau^*})$ on
  $\mathcal{P}_2$, then the estimated source position
  is computed as
  $\bar{\x}=\bs{\tau_2^*}^{-1}(\bar{\bt})=\bs{\tau_2}^{-1}(p_3(\bar{\bt}))$.}
\end{center}
\end{figure}

In Section \ref{sec:p2m} we explained that the relation between $\bs{\tau_2}$ and $\bs{\tau_2^*}$ is given by the projection $p_3$ from the plane $\mathcal{H}\subset\RR^3$ to $\RR^2$ via the equality $ \bs{\tau_2} = p_3 \circ \bs{\tau_2^*}.$ As $p_3$ is invertible, it holds that $ \bs{\tau_2^*} = p_3^{-1} \circ \bs{\tau_2}$ and consequently we have the following result:
\begin{theorem}\label{th:TDOAvsCTDOA}
Im($\bs{\tau_2^*}$)=$p_3^{-1}$(Im($\bs{\tau_2}$)). More precisely, let $\bs{\tau^*}\in\mathcal{H}$, then $\bs{\tau_2^*}^{-1}({\bs{\tau^*}})=\bs{\tau_2}^{-1}({\bs{\tau}}),$ where $\bt=p_3(\bs{\tau^*}).$
\end{theorem}
Theorem \ref{th:TDOAvsCTDOA} allows us to give the explicit description of Im($\bs{\tau_2^*}$), thus reaching one of the main objectives we set ourselves in Section \ref{sec:p2m}. We start by defining the relevant subsets of $\mathcal{H}$.
\begin{definition} \label{complete-subsets}
Assuming $0\leq i,j,k\leq 2$ distinct, in the $ \tau^*$--space we set:
\begin{itemize}
\item
$ \mathcal{P}_2 = \{\bs{\tau}^* \in \mathcal{H} \ \vert \
\Vert\mb{D_{ji}}(\bs{\tau}^*) \Vert^2 \geq 0 \mbox{ for every } i, j\};$
\item
$\mathcal{F}_{k}^+ = \{ \bs{\tau}^* \in \mathcal{P}_2 \ \vert \
\Vert \mb{D_{ji}}(\bs{\tau}^*) \Vert^2 = 0,
\PM{\mb{D_{ji}}(\bs{\tau}^*) , \mb{e_3}} < 0 \};$ \item
$\mathcal{F}_{k}^- = \{ \bs{\tau}^* \in \mathcal{P}_2 \ \vert \
\Vert \mb{D_{ji}}(\bs{\tau}^*) \Vert^2 = 0,
\PM{\mb{D_{ji}}(\bs{\tau}^*) , \mb{e_3}} > 0 \};$
\item
$ \mathcal{E}_k = \{\bs{\tau}^* \in \mathcal{P}_2 \ \vert \
\Vert\mb{D_{ik}}(\bs{\tau}^*) \wedge \mb{D_{jk}}(\bs{\tau}^*) \Vert^2 =0 \}.$
\end{itemize}
\end{definition}
As the above definitions are stated using the exterior
algebra formalism, for completeness we observe that
$\mathcal{H}$ can also be described in similar terms:
\begin{equation*}
\mathcal{H} = \{ \bs{\tau}^* \in \RR^3 \ \vert \
\mb{D_{10}}(\bs{\tau}^*) - \mb{D_{20}}(\bs{\tau}^*)
+ \mb{D_{21}}(\bs{\tau}^*) = \mb{0} \}
\end{equation*}
 In Fig. \ref{fig:taucomimage} we show an example
of Im($\bs{\tau_2^*}$) along with its projection
Im($\bs{\tau_2}$).

As a consequence of Theorem \ref{th:TDOAvsCTDOA},
the structure of Im($\bs{\tau_2^*}$) turns out to be
similar to that of Im($\bs{\tau_2}$), thus we can
omit the proofs and go over the main facts about
$\bs{\tau_2^*}$.
\begin{itemize}
\item
$\bs{\tau_2^*}$ is a local diffeomorphism between the
$x$--plane and $\mathcal{H},$ with the exception of the
degeneracy locus $\cup_{i=0}^2 (r_i^- \cup r_i^+)$,
as described in Theorem \ref{n=2}.
\item
$\mathcal{P}_2$ is the convex polygon given by
$p_3^{-1}(P_2)$, whose facets are
$\mathcal{F}_k^\pm=p_3^{-1}({F_k^\pm})$.
The image of $\bs{\tau_2^*}$ is a proper subset of
$\mathcal{P}_2$ and, in particular, the image of the
degeneracy locus is a subset of the facets.
\item
$ \mathcal{E}_k$ does not depend on $k$. If the
points $ \m{0}$, $\m{1}$ and $\m{2}$ are not aligned,
then we have $\mathcal{E}_k=p_3^{-1}(E)$.
Therefore, $ \mathcal{E}_k$ is the unique ellipse that
is tangent to each facet of the hexagon $\mathcal{P}_2$.
The cardinality of each fiber of $\bs{\tau_2^*}$ is equal
to that of the corresponding fiber of $\bs{\tau_2}$, as
described in Theorem \ref{th:imtau2} and in
Proposition \ref{U0Top}.
\item
If the points $ \m{0}, \m{1}, \m{2}$ are aligned,
then $\mathcal{E}_k$ is one of the diagonals of the
quadrangle $\mathcal{P}_2$. The cardinality of each
fiber of $\bs{\tau_2^*}$ is equal to that of the
corresponding fiber of $\bs{\tau_2}$, as described
in Theorem \ref{th:imspecial}.
\end{itemize}

\begin{remark}\label{rm:symmetry}\rm
In the definition of $\bs{\tau_2^*}$ we notice a natural
symmetry among the points $\m{0}$, $\m{1}$ and $\m{2}$,
which is lost in $\bs{\tau_2}$ as we elected $ \mb{m_0}$
to be the reference microphone.
As noticed in Section \ref{sec:p2m}, by taking
$p_1\circ\bt_2^*$ or $p_2\circ\bt_2^*$ we define different
TDOA maps, with different reference microphones.
Quite obviously, their properties are similar to those
of $\bs{\tau_2}$ studied in Sections \ref{sec:loc-gen}
and \ref{sec:special-conf}, in fact $p_1 \circ p_3^{-1}$
and $ p_2 \circ p_3^{-1} $ are invertible maps between
the images of the TDOA maps, factorizing on
Im($\bs{\tau_2}^*$). Although such maps are, in fact,
equivalent, some of their properties could be more or
less difficult to check depending on the chosen
reference point.
For example, the lines $F_0^\pm$ become parallel
to the reference axes when applying
$p_i \circ p_3^{-1} $ for $i=1$ or $2$.
\end{remark}
\begin{remark}\rm
The previous remark implies that $p_i \circ p_3^{-1}$
sends the ellipse $E$ onto the ellipse associated to
the TDOA map $p_i \circ \bt_2^*$, but this does not
happen for the cubic curve $C$. In fact, both the
cubics associated to $ p_i \circ \bt_2^*$, $i=1,2$
do not contain (the transformations of) 4 of the 11
points characterizing $C$ in Proposition
\ref{th:uniquenessC}: $R^0,R^*,R^0_1,R^*_1$.
This, however, is not an issue for localization
purposes, as the image of any TDOA map only
depends on $ C \cap E=E\cap P_2$.
\end{remark}

\section{Impact assessment}\label{sec:Impact}

As anticipated in the Introduction, a complete
characterization of the TDOA map constitutes an important
building block for tackling a wide range of
more general and challenging problems. For example,
we could optimize sensor placement in terms of robustness
against noise or measuring errors. More generally, we could
embark into a statistical analysis of error propagation or
consider more complex scenarios where the uncertainty lies
with sensor synchronization or spatial sensor placement.
While these general scenarios will be the topic of future
contributions, in this Section we can already show an
example of how to jointly use local and global analysis
to shed light on the uncertainty in localisation
problems.

\begin{figure}[htb]
\begin{center}
\resizebox{7.4cm}{!}{
  \includegraphics
  {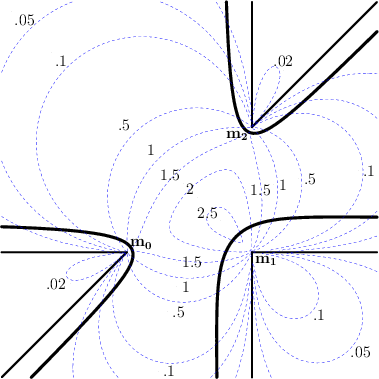}}
\caption{\label{fig:quinticaJac}Level sets of the absolute
   value of the Jacobian determinant $\vert \det(J(\mb{x}))\vert$
   of $ \boldsymbol{\tau_2}$ (values are marked next to them).
   The picture also shows the bifurcation curve $\tilde{E} $ and
   the degeneracy locus $ \det(J(\mb{x})) = 0 $, on the $x$--plane.
   Given the microphones in $\m{0}=(0,0)$, $\m{1}=(2,0)$, and
   $\m{2}=(2,2)$, the best accuracy of the localization is obtained
   in the region that lies the closest to the center of the triangle.
   Notice that, although $\det(J(\mb{x}))$ is not affected by
   $\tilde{E}$, in the proximity to this curve the localization
   fails due to the global properties of $\boldsymbol{\tau_2}$.}
\end{center}
\end{figure}

A possible approach to the study of the accuracy of
localization is based on the linearization of the
TDOA model (see \cite{Compagnoni2012,Raykar2005}).
Usually this analysis is pursued in a statistical context,
but it essentially involves the analysis of the Jacobian
matrix $J(\x)$ of $\bs{\tau_2}$ and its determinant
det$(J(\x))$. In the differential geometry interpretation,
the absolute value of Jacobian determinant is the ratio
between the areas of two corresponding infinitesimal
regions in the $\tau$--plane and in the $x$--plane,
under the action of the map $\bs{\tau_2}$. As a
consequence, the quality of the localization is best in
the regions of maximum of $|$det$(J(\x))|$, where the
TDOAs are highly sensitive to differences of
source position. This local analysis is equivalent,
up to a costant factor, to that of the map $\bs{\tau_2^*}$.
Starting from expression (\ref{eq:Jacobianot2}), in
Fig. \ref{fig:quinticaJac} we display the level sets of
$|$det$(J(\x))|$ along with the geometric configuration
of sensors and with the curves that we displayed earlier
in Fig. \ref{fig:x-plane}.
Fig. \ref{fig:quinticaJac} shows that the local error analysis does
not take count of the global aspects of the localization. In particular,
$\vert \det(J(\mb{x})) \vert$ becomes quite large in the proximity
of the sensors. In these areas, however, localisation is not
accurate because of their proximity to the bifurcation curve $\tilde{E}$
and the overlapping to the sets $\tilde{U}_i$.
Having access to a complete global characterisation of the
TDOA map allows us to predict this behaviour.


\section{Conclusions and perspectives}\label{sec:Conclusions}

In this manuscript we offered an exhaustive mathematical
characterization of the TDOA map in the planar case of
three receivers. We began with defining the
non--algebraic complete TDOA map $\boldsymbol{\tau_2^*}$.
We then derived a complete characterization of both
Im$(\bs{\tau_2^*})\subset \RR^3$ and the various
preimage regions in the $x$--plane.
We found that Im$(\bs{\tau_2^*})$ is a bounded subset
of the plane $\mathcal{H}$ and, in particular, we showed
that the image is contained in the convex polygon
$\mathcal{P}_2$. We also described the subsets of the
image in relation to the cardinality of the fibers, i.e. the
loci where the TDOA map is $1$--to--$1$ or $2$--to--$1$,
which provided a complete analysis of the a-priori identifiability
problem.
On the $x$--plane, we defined the degeneracy locus, where
$\boldsymbol{\tau_2^*} $ is not a local diffeomorphism, and
we described the sets where $\boldsymbol{\tau_2^*} $ is
globally invertible and those where it is not.

We conducted our analysis using various mathematical tools,
including multilinear algebra, the Minkowski space, algebraic
and differential geometry. Indeed, these tools may seem too
sophisticated for a problem as ``simple" as that of TDOA-based
localization. After all, this is a problem that, in the engineering literature,
is commonly treated as consolidated or even taken for granted.
As explained in the Introduction, however, the purpose of this
work was twofold:
\begin{enumerate}
\item to derive analytically and in the most general sense
what was preliminarily shown in a fully simulative fashion
in \cite{Spencer2007}, and to make the analysis valid for
arbitrary sensor geometries;
\item to offer a complete characterisation of the TDOA map,
to be applied to the solution of more general problems.
\end{enumerate}

The first purpose was amply proven throughout the
manuscript (Sections 2 to 8). What remains to be shown
is how this analysis can pave the way to a deeper
understanding of the localization problem in more general
settings, such as in the presence of noisy measurements
(propagation of uncertainty) or even in the presence of
uncertainty in the sensor calibration and/or in their
synchronization.
An early discussion in this direction was offered in Section 9
where we described how errors propagate in a three-sensor
setup based on local analysis and showed that, without
a global perspective on the behaviour of the TDOA map
we could be easily led to drawing wrong conclusions.

The authors are currently working on the extension to arbitrary
distributions of sensors both in the plane and in the 3D Euclidean
space, using similar techniques and notations. In particular, the model
can be encoded as well in a TDOA map $ \boldsymbol{\tau_n^*}$,
whose image is a real surface and a real threefold, respectively.
We expect the bifurcation locus and the $2$--to--$1$ regions to
become thinner as the number of receivers in general position
increase, although they do not immediately disappear (for
example, in the planar case and $n=3,$ there are still curves
in the $x$--plane where the localization is ambiguous).
The precise description of $ \boldsymbol{\tau_n^*}$ is needed
also for the study of the localization with partially synchronized
microphones. In fact, in this scenario not all TDOAs are available
and, in our description, this is equivalent to considering some
kind of projection of Im($\bs{\tau_n^*}$), just like the relationship
between $\bs{\tau_2^*}$ and $\bs{\tau_2}$, explained in
Sections \ref{sec:p2m} and \ref{sec:symmetry}.

In a near future we also want to pursue the study of the
nonlinear statistical model, following a similar gradual approach
to the one of this manuscript. Even in this respect, the knowledge
of the noiseless measurements set Im($\bs{\tau_n^*}$) constitutes
the starting point for any further advances on the study of the
stochastic model. Roughly speaking, a vector of measurements
$\bt^*$ affected by errors corresponds to a point that lies close
to the set Im($\bs{\tau_n^*}$) and the localization is a
two-step procedure: we can first estimate
$\bar{\bt}\in\rm{Im}(\bs{\tau_n^*})$ from $\bt^*$, then we
evaluate the inverse map $\bs{\tau_n^*}^{-1}(\bar{\bt})$.

We can give a real example of this process in the case of the
complete TDOA model defined through the map $\bs{\tau_2^*}$.
In a noisy scenario (e.g. with Gaussian errors), a set of three
TDOAs gives a point $\bs{\tau^*}$ in the three dimensional
$\tau^*$--space, that with probability 1 is not on the plane
$\mathcal{H}$. A standard approach to obtain an estimation
$\bar{\x}$ of the source position is through Maximum Likelihood
Estimator (MLE). With respect to the discussion of the previous
paragraph, it is well known that the estimated
$\bar{\bt}\in\rm{Im}(\bs{\tau_2^*})$ given by MLE is the
orthogonal projection of $\bs{\tau^*}$ on the noiseless
measurements set, i.e. the projection $p_\mathcal{H}(\bs{\tau^*})$
on $\mathcal{H}$ (see \cite{Pazman1993,Amari2000} and
Fig. \ref{fig:taucomimage}) therefore
$\bar{\x}=\bs{\tau_2^*}^{-1}(p_\mathcal{H}(\bs{\tau^*}))$.

A similar reasoning applies to the more complex case of
$\bs{\tau_n^*}$. In particular, any estimator has a geometrical
interpretation and the relative accuracy depends on the
(non trivial) shape of Im($\bs{\tau_n^*}$). Possible techniques
to be applied come from Information Geometry \cite{Amari2000,Kobayashi2013},
an approach that proved successful in similar situations and
that is based on the careful description of Im($\bs{\tau_n^*}$).
With this in mind, notice that our characterization of the TDOA
model in algebraic geometry terms becomes instrumental for
understanding and computing the MLE. Very recently, novel
techniques have been developed and applied to to similar situations,
in cases where scientific and engineering models are
expressed as sets of real solutions to systems of polynomial
equations (see, for example, \cite{thomas2011,Draisma2013,Kobayashi2013}).
The somewhat surprising fact that, although the TDOA map
is not polynomial, all the loci involved in the analysis of
$\bs{\tau_2^*}$ are algebraic or semi--algebraic, is a
promising indicator of the effectiveness of this approach.


\ack
The authors would like to thank E. Schlesinger, F. Belgiorno,
S. Cacciatori, M. Conti and V. Pata for the useful discussions
and suggestions during the preparation of this work and the
anonymous referees for the valuable remarks on the preliminary
version of the paper.
\bigskip


\appendix
\noindent {\bf Appendices}

\renewcommand{\thesection}{A}
\section{The exterior algebra formalism}\label{app:A}

In this appendix, we recall the main definitions and some useful results about the exterior algebra of a real vector space. At the end, we analyze in full detail the two main examples that we use in the paper, namely the $2$--dimensional Euclidean case and the $3$--dimensional Minkowski one. The literature on the subject is wide and we mention \cite{Abraham1988} among the many possible
references.

Let $ V $ be a $ n$--dimensional $ \RR$--vector space. Adopting
a standard notation, $ \wedge V $ is the exterior algebra of $
V $ (hence, $ \wedge^k V = 0 $ for each $ k \geq n+1 $ while $
\wedge^k V $ has dimension ${n \choose k}$ for $ k=0, \dots, n$). Roughly speaking, the symbol $ \wedge $ is skew--commutative, and linear with respect to each factor. Hence, given the basis $ (\mb{e_1} , \dots, \mb{e_n}) $ of $ V,$ the reader can simply think at $ \wedge^k V $ as the $\RR$--vector space spanned by $ \mb{e_{i_1}} \wedge \dots \wedge \mb{e_{i_k}} $ for $ 1 \leq i_1 < \dots < i_k \leq n.$

We choose a non--degenerate, symmetric bilinear
form $ b: V \times V \to \RR,$ and
an orthonormal basis $ B = (\mb{e_1}, \dots,
\mb{e_n}) $ with respect to $ b$, which means
$$ b(\mb{e_i},\mb{e_j}) = \left\{ \begin{array}{cl} 1 & \mbox{ if } i=j=1, \dots, r, \\ -1 & \mbox{ if } i = j = r+1, \dots, n, \\ 0 & \mbox{ if } i \not= j. \end{array} \right. $$ The couple $ (r, n-r) $ is the signature of $ b.$ By setting $ \langle \mb{u}, \mb{v} \rangle =
b(\mb{u}, \mb{v}) $ and $
\parallel \mb{u}
\parallel^2 = b(\mb{u}, \mb{u}),$ we can easily compute their
expression in coordinates with respect to $ B,$ and we have
\begin{equation} \label{minkowski1} \fl\qquad \begin{array}{l}
\PM{\sum_{i=1}^n u_i\mb{e_i}, \sum_{j=1}^n v_i \mb{e_i}}=
u_1 v_1 + \dots + u_r v_r - u_{r+1} v_{r+1} - \dots - u_n v_n \\ \\
\parallel \sum_{i=1}^n u_i \mb{e_i} \parallel^2 =
u_1^2 + \dots + u_r^2 - u_{r+1}^2 - \dots - u_n^2 \end{array}
\end{equation}
The inner product in $ \wedge^k V$ is defined by
\begin{equation} \label{minkowski2}\fl\qquad\qquad
\langle \mb{u_1} \wedge \dots \wedge \mb{u_k},
\mb{v_1} \wedge \dots \wedge \mb{v_k} \rangle = \det \left(
\begin{array}{ccc} \langle \mb{u_1}, \mb{v_1} \rangle & \dots & \langle
\mb{u_1}, \mb{v_k} \rangle \\ \vdots & \ & \vdots \\ \langle
\mb{u_k}, \mb{v_1} \rangle & \dots & \langle \mb{u_k}, \mb{v_k}
\rangle \end{array} \right) \end{equation} and extended by
linearity. For example, $ (\mb{e_1} \wedge \mb{e_2}, \mb{e_1}
\wedge \mb{e_3}, \dots, \mb{e_{n-1}} \wedge \mb{e_n} ) $ is an orthonormal
basis of $ \wedge^2 V,$ while $
(\boldsymbol{\omega} = \mb{e_1} \wedge \dots \wedge \mb{e_n} )
$ is an orthonormal basis of $ \wedge^n V $ with $ \parallel
\boldsymbol{\omega} \parallel^2 = (-1)^{n-r}.$

Finally, from the choice of $ \boldsymbol{\omega} $ as positive
basis of $ \wedge^n V,$ and from the fact that the natural
concatenation of a $k$--form and a $(n-k)$--form gives a
$n$--form, one recovers the classical Hodge $\ast$ operator
$\RR \to \wedge^n V$, $V \to \wedge^{n-1} V, \dots, \wedge^{n-1} V \to V$,
and $\wedge^n V \to \RR,$ that are all isomorphisms.
\begin{definition} \label{hodge}
Given $ \bs{\omega} \in \wedge^n V, \bs{\omega} \not= \mb{0},$ there exists a unique linear map $ \ast: \wedge^k V \to \wedge^{n-k} V $ that verifies the condition
$$ \mb{x}\wedge\ast\mb{y}=\PM{\mb{x}, \mb{y}}\bs{\omega}$$ for every $ \mb{x}, \mb{y} \in \wedge^{k}V.$
\end{definition}

\begin{theorem} The map $ * : \wedge^k V \to \wedge^{n-k} V $ satisfies both
$ * \circ * = (-1)^{n-r + k(n-k)} \id_{\wedge^k V} $ and $ \langle *\mb{x}, *\mb{y} \rangle
= (-1)^{n-r} \langle \mb{x}, \mb{y} \rangle$ for every $\mb{x},\mb{y}$
in $\wedge^k V,$ and for any $k=0,\ldots, n$.
\end{theorem}

We now consider the dual space $ V^* $ of $ V $, i.e. the $\RR$-vector space of the $\RR$--linear maps from $ V $ to $ \RR.$ Given the basis $ (\mb{e_1} , \dots, \mb{e_n}) $ of $ V,$ the dual space can be identified with the $ n \times 1 $ row matrices whose entries are the values that the map takes at $ \mb{e_i}, i=1, \dots, n.$

We use
the form $b$ to construct an isomorphism between $V$ and
$V^*$. Given $\mb{u}\in V$, we define $\mb{u}^\flat \in V^*$
by setting $ \mb{u}^\flat (\mb{v}) = \langle \mb{u}, \mb{v}
\rangle$. It is easy to prove that $^\flat: V \to V^* $ is an
isomorphism, and so $ B^\flat = (\mb{e_1}^\flat, \dots,
\mb{e_n}^\flat) $ is a basis of $ V^*.$ We want $ V $ and $ V^* $
to be isometric. Therefore we choose the non--degenerate, symmetric,
bilinear form $ b^\flat $ on $ V^* $ as $$ b^\flat(\mb{u}^\flat, \mb{v}^\flat) = b(\mb{u}, \mb{v}) \qquad \mbox{ for every } \mb{u}^\flat , \mb{v}^\flat \in V^*.$$
In such a way, $B^\flat $ is
orthonormal with the same signature as $B$.
We define $ ^\sharp : V^* \to V $ to be the inverse
isomorphism of $ ^\flat,$ i.e. $ (\sum_{i=1}^n u_i
\mb{e_i}^\flat)^\sharp = \sum_{i=1}^n u_i \mb{e_i}.$ We can estend
$ ^\flat $ and $ ^\sharp $ to the associated exterior algebra $ \wedge V^* $, obtaining
the isomorphisms $ \wedge^k V \to \wedge^k V^* $ and $ \wedge^k V^* \to \wedge^k V:$
$$ (\mb{u_1} \wedge \dots \wedge
\mb{u_k})^\flat = \mb{u_1}^\flat \wedge \dots \wedge
\mb{u_k}^\flat \quad \mbox{ and } \quad (\boldsymbol{\alpha_1}
\wedge \dots \wedge \boldsymbol{\alpha_k})^\sharp =
\boldsymbol{\alpha_1}^\sharp \wedge \dots \wedge
\boldsymbol{\alpha_k}^\sharp.$$
As for $ \wedge^k V$, we follow a similar procedure, and extend
$ b^\flat $ to $\wedge^k V^*$. Finally, after choosing
$ \bs{\omega}^\flat $ as positive basis of $ \wedge^n V^*$, we
define the Hodge $ \ast $ operator on $ \wedge^k V^*$, as $$ \ast(\mb{x}) = \left( \ast(\mb{x}^\sharp) \right)^\flat \qquad \mbox{ for each } \mb{x} \in \wedge^k V^*.$$

As last general topic, we consider the evaluation of a $ k$--form in $ \wedge^k V^* $ on $ \mb{u} \in
V.$ Such operation gives rise to the linear map $ \mb{i}_{\mb{u}} : \wedge^k V^*
\to \wedge^{k-1} V^* $ defined as
\begin{equation}\label{minkowski3}
\fl\qquad\qquad\mb{i}_{\mb{u}}( \boldsymbol{\alpha_1}
\wedge \dots \wedge \boldsymbol{\alpha_k}) = \sum_{i=1}^k (-1)^{i-1}
\boldsymbol{\alpha_i}(\mb{u}) \ \boldsymbol{\alpha_1} \wedge \dots \wedge \widehat{\boldsymbol{\alpha_i}}
\wedge \dots \wedge \boldsymbol{\alpha_k}
\end{equation}
where $\widehat{\boldsymbol{\alpha_i}} $ means that the item is missing.

\subsection{The Euclidean vector space of dimension $ 2 $}

Let $ V $ be a $ 2$--dimensional vector space on $ \RR,$ let $ b $ a non--degenerate bilinear form with signature $ (2,0),$ and let $ B = ( \mb{e_1}, \mb{e_2} ) $ be an orthonormal basis with respect to $ b.$ Then, $ \wedge^k V = 0 $ for $ k \geq 3,$ and $ \wedge^2 V $ has dimension $ 1 $ with $ (\bs{\omega} = \mb{e_1} \wedge \mb{e_2} ) $ as orthonormal basis. On the natural bases, the Hodge operator is defined as:
$$
\ast(1) = \boldsymbol{\omega},\  \ast(\mb{e_1}) = \mb{e_2},\ \ast(\mb{e_2})=-\mb{e_1},\ \ast(\boldsymbol{\omega}) = 1.
$$
Analogously, $ V^* $ has dimension $ 2,$ with basis $ ( \mb{e_1}^\flat, \mb{e_2}^\flat ),$ where $$ \mb{e_i}^\flat ( u_1 \mb{e_1} + u_2 \mb{e_2} ) = b(\mb{e_i}, u_1 \mb{e_1} + u_2 \mb{e_2} ) = u_i \qquad \mbox{ for } i = 1, 2,$$ and
$$
\ast(1) = \boldsymbol{\omega}^\flat,\  \ast(\mb{e_1}^\flat) = \mb{e_2}^\flat,\ \ast(\mb{e_2}^\flat)=-\mb{e_1}^\flat,\ \ast(\boldsymbol{\omega}^\flat) = 1.
$$
\begin{proposition} \label{hodge-2-dim}
Let $ \mb{u} = u_1 \mb{e_1} + u_2 \mb{e_2}, \mb{v} = v_1 \mb{e_1} + v_2 \mb{e_2} \in V $ and
$ \mb{u}^\flat = u_1 \mb{e_1}^\flat + u_2 \mb{e_2}^\flat, \mb{v}^\flat = v_1 \mb{e_1}^\flat + v_2 \mb{e_2}^\flat \in V^*.$ Then, $$ \ast(\mb{u} \wedge \mb{v}) = \det \left( \begin{array}{cc} u_1 & v_1 \\ u_2 & v_2 \end{array} \right) \qquad \mbox{ and } \qquad \ast(\mb{u}^\flat \wedge \mb{v}^\flat) = \det \left( \begin{array}{cc} u_1 & u_2 \\ v_1 & v_2 \end{array} \right).$$
\end{proposition}

We adopt the usual convention that the components of a vector in $ V $ are written as columns, while the components of a vector in $ V^* $ are written as rows. Of course, the images of the two $ 2$--form are equal because of the properties of the determinant of a matrix. The proof is an easy computation and we do not write the details.

\subsection{The Minkowski vector space of dimension $ 3 $}\label{app:AMink}

Let $ V $ be a $ 3$--dimensional vector space on $ \RR,$ let $ b $ a non--degenerate bilinear form with signature $ (2,1),$ and let $ B = ( \mb{e_1}, \mb{e_2}, \mb{e_3} ) $ be an orthonormal basis with respect to $ b.$ Then, $ \wedge^k V = 0 $ for $ k \geq 4,$ and $ \wedge^2 V $ has dimension $ 3 $ with $ (\mb{e_1} \wedge \mb{e_2}, \mb{e_1} \wedge \mb{e_3}, \mb{e_2} \wedge \mb{e_3} ) $ as orthonormal basis with signature $ (1,2),$ while $ \wedge^3 V $ has dimension $ 1 $ and
$ (\bs{\omega} = \mb{e_1} \wedge \mb{e_2} \wedge \mb{e_3} ) $ is an orthonormal basis of this last space. On the natural bases the $ \ast $ operator acts as follows:
$$ \ast(1) = \boldsymbol{\omega},\ \ast(\boldsymbol{\omega}) = -1,
$$ $$ \ast(\mb{e_1}) = \mb{e_2} \wedge \mb{e_3},\
\ast(\mb{e_2}) = -\mb{e_1} \wedge \mb{e_3},\ \ast(\mb{e_3}) = -\mb{e_1}
\wedge \mb{e_2},$$ $$ \ast(\mb{e_1} \wedge \mb{e_2}) = \mb{e_3},\
\ast(\mb{e_1} \wedge \mb{e_3}) = \mb{e_2},\ \ast(\mb{e_2} \wedge
\mb{e_3}) = -\mb{e_1}.$$
As before, we compute the images of the elements of the bases of $ \wedge^k V^* $ via the Hodge $ \ast $ operator, and we get:
$$ *(1) = \boldsymbol{\omega}^\flat, *(\boldsymbol{\omega}^\flat) = -1,
$$ $$ *(\mb{e_1}^\flat) = \mb{e_2}^\flat \wedge \mb{e_3}^\flat,
*(\mb{e_2}^\flat) = -\mb{e_1}^\flat \wedge \mb{e_3}^\flat,
*(\mb{e_3}^\flat) = -\mb{e_1}^\flat \wedge \mb{e_2}^\flat,$$ $$
*(\mb{e_1}^\flat \wedge \mb{e_2}^\flat) = \mb{e_3}^\flat,
*(\mb{e_1}^\flat \wedge \mb{e_3}^\flat) = \mb{e_2}^\flat,
*(\mb{e_2}^\flat \wedge \mb{e_3}^\flat) = -\mb{e_1}^\flat.$$

Now we state some results that we use in the body of the paper.
\begin{lemma}
\label{star-u}
Let $ \mb{u}, \mb{v}, \mb{w} \in V $ be linearly independent, so $ \bs{\Omega} = \mb{u} \wedge \mb{v} \wedge \mb{w} \not= \mb{0}.$ Then,
$$ \ast (\mb{u}) = -\frac 1{\ast (\bs{\Omega})} \left( \langle \mb{u}, \mb{w} \rangle \mb{u} \wedge \mb{v} + \Vert \mb{u} \Vert^2 \mb{v} \wedge \mb{w} + \langle \mb{u}, \mb{v} \rangle \mb{w} \wedge \mb{u} \right) $$
and
$$ \ast \left( \mb{u} \wedge \mb{v} \right) = -\frac 1{\ast (\bs{\Omega})} \left( \langle \mb{u} \wedge \mb{v}, \mb{v} \wedge \mb{w} \rangle \mb{u} + \langle \mb{u} \wedge \mb{v}, \mb{w} \wedge \mb{u} \rangle \mb{v} + \Vert \mb{u} \wedge \mb{v} \Vert^2 \mb{w} \right).$$
\end{lemma}

\noindent\emph{Proof.}
From the linear independence of $\mb{u}$, $\mb{v}$, and $\mb{w}$, it
follows that $ \mb{u} \wedge \mb{v}, \mb{v} \wedge \mb{w}, \mb{w}
\wedge \mb{u} $ is a basis of $ \wedge^2 V$. Hence, there exist
elements in $ \RR $ such that $ \ast (\mb{u}) = a \mb{u} \wedge
\mb{v} + b \mb{v} \wedge \mb{w} + c \mb{w} \wedge \mb{u}$.
From the definition of $ \ast$ follows that $ \mb{u} \wedge \ast
(\mb{u}) = \Vert \mb{u} \Vert^2 \bs{\omega}$. By substituting the
expression of $\ast (\mb{u})$, and using the properties of $\wedge$
we obtain $ b \, \bs{\Omega} = \Vert \mb{u} \Vert^2
\bs{\omega}$ or, equivalently, $ b \ast (\bs{\Omega}) = -\Vert
\mb{u} \Vert^2$, which gives $b = - \Vert \mb{u} \Vert^2 / \ast
(\bs{\Omega})$. Through a similar computation we can
derive $a$ and $c$, therefore the first claim follows.
The second claim can be proven through the same arguments,
therefore we can skip the details.
\hfill$\square$\vspace{1mm}

\begin{lemma} Let $ \mb{u}, \mb{v} \in V,$ and $
\boldsymbol{\gamma}\in V^*.$ Then, $ \boldsymbol{\gamma}(*(\mb{u}
\wedge \mb{v})) = *(\mb{u} \wedge \mb{v} \wedge
\boldsymbol{\gamma}^\sharp).$
\end{lemma}

\noindent\emph{Proof.} We can verify the equality by using
components with respect to $ B, B^\flat$.
\hfill$\square$

\begin{corollary} \label{2-forme}
Let $ \boldsymbol{\alpha}, \boldsymbol{\beta} \in V^* $ be
linearly independent. Then $ \mb{i}_{\mb{u}}(\boldsymbol{\alpha}
\wedge \boldsymbol{\beta} ) = \mb{0} $ if, and only if, $ \mb{u}
\in \mathcal{L}\left( \left(*(\boldsymbol{\alpha} \wedge
\boldsymbol{\beta}) \right)^\sharp \right),$ where $
\mathcal{L}(\dots) $ is the subspace generated by the vectors in
parenthesis.
\end{corollary}

\noindent\emph{Proof.} Assume $ \mb{u} = t \left( *(\boldsymbol{\alpha}
\wedge \boldsymbol{\beta}) \right)^\sharp ,$ for some $ t \in
\RR.$ Then
$$
\boldsymbol{\alpha}(\mb{u}) = t
\boldsymbol{\alpha}\left( \left(*(\boldsymbol{\alpha} \wedge
\boldsymbol{\beta})\right)^\sharp \right) =
*(\boldsymbol{\alpha}^\sharp \wedge \boldsymbol{\beta}^\sharp
\wedge \boldsymbol{\alpha}^\sharp) = 0.
$$
A similar argument proves that $ \boldsymbol{\beta}(\mb{u}) = 0$.
By definition,
$ \mb{i}_{\mb{u}}(\boldsymbol{\alpha} \wedge \boldsymbol{\beta}) =
\boldsymbol{\alpha}(\mb{u}) \boldsymbol{\beta} -
\boldsymbol{\beta}(\mb{u}) \boldsymbol{\alpha} $ therefore the claim
follows.\\
Conversely, assume that $\boldsymbol{\alpha}(\mb{u})
\boldsymbol{\beta} - \boldsymbol{\beta}(\mb{u})
\boldsymbol{\alpha} = \mb{0}$. Then $ \boldsymbol{\alpha}(\mb{u})
= \boldsymbol{\beta}(\mb{u}) = 0 $ because $ \boldsymbol{\alpha},
\boldsymbol{\beta} $ are linearly independent. Hence, $ \langle
\boldsymbol{\alpha}^\sharp, \mb{u} \rangle = \langle
\boldsymbol{\beta}^\sharp, \mb{u} \rangle = 0 $. This implies that
$\mb{u} = t \left( *(\boldsymbol{\alpha}\wedge
\boldsymbol{\beta})\right)^\sharp $ for some $ t \in \RR $, which completes the proof.
\hfill$\square$

\begin{lemma}\label{lemma:3forma} Let $ \boldsymbol{\Theta} \in \wedge^3 V^* $ be
a non--zero $ 3$--form. Then, $ \mb{i}_{\mb{u}}(\boldsymbol{\Theta}) =
\boldsymbol{\alpha}\wedge \boldsymbol{\beta} $ if, and only if,
$\mb{u} = \frac 1{*(\boldsymbol{\Theta})} \left(
*(\boldsymbol{\alpha}\wedge \boldsymbol{\beta})\right)^\sharp.$
\end{lemma}

\noindent\emph{Proof.} There exists $ t \in \RR$, $t \not= 0$,
such that $ \boldsymbol{\Theta} = t \boldsymbol{\omega}^\flat$, therefore,
$$
\mb{i}_{\mb{u}}(\boldsymbol{\Theta}) = t \left(
\mb{e_1}^\flat(\mb{u}) \mb{e_2}^\flat \wedge \mb{e_3}^\flat -
\mb{e_2}^\flat(\mb{u}) \mb{e_1}^\flat \wedge \mb{e_3}^\flat +
\mb{e_3}^\flat(\mb{u}) \mb{e_1}^\flat \wedge \mb{e_2}^\flat
\right) = \boldsymbol{\alpha} \wedge \boldsymbol{\beta}.
$$
This implies
$$*(\boldsymbol{\alpha} \wedge
\boldsymbol{\beta}) = t \left( -\langle \mb{e_1}, \mb{u} \rangle
\mb{e_1}^\flat - \langle \mb{e_2}, \mb{u} \rangle \mb{e_2}^\flat +
\langle \mb{e_3}, \mb{u} \rangle \mb{e_3}^\flat \right) = -t
\mb{u}^\flat
$$
and $ \mb{u} = - \frac 1t
*(\boldsymbol{\alpha} \wedge \boldsymbol{\beta})^\sharp.$
Moreover, $ *(\boldsymbol{\Theta}) =
t*\!(\boldsymbol{\omega}^\flat) = -t,$ therefore one side of the
claim follows. The converse is proven through a straightforward
computation.
\hfill$\square$\vspace{1mm}


\renewcommand{\thesection}{B}
\section{A brief introduction to plane algebraic geometry}\label{app:B}

In this Appendix, we recall the main definitions and results
concerning curves in the affine or projective plane.

\subsection{Affine spaces and algebraic subsets}

\begin{definition} \label{aff-space}
Let $ \mathbb{K} $ be a field and let $ V $ be a $\mathbb{K}$--vector space.
Let $ \Sigma $ be a non--empty set. A map $ \phi: \Sigma
\times \Sigma \to V $ that verifies
\begin{itemize}
\item[i.] $ \phi(A,B) + \phi(B,C) = \phi(A,C) $ for every $ A,B,C
\in \Sigma$
\item[ii.] $ \phi_A: \Sigma \to V $ defined as $
\phi_A(X) = \phi(A,X) $ is $ 1$--to--$1 $ for every $ A \in
\Sigma$
\end{itemize}
is an affine structure on $ \Sigma,$ and the couple $ (\Sigma,
\phi) $ is called affine space and named $ \mathbb{A}(V).$
\end{definition}

The main example we use is the following: let $ \Sigma = V,$ and
define $ \phi(\mb{u},\mb{v}) = \mb{v} - \mb{u}.$ $ \phi $ is an
affine structure on $ V $ and so we get the affine space $
\mathbb{A}(V).$ If $ \dim(V) = n,$ we say that $ \mathbb{A}(V) $
has dimension $ n,$ as well. The advantage to have an affine
structure on a set of points is that we can easily define the
coordinates of the points.
\begin{definition} \label{aff-ref}
Let $ \mathbb{A}(V) $ be an affine space of dimension $ n.$ A
reference frame is a couple $ \mathcal{R} = (O, B),$ where $ O $
is a point, and $ B = (\mb{v_1}, \dots, \mb{v_n}) $ is a basis of
$ V.$ Given $ P \in \mathbb{A}(V),$ its coordinates in the frame $ \mathcal{R} $ are the components of $ \phi(O,P) $ with
respect to $ B.$
\end{definition}

Thanks to the properties of the affine structure, once $
\mathcal{R} $ is given, there is a $1$--to--$1$ correspondence
between points in $ \mathbb{A}(V) $ and elements in $
\mathbb{K}^n.$ So, usually, the two spaces are identified. When
this happens, one denotes $ \mathbb{K}^n $ as $
\mathbb{A}^n_{\mathbb{K}}$. We remark that the identification
imply the choice of the reference frame, and so some care has to
be taken if one works with more than one reference frame.

If $ \mathbb{K} = \RR,$ and $ V $ is an Euclidean vector space,
then $ \mathbb{A}(V) $ is referred to as Euclidean space, and
indicated with $ \mathbb{E}(V),$ or $ \mathbb{E}^n $ emphasizing
just the dimension of $ V.$ In this setting, the set--theoretical equality between $ \mathbb{E}^n $ and $ \mathbb{A}^n_{\RR} $ is evident. However, if one switches from $ \mathbb{E}^n $ to $ \mathbb{A}^n_{\RR},$ then
one is not allowed to use distances and angles.

Another standard construction is the following. Given the real
vector space $ V,$ one can consider the complex vector space $
\bar{V} = \mathbb{C} \otimes_{\RR} V.$ Roughly speaking, we allow
complex numbers to multiply the vectors of $ V.$ As example of the previous construction, we remark that $ \mathbb{C}^n = \mathbb{C} \otimes_{\RR} \RR^n.$ It holds $
\dim_{\mathbb{C}} \bar{V} = \dim_{\RR} V,$ and $ \dim_{\RR}
\bar{V} = 2 \dim_{\RR} V.$ Of corse, we have a set--theoretical
inclusion $ V \subset \bar{V},$ that is not a linear map. The
inclusion of vector spaces provides an inclusion of the corresponding affine
spaces, that can be written as $ \mathbb{A}^n_{\RR} \subset
\mathbb{A}^n_{\mathbb{C}},$ up to the choice of a
reference frame with the same origin $ O \in \mathbb{A}^n_{\RR},$ and the same basis $ B \subset V $ both for $ V $ and for $ \bar{V}$.

In the paper, we mainly use the affine space with $ n=2,$ namely
the affine plane. The geometrical objects in $
\mathbb{A}^2_{\mathbb{K}} $ that are studied in the algebraic
geometry framework are (algebraic) curves and their intersections.
We recall the definition of algebraic curve.
\begin{definition} \label{van-loc}
Let $ f_1, \dots, f_r \in \mathbb{K}[x,y] $ be polynomials. The
vanishing locus $ V(f_1, \dots, f_r) $ of the given polynomials is
$$ V(f_1, \dots, f_r) = \{ P \in \mathbb{A}^2_{\mathbb{K}} \ \vert \ f_i(P) =
0 \mbox{ for every } i=1, \dots, r \}.$$
\end{definition}
The evaluation of a polynomial $ f $ at $ P, f(P),$ simply
consists in substituting the coordinates of $ P $ in the
expression of $ f.$
\begin{definition} \label{alg-curve}
A non--empty subset $ C \subset \mathbb{A}^2_{\mathbb{K}} $ is an
algebraic curve if there exists a polynomial $ f \in
\mathbb{K}[x,y] $ of degree $ \geq 1 $ such that $ C = V(f).$
\end{definition}
We get a line when the degree of $ f $ is $ 1,$ a conic when the
degree of $ f $ is $ 2.$ From degree $ 3 $ on, a curve is named
according to the degree of $ f,$ e.g. there are cubic curves,
quartic ones, and so on.

The advantage of considering curves in $ \mathbb{A}^2_{\mathbb{C}}
$ is that some unpleasant phenomenon do not happen: the vanishing
locus of $ x^2+y^2 $ is a single point in $ \mathbb{A}^2_{\RR} $
and a couple of lines in $ \mathbb{A}^2_{\mathbb{C}},$ the
vanishing locus of $ x^2+y^2+1 $ is empty in $
\mathbb{A}^2_{\RR},$ and a conic in $ \mathbb{A}^2_{\mathbb{C}}.$

In greater generality, we can consider algebraic subsets.
\begin{definition} \label{alg-set}
A subset $ X \subset \mathbb{A}^2_{\mathbb{K}} $ is algebraic if
there exist $ f_1, \dots, f_r \in \mathbb{K}[x,y] $ such that $ X
= V(f_1, \dots, f_r).$
\end{definition}

For example, the intersection of the curves $ C_i = V(f_i), i=1,
\dots, r,$ is the algebraic set $ X = V(f_1, \dots, f_r).$ It is
possible to prove that algebraic sets are the closed sets of a
topology, the Zariski topology, on $ \mathbb{A}^2_{\mathbb{K}}.$

\subsection{Projective spaces and algebraic subsets}

Roughly speaking, the
points of a projective space are the $1$--dimensional subspaces of
a vector space. Hence, we have to identify all the vectors
belonging to the same subspace. The mathematical machinery is the
following one.
\begin{definition} \label{rel-eq}
Let $ V $ be a $ n+1$--dimensional vector space over the ground
field $ \mathbb{K}.$ We define the relation $ \sim $ in $ V
\setminus \{ \mb{0} \} $ as $$ \mb{u} \sim \mb{v} \quad \mbox{ if
there exists } t \in \mathbb{K}, t \not= 0, \mbox{ such that }
\mb{u} = t \mb{v}.$$
\end{definition}
It is easy to check the following.
\begin{proposition}
$ \sim $ is an equivalence relation.
\end{proposition}

\begin{definition} \label{proj-space}
The projective space of dimension $ n $ over $ V $ is the set of
equivalence classes of $ V \setminus \{ \mb{0} \} $ modulo $
\sim,$ that is to say, $$ \mathbb{P}(V) = (V \setminus \{ \mb{0}
\}) / \sim.$$
\end{definition}

As for the affine space, we define a reference frame in the
projective space.
\begin{definition} \label{proj-frame}
Let $ \mathbb{P}(V) $ be a projective space of dimension $ n.$ A
reference frame is $ \mathcal{R} = (B),$ where $ B = (\mb{v_0},
\dots, \mb{v_n}) $ is a basis of $ V.$ Given $ P \in
\mathbb{P}(V),$ its homogeneous coordinates with respect to $
\mathcal{R} $ are the components of $ \mb{v} \in P $ with respect
to $ B,$ and we set $ P = (x_0: \dots : x_n).$
\end{definition}

The homogeneous coordinates of a point $ P \in \mathbb{P}(V) $ are
not unique. In fact, if $ \mb{v} \in P,$ then $ P $ contains also
$ t \mb{v} $ for every $ t \in \mathbb{K},$ $ t \not= 0.$ The
components of $ t \mb{v} $ are the ones of $ \mb{v} $ times $ t,$
and so the homogeneous coordinates of a point are unique up to a
scalar factor, i.e. if $ (x_0: \dots : x_n) $ are the homogeneous
coordinates of $ P,$ then also $ (tx_0: \dots : tx_n) $ are so,
for every $ t \not= 0.$

A first property is the following one.
\begin{proposition}
Let $ V $ be a real vector space of dimension $ n+1.$ Then $$
\mathbb{P}^2(V) \subset \mathbb{P}^2(\bar{V} = \mathbb{C}
\otimes_{\RR} V).$$
\end{proposition}

\noindent\emph{Proof.} From the definition of $ \bar{V},$ it
follows that vectors that are proportional in $ V $ are
proportional also in $ \bar{V},$ and so there is a natural way to
identify a point $ P \in \mathbb{P}(V) $ with a point in $
\mathbb{P}(\bar{V}).$ This identification gives an inclusion. We
remark that the two spaces are not equal for $ n \geq 1.$
\hfill$\square$\vspace{1mm}

Hence, we restrict to $ \mathbb{P}(V) $ where $ V $ is a vector
space over the complex field and we stress the properties that
behave differently in a projective space over a real vector space.
Moreover, once a reference frame is given, we identify the points
with their homogeneous coordinates. In this case, we simply write
$ \mathbb{P}^n_{\mathbb{C}} $ or $ \mathbb{P}^n_{\RR} $ to stress
the dimension and the ground field. Motivated again from the case
considered in the paper, we focus on the projective space of
dimension $ 2,$ i.e. on the projective plane $
\mathbb{P}^2_{\mathbb{C}}.$

In the projective setting, the polynomial to be considered are the homogeneous ones.
In fact, due to the construction of the homogeneous coordinates, the evaluation of a polynomial at a points is in general a meaningless concept. However, it is meaningful to check if a polynomial vanishes at a projective point $P.$
\begin{proposition}
Let $ f \in \mathbb{K}[x_0,x_1,x_2]$ be equal to $ f = f_0 + f_1 + \dots + f_d $ where $ f_i $ is homogeneous of degree $ i.$ Let $P\in\PP^2_\mathbb{K}$ be the point
with homogeneous coordinates $ (x_0: x_1: x_2).$ Then, $f(P)=0$ if, and only if, $f_i(P)=0$ for each $i$.
\end{proposition}
\noindent\emph{Proof.}
We have $f(P) = f(x_0:x_1:x_2) = f_0(x_0:x_1:x_2) + f_1(x_0:x_1:x_2) + \dots +
f_d(x_0:x_1:x_2).$ The point $ P,$ however, is represented also from the coordinates $ (tx_0: tx_1 : tx_2) $ for every $ t \not=0,$ and so we have $ f(tx_0:tx_1:tx_2) =f_0(tx_0:tx_1:tx_2) + f_1(tx_0:tx_1:tx_2) + \dots +
f_d(tx_0:tx_1:tx_2) = f_0(x_0:x_1:x_2) + t f_1(x_0:x_1:x_2) +
\dots + t^d f_d(x_0:x_1:x_2).$ So, if $\mathbb{K} $ contains infinitely many elements, then $f_i(x_0:x_1:x_2) = 0 $ for every $ i = 0, \dots, d.$
\hfill$\square$\vspace{1mm}

\noindent This proposition justifies the fact that we restrict to homogeneous polynomials.
\begin{definition} \label{proj-van}
Let $ f_1, \dots, f_r \in \mathbb{K}[x_0,x_1,x_2] $ be homogeneous
polynomials. Then, their vanishing locus is $$ V(f_1, \dots, f_r)
= \{ P \in \mathbb{P}^2_{\mathbb{K}} \ \vert \ f_i(P) = 0 \mbox{
for every } i = 1, \dots, r \}.$$
\end{definition}
We are now ready to define the projective algebraic sets, and
projective curves in particular.
\begin{definition} \label{proj-set}
A non--empty subset $ C \subset \mathbb{P}^2_{\mathbb{K}} $ is a
projective plane curve if there exists a homogeneous polynomial $
f $ such that $ C = V(f).$\\ \noindent A subset $ X \subseteq
\mathbb{P}^2_{\mathbb{K}} $ is a projective algebraic set if there
exist homogeneous polynomials $ f_1, \dots, f_r $ such that $ X =
V(f_1, \dots, f_r).$
\end{definition}

As in the case of the affine plane, it is possible to prove that
the projective algebraic sets are the closed sets of a topology,
the Zariski topology, on the projective plane. A line in $
\mathbb{P}^2_{\mathbb{K}} $ is the vanishing locus of a
degree $ 1 $ homogeneous polynomial, a conic is the vanishing
locus of a degree $ 2 $ homogeneous polynomial, and so on.

As further step, we show that it is possible to include the affine
plane in the projective one, as an open subset.
\begin{proposition} \label{aff-proj}
Let $ H_0 = V(x_0) \subset \mathbb{P}^2_{\mathbb{K}} $ be a line,
and let $ U_0 $ be the open complement of $ H_0.$ Then, $ U_0 $
can be identified with the affine plane, and the identification
preserves the algebraic sets.
\end{proposition}

\noindent\emph{Proof.} Let $ P = (x_0:x_1:x_2) \in U_0.$ From the
definition of $ U_0,$ it follows that $ x_0 \not= 0,$ and so we
can take $ (1: x_1/x_0 : x_2/x_0) $ as the homogeneous coordinates
of $ P.$ If we set $ x = x_1/x_0, y = x_2/x_0,$ then we can
identify $ P $ with the point $ Q \in \mathbb{A}^2_{\mathbb{K}} $
whose coordinates are $ (x,y).$ Conversely, each point $ Q (x,y)
\in \mathbb{A}^2_{\mathbb{K}} $ can be identified with the point $
P \in U_0 $ whose homogeneous coordinates are $ (1:x:y).$ Hence,
there is a $1$--to--$1$ correspondence between $ U_0 $ and $
\mathbb{A}^2_{\mathbb{K}}.$ To prove the remaining part of the
statement, we start considering curves. So, let $ C \subset
\mathbb{P}^2_{\mathbb{K}} $ be a curve different from $ H_0,$ and
let $ f(x_0,x_1,x_2) \in \mathbb{K}[x_0,x_1,x_2] $ be a
homogeneous polynomial such that $ C = V(f).$ Let $ g(x,y) =
f(1,x,y) \in \mathbb{K}[x,y],$ and let $ D \subset
\mathbb{A}^2_{\mathbb{K}} $ be the affine curve it defines. Then,
the previous correspondence maps the points in $ C \cap U_0 $ to
points in $ D,$ and conversely. So a curve in the projective plane
is transformed in a curve in the affine plane. Conversely, let $ D
= V(g) $ be a curve in $ \mathbb{A}^2_{\mathbb{K}},$ with $ g \in
\mathbb{K}[x,y].$ Let $ d $ be the degree of $ g,$ and let $
f(x_0, x_1, x_2) = x_0^d g(x_1/x_0, x_2/x_0).$ It is easy to check
that $ f \in \mathbb{K}[x_0, x_1, x_2] $ is a degree $ d $
homogeneous polynomial. Let $ C = V(f) $ be the corresponding
curve in the projective plane. Then, it is straightforward to
prove that the points of $ D $ are mapped to points in $ C \cap
U_0.$ As the algebraic sets are union of finitely many curves or
intersection of curves, the proof is complete, because it holds on
curves. \hfill$\square$\vspace{1mm}

Notice that a consequence of the previous proof is that the
complement of whatever line in $ \mathbb{P}^2_{\mathbb{K}} $ is an
affine plane. Conversely, the projective plane is the union of an
affine plane and a projective line. The points of the projective
line are called ideal points of the affine plane and are thought
to as directions of the lines in the affine plane.

From now on, we deal with the geometry of algebraic sets in the
projective plane $ \mathbb{P}^2_{\mathbb{C}} $ because we have the
chains of inclusions $$ \mathbb{E}^2 = \mathbb{A}^2_{\RR}
\subset \mathbb{A}^2_{\mathbb{C}} \subset
\mathbb{P}^2_{\mathbb{C}} \qquad \mbox{ and } \qquad \mathbb{E}^2
= \mathbb{A}^2_{\RR} \subset \mathbb{P}^2_{\mathbb{R}}
\subset \mathbb{P}^2_{\mathbb{C}},$$ and we underline the problems
when restricting to smaller ambient spaces.

\subsection{Intersection of curves and singular points}

A line $ L $ in $ \mathbb{P}^2_{\mathbb{C}} $ is the vanishing
locus of a linear homogeneous equation $ a_0 x_0 + a_1 x_1 + a_2
x_2 = 0,$ and so it has parametric equation $ x_i = b_i s + c_i t,
s, t \in \mathbb{C}, i=0,1,2.$ Given a curve $ C = V(f),$ the intersection
$ L \cap C $ is given by solving the equation $ f(b_0 s + c_0 t,
\dots, b_2s + c_2 t) = F(s,t) $ homogeneous of the same degree of
$ f.$ Hence, we have proved
\begin{proposition} \label{line-curve}
A line $ L $ intersects a degree $ d $ curve $ C $ at exactly $ d
$ points, up to count each point with its multiplicity.
\end{proposition}

\noindent\emph{Proof.} The Fundamental Theorem of Algebra states
that a degree $ d $ equation in one variable has exactly $ d $
roots over $ \mathbb{C},$ up to count each root with its
multiplicity. We can apply it to $ F(1,t),$ after computing the
largest power of $ s $ that divides $ F.$
\hfill$\square$\vspace{1mm}

Of course, when restricting to $ \mathbb{P}^2_{\mathbb{R}},$ the
complex roots do not give contribution, and so a line meets a
degree $ d $ curve in at most $ d $ points, up to count each one
of them with multiplicity. When considering the
intersection in $ \mathbb{A}^2_{\mathbb{C}},$ the roots that correspond to ideal points give no contribution, and so again a line meets a degree $ d $ curve in at most $ d $ points. When restricting to $ \mathbb{A}^2_{\RR},$ one has to take care of both problems. As an example, we consider the conic $ C = V(x_1^2 - x_0x_2) \subset \mathbb{P}^2_{\mathbb{C}},$ and the line $ L_1 = V(x_0+x_2).$ They meet at $ A_1(i:1:-i) $ and $ B_1(-i:1:i) $ where $ i^2=-1.$ So, $ C \cap L_1 $ is empty, when the intersection is considered in $ \mathbb{P}^2_{\RR}.$ If we consider the line $ L_2 = V(x_0 - x_1),$ the intersection of $ C $ and $ L_2 $ consists of the points $ A_2(0:0:1) $ and $ B_2(1:1:1).$ Let $ \mathbb{A}^2_{\mathbb{C}} $ be identified with $ U_1 $ complement of the line $ H_1 = V(x_1).$ Then the conic $ C $ (the line $ L_2,$ respectively) has equation $ xy-1=0 $ ($ x = 1,$ respectively). The intersection contains the point $ (1,1),$ only. In fact, $ A_2 $ is an ideal point for the identification of the affine plane with $ U_1.$

The previous result can be generalized to the intersection of two
curves of arbitrary degree.
\begin{theorem}[B\'{e}zout's Theorem] \label{bezout}
Let $ C, C' $ be projective plane curves of degree $ d $ and $
d',$ respectively. If $ C $ and $ C' $ have a finite number of
common points, then there are exactly $ d d' $ intersection
points, up to count them with their multiplicity.
\end{theorem}
The proof of B\'{e}zout's Theorem can be found in \cite[Ch.I, Corollary 7.8]{hart}, and goes beyond the scopes of this
introduction.

Now, we can define a smooth point on a curve.
\begin{definition} \label{sing-point}
Let $ C \subset \mathbb{P}^2_{\mathbb{C}} $ be a curve and let $ P \in C $ be a point. $ P $ is a smooth point of $ C $ if there exists a line $ L $ containing $ P $ that intersects $ C $ at $ P $ with multiplicity $ 1.$ A curve $ C $ whose points are all smooth is said to be smooth, singular otherwise.
\end{definition}

The property is local, so we can reduce to an affine plane, by taking the complement of a line that do not contain $ P.$ Moreover, we choose a reference frame in such affine plane so that $ P $ is the origin. Hence, $ C = V(f) $ for a suitable $ f \in \mathbb{K}[x,y],$ with $ f(0,0) = 0.$ The lines through the origin have parametric equation $ x=lt, y=mt,$ and the intersection between $ C $ and one of such line is described by $ f(lt, mt) = 0.$ By McLaurin expansion, we have $ 0 = (l f_x(0,0) + m f_y(0,0)) t + $ higher degree terms. Hence, $ P $ is smooth if there exists $ l, m $ such that $ l f_x(0,0) + m f_y(0,0) \not= 0,$ or equivalently, the gradient $ \nabla f(0,0) $ is not zero. We have then proved
\begin{proposition} \label{char-sing-point}
A point $ P \in C = V(f) $ is smooth for $ C $ if $ \nabla f(P) \not= \mb{0},$ where $ f $ is a homogeneous polynomial that defines $ C.$
\end{proposition}

Proposition \ref{char-sing-point} allows us to prove that the singular locus $ \mbox{Sing}(C) $ of $ C = V(f) $ is algebraic and it holds $ \mbox{Sing}(C) = V(f, f_{x_0}, f_{x_1}, f_{x_2}).$ Thanks to Euler formula for homogeneous functions
\begin{equation} \label{Euler-form}
\fl\qquad d F(x_0,x_1,x_2) = x_0 F_{x_0}(x_0,x_1,x_2) +  x_1
F_{x_1}(x_0,x_1,x_2) + x_2 F_{x_2}(x_0,x_1,x_2)
\end{equation} where $ d $ is the degree of $ F,$ we have that $ \mbox{Sing}(C) = V(f_{x_0}, f_{x_1}, f_{x_2}).$ In the affine plane, if $ C = V(f),$ we have $ \mbox{Sing}(C) = V(f, f_x, f_y).$

Also if intuition suggests that the singular points on a curve are finitely many special points, there are examples of curves with a subcurve of singular points. For example, consider the curve $ C = V(x_0x_1^2) $ in the projective plane $ \mathbb{P}^2_{\RR}.$ Then, $ \mbox{Sing}(C) = V(2x_0x_1, x_1^2) = V(x_1).$ Hence, $ C $ has the line $ V(x_1) $ as its singular locus. The curve $ C $ is the union of the line $ V(x_0) $ and of the conic $ V(x_1^2).$ The conic, however, is a double line (twice the line $ V(x_1)$), and so the singular locus of $ C $ is equal to the double line. This phenomenon can be easily generalized. Before giving the definitions on curves to handle it, we recall some properties of polynomials. We state them for polynomials in two variables but they can be extended without effort to homogeneous polynomials in $ 3 $ variables.
\begin{definition} \label{irred-pol}
A polynomial $ f \in \mathbb{K}[x,y] $ is irreducible if it cannot be written as product of two non--constant polynomials.
\end{definition}

\begin{theorem} \label{fac-pol}
Every polynomial $ f \in \mathbb{K}[x,y] $ can be written as product of powers of irreducible polynomials, in a unique way, up to some non--zero constant.
\end{theorem}
Now, we translate the previous results in geometrical terms.
\begin{definition} \label{red-irred-curve}
A curve $ C $ is
\begin{enumerate}
\item reduced and irreducible if $ C = V(f) $ with $ f $
irreducible; \item irreducible and non--reduced if $ C = V(f^m) $
with $ f $ irreducible and $ m \geq 2;$ \item reduced,
non--irreducible if $ C = V(f_1 \cdots f_r) $ with $ f_i $
irreducible for every $ i;$ \item non--reduced and
non--irreducible if $ C = V(f_1^{m_1} \cdots f_r^{m_r}) $ with $
f_i $ irreducible and $ m_1 + \dots + m_r \geq r+1.$
\end{enumerate}
\end{definition}
Going back to the study of the singular locus of a curve, we have the following result.
\begin{theorem} \label{bound-sing}
The singular points of a curve $ C $ are finitely many, or $ C $ is smooth if and only if $ C $ is reduced. Moreover, if $ C $ is reduced and irreducible (reduced, respectively), there are at most $ \left( \begin{array}{c} d-1 \\ 2 \end{array} \right) $ ($ \left( \begin{array}{c} d \\ 2 \end{array} \right),$ respectively) singular points on $ C,$ where $ d $ is the degree of $ C.$
\end{theorem}

For example, a reduced and irreducible conic is smooth, while a
reduced non--irreducible conic has exactly one singular point (the
point where the two lines, whose union is the conic, meet).
Furthermore, a reduced and irreducible cubic can have at most one
singular point, a reduced and irreducible quartic curve can have
at most $ 3 $ singular points, and so on.

A notion, apparently non related to the singular locus of a curve,
is the rationality of a curve.
\begin{definition} \label{rat-curve}
A curve $ C = V(f) \subset \mathbb{P}^2_{\mathbb{C}} $ is rational
if there exist $ g_0(s,t), g_1(s,t), g_2(s,t) \in
\mathbb{C}[s,t],$ homogeneous of degree equal to the one of $ f,$
and without common factors of positive degree, such that $ f(g_0,
g_1, g_2) $ is identically zero.
\end{definition}

A rational curve is then a curve whose points have coordinates
that can be expressed via the parameter functions $ g_i(s,t),
i=0,1,2.$ It is possible to prove that a reduced irreducible curve
is rational if it has as much singular points as its degree
allows. E.g., smooth conic, cubic with one singular point, quartic
with three singular points, quintic with $ 6 $ singular points,
are all examples of rational curves. From one hand, more than the
number of singular points, the rationality depends on the kind of
singularities of the curve itself, on the other hand, a deeper
study of the singular points of a curve goes further the scope of
this introduction, and so we do not go on along these lines.

\subsection{Dual projective plane}

As explained earlier, a line in the projective plane is the
vanishing locus of a non--zero degree $ 1 $ homogeneous polynomial
$ a_0 x_0 + a_1 x_1 + a_2 x_2 = 0.$ The coefficients $ a_0, a_1,
a_2 $ are defined up to a scalar. In fact, for each $ k \not= 0,$
$ (k a_0) x_0 + (k a_1) x_1 + (k a_2) x_2 = 0 $ defines the same
line as the previous equation. Hence, the coefficients can be
interpreted as points of a projective plane.
\begin{definition} \label{dual-plane}
Given the projective plane $ \mathbb{P}(V),$ the dual plane is
defined as $ \mathbb{P}(V^*),$ where $ V^* $ is the dual vector
space of $ V.$ Once a reference frame $ \mathcal{R} = (B) $ is
given in $ \mathbb{P}(V),$ the dual basis $ B^* $ defines the dual
reference frame $ \mathcal{R}^* $ in $ \mathbb{P}(V^*).$ With this
in mind, we set $ \check{\mathbb{P}}^2_{\mathbb{K}} =
\mathbb{P}(V^*),$ and the points of $
\check{\mathbb{P}}^2_{\mathbb{K}} $ are the coefficients of the
lines of $ \mathbb{P}^2_{\mathbb{K}},$ or the lines of $
\mathbb{P}^2_{\mathbb{K}},$ for short.
\end{definition}

As $ (V^*)^* \cong V,$ the dual of the dual projective plane in the initial one.
We can now define the dual curve.
\begin{definition} \label{dual-curve}
Let $ C \subset \mathbb{P}^2_{\mathbb{K}} $ be a reduced and
irreducible curve. The dual curve $ \check{C} \subset
\check{\mathbb{P}}^2_{\mathbb{K}} $ is the unique algebraic curve
that contains the tangent lines at the points of $ C.$
\end{definition}
Assume that $ C $ is smooth of degree $ d.$ Then, $ \check{C} $
has degree $ d(d-1).$ We can assume $ \mathbb{K} = \mathbb{C} $
because the degree does not depend on the ground field. A line in
$ \check{\mathbb{P}}^2_{\mathbb{C}} $ is a point in $
\mathbb{P}^2_{\mathbb{C}}.$ Hence, we have to compute how many
tangent lines to $ C $ pass through the same point $ A(x_0^A:
x_1^A: x_2^a) $ in $ \mathbb{P}^2_{\mathbb{C}}.$ A tangent line
contains $ A $ if, and only if, $ x_0^A f_{x_0} + x_1^A f_{x_1} +
x_2^A f_{x_2} = 0.$ This last curve has degree $ d-1 $ and is
called the polar curve to $ C $ with respect to the pole $ A.$ The
intersection points of $ C $ and the polar curve are $ d(d-1) $ by
B\'{e}zout Theorem, and so the degree of $ \check{C} $ is $
d(d-1),$ as claimed.
\begin{proposition} \label{deg-dual-conic}
Let $ C $ be a smooth curve. Then, $ C $ and $ \check{C} $ have
the same degree if and only if $ C $ is a conic.
\end{proposition}

\noindent\emph{Proof.} The solutions of the equation $ d(d-1) = d
$ are $ d=0 $ and $ d=2,$ and $ d=0 $ cannot be accepted.
\hfill$\square$\vspace{1mm}

\subsection{Hilbert function and the geometry of a set of points}

\begin{definition} \label{coord-ring}
Let $ X \subset \mathbb{P}^2_{\mathbb{K}} $ be an algebraic set.
We set $$ I_X = \{ f \in \mathbb{K}[x_0,x_1,x_2] \ \vert \ f
\mbox{ is homogeneous and } f(P) = 0 \mbox{ for every } P \in X
\}.$$ Moreover, we call $$ S(X) = \frac{\mathbb{K}[x_0, x_1,
x_2]}{I_X} $$ the homogeneous coordinate ring of $ X.$ The
function $$ H_X: t \in \mathbb{Z} \to \dim_{\mathbb{K}} \left(
\frac{\mathbb{K}[x_0, x_1, x_2]}{I_X} \right)_t \in \mathbb{Z} $$
is called the Hilbert function of $ X.$
\end{definition}
The homogeneous elements of $ I_X $ can be interpreted as the
curves that vanish at all the points of $ X.$ Hence, the
homogeneous elements of the quotient ring $ S(X) $ are the curves
that do not vanish at all the points of $ X.$ Finally, if we fix
the degree $ t,$ $ S(X)_t $ is a $ \mathbb{K}$--vector space,
whose dimension is equal to $ \dim_{\mathbb{K}} \left(
\mathbb{K}[x_0, x_1, x_2] \right)_t - \dim_{\mathbb{K}} \left( I_X
\right)_t.$

To illustrate the importance of the Hilbert function of an
algebraic subset, we connect it to some known results. At first,
we recall without proof a general result on the Hilbert function
of a finite set of points.
\begin{theorem} \label{points-on-curves}
Let $ X $ be a finite subset of points, eventually with
multiplicities, and assume that the sum of the multiplicities of
all the points of $ X $ is $ d.$ Then,
\begin{enumerate}
\item $ H_X(t-1) \leq H_X(t) $ for each $ t \geq 1;$ \item if $
H_X(t) = H_X(t+1) $ for a suitable $ t,$ then $ H_X(t) = H_X(t+j)
$ for every $ j \geq 0;$ \item $ H_X(t) = d $ for $ t \geq d-1.$
\end{enumerate}
\end{theorem}

Now, we can state the announced results.
\begin{corollary} \label{uni-conic}
Given five distinct points, there exists at least a conic that
contains them. Moreover, it is unique if, and only if, no four
points of the given five ones are collinear.
\end{corollary}

\noindent\emph{Proof.} Let $ X $ be a set of five distinct points.
From Theorem \ref{points-on-curves}, it follows that $ H_X(t) \leq
5 $ for every $ t,$ and so, in particular, $ H_X(2) \leq 5.$ Then,
$ \dim_{\mathbb{C}} (I_X)_2 = \dim_{\mathbb{C}}
(\mathbb{C}[x_0,x_1,x_2])_2 - H_X(2) \geq 1,$ and the first claim
is proved. Assume now that there are two different conics $ C_1,
C_2 $ through $ X.$ Then, by B\'{e}zout Theorem, the intersection
$ C_1 \cap C_2 $ contains infinitely many points, and so we have $
C_1 = L \cup L_1, C_2 = L \cup L_2,$ with $ L, L_1, L_2 $ lines,
and eventually $ L = L_1,$ or $ L = L_2.$ So, at most one point in
$ X $ is $ L_1 \cap L_2,$ and then at least four ones belong to $
L.$ Conversely, if four points in $ X $ are contained in a line $
L,$ and $ L_1 \cap L_2 $ is the fifth point, then $ L \cup L_1 $
and $ L \cup L_2 $ are two distinct conics containing $ X.$
\hfill$\square$\vspace{1mm}

Similarly, it is possible to prove also results
on the generators of an ideal.
\begin{theorem} \label{ideal-of-points}
Let $ X $ be a set of $ 6 $ distinct points lying on exactly one
conic $ C = V(f).$ Then, $ I_X = \langle f, g \rangle $ where $
V(g) $ is a cubic curve, and $ f, g $ without common factors.

Let $ X $ be a set of $ 5 $ distinct points, lying on exactly one
conic $ C = V(f).$ Then, $ I_X = \langle f, g_1, g_2 \rangle $
where $ g_1, g_2 $ are homogeneous polynomials of degree $ 3,$
such that $ g_1, g_2 $ are linearly independent in $ S(C).$
\end{theorem}

It is possible to prove that $ I_X $ is an ideal in $
\mathbb{K}[x_0, x_1, x_2],$ and it is easy to prove that, if $ X $
and $ Y $ are algebraic sets, then $ I_{X \cup Y} = I_X \cap I_Y.$
Moreover, $ I_X + I_Y \subseteq I_{X \cap Y},$ and it is possible
to prove that $ F \in I_X + I_Y $ if, and only if, $ F \in I_{X
\cap Y} $ under the assumption that the degree of $ F $ is large
enough. The coordinate rings of $ X, Y, X \cup Y$ and $ X \cap Y $
are related each other from the short exact sequence of vector
spaces
\begin{equation} \label{M-V-seq}
0 \to S(X \cup Y)_t \stackrel{\alpha}{\longrightarrow} S(X)_t
\oplus S(Y)_t \stackrel{\beta}{\longrightarrow} \left(
\frac{\mathbb{K}[x_0, x_1, x_2]}{I_X + I_Y} \right)_t \to 0
\end{equation}
where the first linear map $ \alpha $ is defined as $ \alpha(F) =
(F, F),$ and the second linear map $ \beta $ is defined as $
\beta(F,G) = F - G.$ A direct consequence of the exactness of (\ref{M-V-seq}) is that $$ \dim_{\mathbb{K}} \left(
\frac{\mathbb{K}[x_0, x_1, x_2]}{I_X + I_Y} \right)_t = H_X(t) +
H_Y(t) - H_{X \cup Y}(t).$$

\subsection{Topology of real algebraic curves}

The study of  the topological properties of algebraic curves in the real projective
plane in full generality is outside the scope of this appendix,
so we consider only the cases of smooth conic and cubic
curves. In particular we focus on the problem of connected components of a curve with respect to the Euclidean topology. The main result is by Harnack, that found a bound on the number of such connected components.
\begin{theorem} [Harnack's Theorem] \label{harnack}
Let $ C \subset \mathbb{P}^2_{\RR} $ be a smooth algebraic curve.
If $ C $ is a conic, then either $ C $ is empty, or $ C $
is a closed connected curve. If $ C $ is a cubic curve,
then either $ C $ is connected, or $ C $ is the disjoint union of
two connected components.
\end{theorem}

Let $ H $ be a line in $ \mathbb{P}^2_{\RR},$ and let  $
\mathbb{A}^2_{\RR} $ be the complement of $ H.$ Moreover, let $ C
\subset \mathbb{P}^2_{\RR} $ be a smooth conic with a real point,
so that $ C $ is a connected curve. If $ H $ meets $ C $ at two
non--real points ($ H $ is tangent to $ C $ or $ H \cap C $
consists of two real distinct points, respectively), then $ C $ is
an ellipse (a parabola or an hyperbola, respectively).

For cubic curves, the picture is as follows. We call oval each
connected component of $ C.$ Then, $ C $ has either one oval or
two ones. In both cases, one of the two ovals, the only one if $ C
$ is connected, meets all the lines of $ \mathbb{P}^2_{\RR} $ at $
1 $ or $ 3 $ points, counted with their multiplicities, and so it
is called the odd oval $ C_o.$ The second oval, if it exists,
meets all the lines at an even number of points, eventually the
intersection with a line is empty, and so it is called the even
oval $ C_e.$ The oval $ C_o,$ when we restrict to $
\mathbb{A}^2_{\RR},$ is either connected and unbounded (if the
ideal line meets $ C_o $ at $ 1 $ point), or the union of three
unbounded arcs (if the ideal line meets $ C_o $ at $ 3 $ distinct
points). The even oval $ C_e $ does not contain real inflectional
points, i.e. smooth points $ P \in C $ with the property that the
tangent line at $ P $ to $ C $ meets $ C $ at $ P $ with
multiplicity $ 3.$ So, it behaves like conics: if
the ideal line meets $ C_e $ at $ 2 $ non--real points, then $ C_e
$ is topologically like an ellipse, if the ideal line meets $ C_e
$ at two real distinct points, then $ C_e $ is topologically like
a hyperbola, and finally, if the ideal line is tangent to $ C_e,$
then $ C_e $ is topologically like a parabola (we remind that
two curves behave topologically the same if the first
one can be deformed with continuity to the second one).


\renewcommand{\thesection}{C}
\section{An algorithm for the bifurcation curve}

This Appendix lists the source code in Singular
language \cite{DGPS} for computing the Cartesian
equation of the bifurcation curve $ \tilde{E}$ (see
Definition \ref{bif-curve} and Theorem \ref{bif-par}).
It requires specifying the location of the sensors
and assigning the components of the displacement
vectors $\mb{d_{10}}$ and $\mb{d_{20}}$.\\[-2mm]

\tt

\noindent ring r=0,(x,y,z,m1,m2),dp; \\
LIB"linalg.lib";\\
matrix d1[2][1];\\
matrix d2[2][1];\\
poly p0=(m2*m2*transpose(d1)*d1-2*m1*m2*transpose(d1)*d2+m1*m1*transpose(d2)*d2)[1,1];\\
poly q1=(m1*transpose(d2)*d1-m2*transpose(d1)*d1)[1,1];\\
poly p1=q1*q1;\\
poly q2=(m1*transpose(d2)*d2-m2*transpose(d1)*d2)[1,1];\\
poly p2=q2*q2;\\
poly pv=det(concat(d1,d2));\\
poly p3=(p1*p1*transpose(d2)*d2-2*p1*p2*transpose(d1)*d2+p2*p2*transpose(d1)*d1)[1,1];\\
poly den=4*pv*p0*q1*q2*(q1-q2);\\
poly numx=2*q1*q2*(q1-q2)*(p1*d2[2,1]-p2*d1[2,1])+p3*(m2*d1[2,1]-m1*d2[2,1]);\\
poly numy=2*q1*q2*(q1-q2)*(-p1*d2[1,1]+p2*d1[1,1])-p3*(m2*d1[1,1]-m1*d2[1,1]);\\
ideal ii=x-numx, y-numy, z-den;\\
ideal jj=elim(ii,m1*m2);\\
jj=reduce(jj,std(z-1));

\rm


\section*{References}
\bibliographystyle{hplain}   
\bibliography{biblio}       

\end{document}